\documentclass{article}

\usepackage{arxiv}

\usepackage[utf8]{inputenc} % allow utf-8 input
\usepackage[T1]{fontenc}    % use 8-bit T1 fonts
\usepackage{hyperref}       % hyperlinks
\usepackage{url}            % simple URL typesetting
\usepackage{booktabs}       % professional-quality tables
\usepackage{amsfonts}       % blackboard math symbols
\usepackage{nicefrac}       % compact symbols for 1/2, etc.
\usepackage{microtype}      % microtypography
\usepackage{lipsum}		% Can be removed after putting your text content
\usepackage{graphicx}
\usepackage{natbib}
\usepackage{doi}

\usepackage{appendix}
\usepackage{algorithm}
\usepackage{algpseudocode}
\usepackage{pdfpages}
\usepackage{amsmath}
\usepackage{threeparttable} 

\newcommand{\Ima}{\textsc{Image Input}}
\newcommand{\Lo}{\textsc{Paddle\&Ball Position Input}}
\newcommand{\twod}{\textsc{Ball Position Input}}

\title{Biological Neurons Compete with Deep Reinforcement Learning in Sample Efficiency in a Simulated Gameworld}

\date{} 					% Or removing it

\author{  
  Moein Khajehnejad $^*$ \\
  Turner Institute for Brain and Mental Health,\\ Monash University, Clayton, Australia\\
  \And
  Forough Habibollahi \thanks{Equal contribution.}\\
  Cortical Labs, \\
  Melbourne, Australia \\
   \And
  Aswin Paul \\
  Turner Institute for Brain and Mental Health,\\ Monash University, Clayton, Australia\\
  IITB-Monash Research Academy,\\
  Mumbai, India
    \And
    Adeel Razi\\
    Turner Institute for Brain and Mental  Health,\\ Monash University, Clayton, Australia\\
    Wellcome Centre for Human Neuroimaging,\\ University College London, United Kingdom
    \AND
  Brett J. Kagan \thanks{Corresponding author:
  \href{mailto:brett@corticallabs.com}{brett@corticallabs.com}}\\
  Cortical Labs,\\
  Melbourne, Australia
  }

\renewcommand{\headeright}{}
\renewcommand{\undertitle}{}
\renewcommand{\shorttitle}{Biological Neurons Vs. Deep Reinforcement Learning }

\hypersetup{
pdftitle={A template for the arxiv style},
pdfsubject={q-bio.NC, q-bio.QM},
pdfauthor={David S.~Hippocampus, Elias D.~Striatum},
pdfkeywords={First keyword, Second keyword, More},
}

\begin{document}
\maketitle

\begin{abstract}
How do biological systems and machine learning algorithms compare in the number of samples required to show significant improvements in completing a task? We compared the learning efficiency of \textit{in vitro} biological neural networks to the state-of-the-art deep reinforcement learning (RL) algorithms in a simplified simulation of the game `Pong'. Using \textit{DishBrain}, a system that embodies \textit{in vitro} neural networks with \textit{in silico} computation using a high-density multi-electrode array, we contrasted the learning rate and the performance of these biological systems against time-matched learning from three state of the art deep RL algorithms (i.e., DQN, A2C, and PPO) in the same game environment. This allowed a meaningful comparison between biological neural systems and deep RL. We find that when samples are limited to a real-world time course, even these very simple biological cultures outperformed deep RL algorithms across various game performance characteristics, implying a higher sample efficiency. Ultimately, even when tested across multiple types of information input to assess the impact of higher dimensional data input, biological neurons showcased faster learning than all deep reinforcement learning agents.
\end{abstract}

% keywords can be removed
\keywords{In Vitro \and Neural Cultures \and Deep Reinforcement Learning \and Synthetic Biological Intelligence \and Sample Efficiency \and Electrophysiology \and Biocomputing \and Learning \and Intelligence}

\section{Introduction}\label{sec:intro}
Both biological and machine intelligence systems demonstrate the ability to learn and achieve goals. Although the complexity of, and drivers behind, these tasks may differ, comparisons between these types of systems can yield valuable insights \citep{neftci_reinforcement_2019}. Even definitions of what traits artificial intelligence should demonstrate are heavily informed by traits observed in biological intelligence \citep{lake_building_2017, hassabis_neuroscience-inspired_2017}. Yet comparisons between biological and machine intelligence have been notoriously difficult, as the scale of connections in even simple biological organisms far exceeds that found in artificial neural networks or comparable Machine Learning (ML) algorithms \citep{richards_deep_2019,hasson_direct_2020}. However, by taking a system-based approach, we aimed to compare data gathered from a biological neural network (BNN) using the recently validated \textit{DishBrain} system \citep{kagan2022vitro} against time-matched learning from deep reinforcement learning (RL) algorithms - DQN, A2C and PPO. Despite the inherent differences between silicon and biological systems - such as power consumption and network size - this approach makes it possible to explore learning performance and sample efficiency in these different systems. Should compelling differences be found, it would further support extended efforts to understand key differences in the information processing dynamics unique to each system.  

RL has become increasingly popular in the fields of ML and artificial intelligence by offering a way of programming agents through reward and punishment cues without having to specify how the task is to be accomplished. However, to deliver on this promise, formidable computational obstacles must be overcome. RL implies learning the best policy to maximize an expected cumulative long-term reward throughout many steps in order to achieve objectives (goals) \citep{sutton2018reinforcement}. A deep RL approach integrates artificial neural networks with an RL framework that helps the system to achieve its goals \citep{hessel2018rainbow}. It maps states and actions to the rewards they bring, combining function approximation and target optimization. Reinforcement algorithms that incorporate deep neural networks have been developed to beat human experts in multiple game settings including: poker \citep{moravvcik2017deepstack}, multiplayer contests \citep{jaderberg2018human}, complex board games, including go and chess \citep{silver2017mastering,silver2017mastering2,silver2018general} and numerous Atari video games \citep{mnih2015human}. Nevertheless, RL still faces real challenges including but not limited to: complexities in the selection of hyper-parameters and reward structure, sample inefficiency \citep{tsividis2017human,marcus2018deep}, reproducibility issues \citep{gibney2020ai}, and catastrophic forgetfulness \citep{kirkpatrick_overcoming_2017, fan_fragility_2022}. Furthermore, to allow RL algorithms to train quickly requires considerable levels of computing power \citep{mousavi2018deep} with notable associated environmental impacts \citep{freitag_real_2021}. Finally, RL algorithms are typically trained for narrow tasks in static environments; where training and performance phases are separate \citep{neftci_reinforcement_2019, fan_fragility_2022}. 

Holistically, these traits suggest that although deep RL algorithms are highly functional, their learning mechanisms almost certainly differ fundamentally from biological learning\citep{marcus2018deep, neftci_reinforcement_2019, whittington_theories_2019}. It is noted that RL as a mechanism has been found to elicit rapid and adaptable learning in animals \citep{hamid_mesolimbic_2016, costa_amygdala_2016}. Yet it seems unlikely that similar underlying statistical mechanisms that support RL, such as back-propagation and gradient descent, have biological parallels in the brain \citep{whittington_theories_2019, friston_reinforcement_2009}. Ultimately, these mechanisms are likely too inefficient to be accepted as plausible models of human learning \citep{tsividis2017human, song2022inferring, whittington2017approximation}. This is especially true when considering how intelligence may arise from cells without established pathways of motivation.  Early work investigating how cells respond to stimulation that can be modified through their own activity showed rapid adaptation displayed through synaptic plasticity \citep{tessadori_modular_2012,bakkum_spatio-temporal_2008, muller2013sub}. Furthermore, it was recently demonstrated that by using electrophysiological stimulation and recording in a real-time closed-loop system with a monolayer of living biological neurons, biological neural cells could be trained to significantly improve performance in a simulated \textit{`pong'} game-world  \citep{kagan2022vitro}. 
The question arises as to whether the observed performance in these simple BNNs is notable compared to that of RL at the same task, especially regarding sample efficiency.

\begin{figure*}[!ht]
  \centering
  \hspace{-1cm}
  {\includegraphics[width=0.9\textwidth]{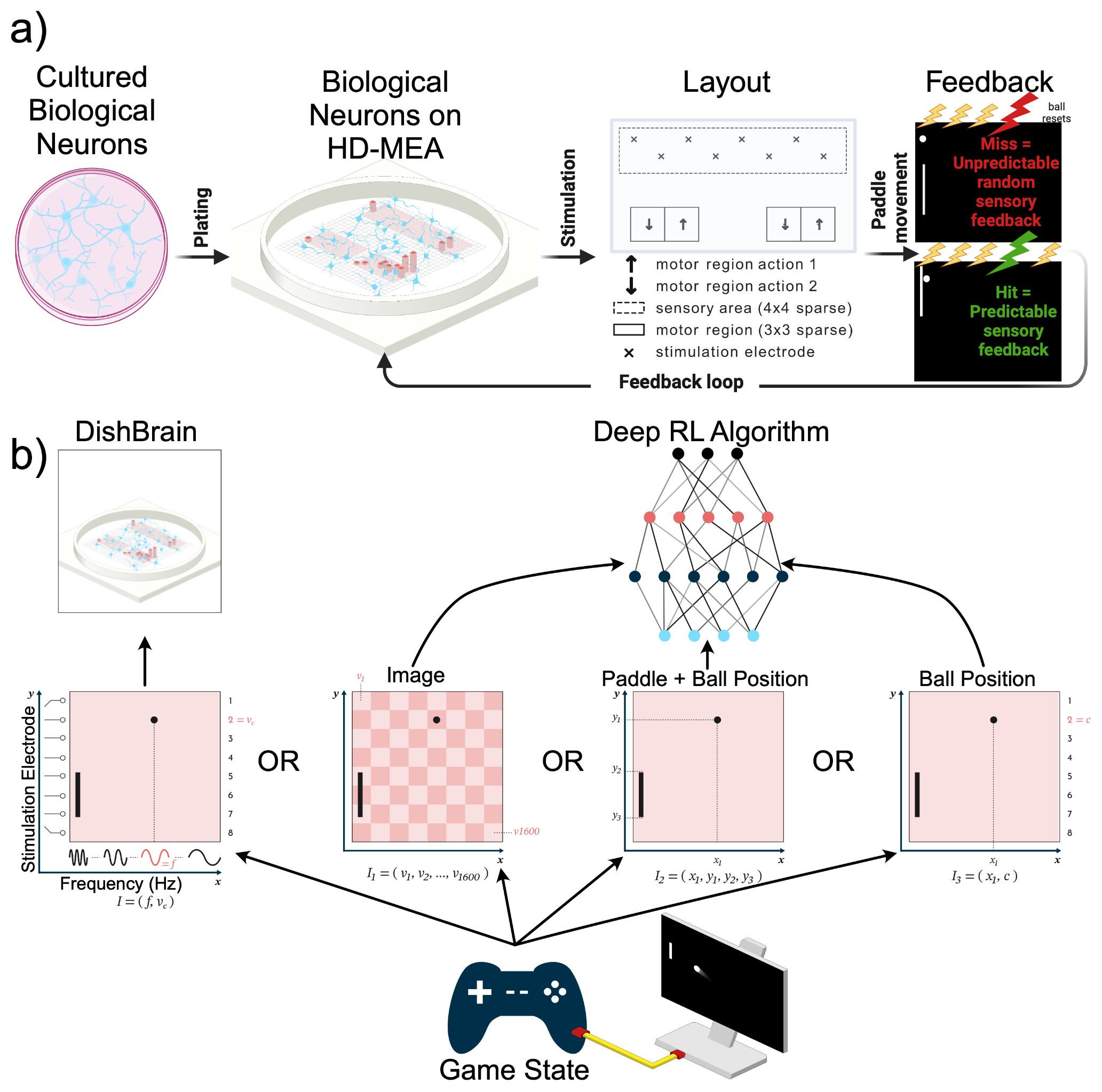}}
  \caption{\textbf{DishBrain system and Various input designs to RL algorithms.} \textbf{a)} \textit{DishBrain} feedback loop setup and Electrode configuration and predefined sensory and motor regions. Figures adapted and modified from \citep{kagan2022vitro}. \textbf{b)} Schematic comparing the information input routes in the \textit{DishBrain} system (left) and the three implementations of the deep RL algorithms (right). In each design, the input information to the computing module (deep RL algorithms or \textit{DishBrain}) is denoted by a vector $I$.}
  \label{fig:dish_feed&diagram}
\end{figure*}

\textit{DishBrain} is a novel system shown to display simple biological intelligence by harnessing inherent adaptive properties of neurons. In \textit{DishBrain}, \textit{in vitro} neuronal networks are integrated with \textit{in silico} computing via high-density multi-electrode arrays (HD-MEAs). These cultured neuronal networks showcase biologically-based adaptive intelligence within a simulated gameplay environment in real-time through closed-loop stimulation and recordings \citep{kagan2022vitro}. Specifically, BNNs exhibited self-organised adaptive electrophysiological activity that was consistent with an innate ability to learn and showcase an intelligent response to limited - although biologically plausible \citep{harrell_elaborate_2020} - structured external information. Data was generated from cortical cells obtained from either embryonic rodent or human induced pluripotent stem cell (hiPSC) sources. These cell types were compared to assess reproducibility of learning effects across species and preparations. Here, we investigate whether these elementary learning systems achieve performance levels that can compete with state-of-the-art deep RL algorithms. Additionally, by varying the input information density presented during training of the RL algorithms, we can determine the impact of information sparsity and ensure suitable comparisons to the biological system. This is the first comparison between a Synthetic Biological Intelligence (SBI) system \citep{kagan2023technology} and state-of-the-art RL algorithms. Figure \ref{fig:dish_feed&diagram}.a,b illustrate the input information, feedback loop setup, and electrode configurations in the \textit{DishBrain} system. This research aims to investigate whether simple biological systems can demonstrate characteristics compared to established RL methods to justify further research in this area, either where SBI systems are standalone learning devices, or inform further algorithm development in the ML space. We anticipate that SBI systems will exhibit greater sample efficiency than RL models, as suggested by prior research. However, this entails constraining training to a real-time approximate sample count for RL algorithms. Figure \ref{fig:dish_feed&diagram}.c illustrates the comparison between input information in the \textit{DishBrain} system and deep RL algorithms. \\

% \begin{figure*}[!ht]
%   \centering
%   \hspace{-1cm}
%   {\includegraphics[width=1.1\textwidth]{Figures/diagram.pdf}}
%   \caption{\textbf{Various input designs to RL algorithms.} Schematic comparing the information input routes in the \textit{DishBrain} system (bottom) and the three implementations of the deep RL algorithms (top). In each design, the input information to the computing module (deep RL algorithms or \textit{DishBrain}) is denoted by a vector $I$.}
%   \label{fig:diagram}
% \end{figure*}

\section{Results}
Game performance of human cortical cells (HCCs; 174 sessions) and mouse cortical cells (MCCs; 110 sessions) was compared with three RL baseline methods. To determine how learning arises both in cultures and in baseline methods, three key gameplay characteristics were examined. These include: mean hit-to-miss ratio (average hits-per-rally), number of times the paddle failed to intercept the ball on the initial serve (aces), and number of long rallies or episodes ($\geq$~3 consecutive hits). 

\begin{figure*}[htbp]
  \centering
%   \hspace{-0.6cm}
  {\includegraphics[width=0.89\textwidth]{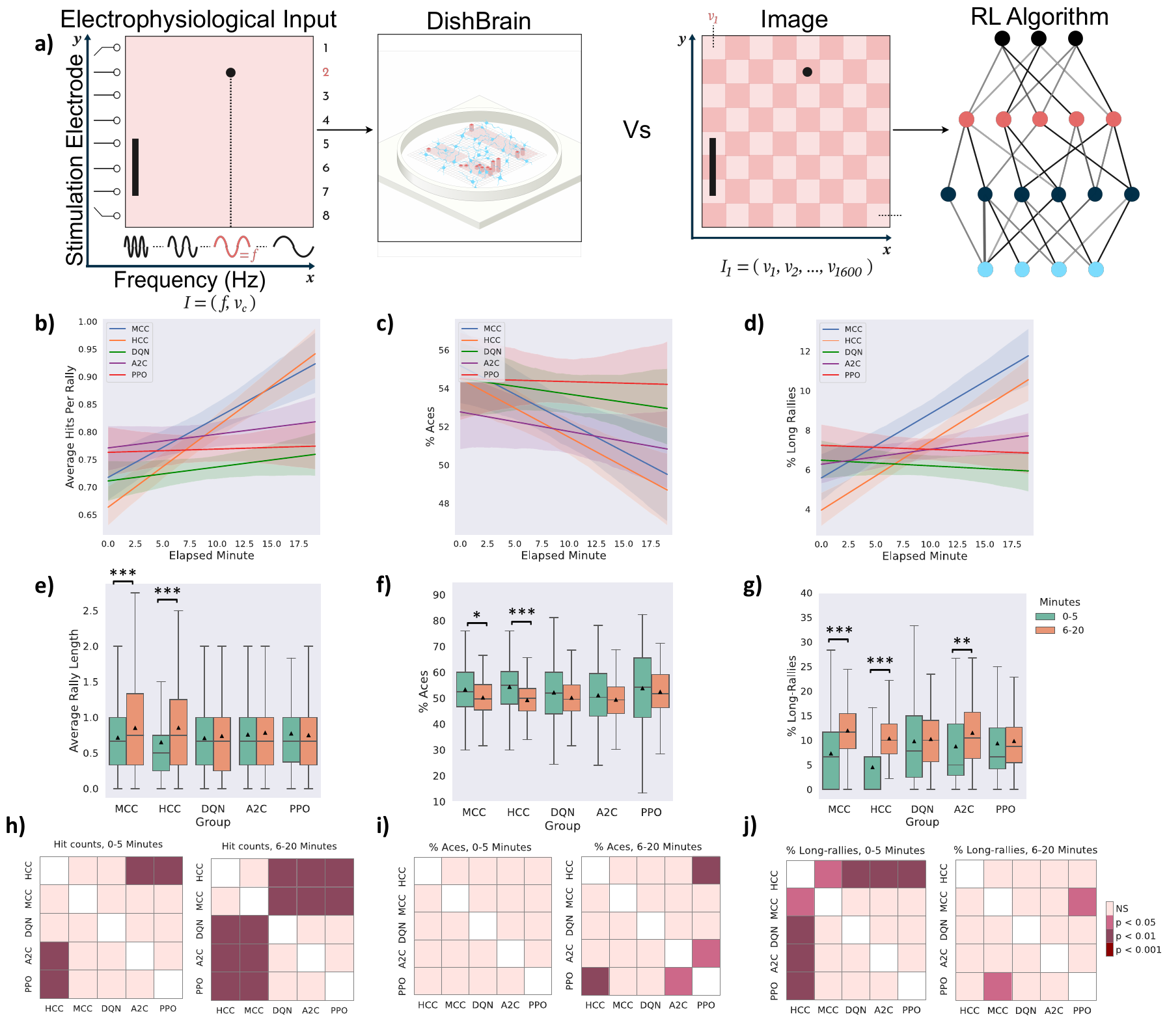}}
  \caption{ \textbf{\Ima{} to the deep RL algorithms.} \textbf{a)} Schematic highlighting figure comparisons are between biological DishBrain system and an pixel-based information input to te RL algorithms. Average number of  \textbf {b)} hits-per-rally, \textbf {c)} $\%$ of aces, and \textbf {d)} $\%$ of long rallies over 20 minutes real-time equivalent of training DQN, A2C, PPO, and MCC, HCC cultures. A regressor line on the mean values with a 95\% confidence interval highlights the learning trends. Comparing the performance amongst all groups, the highest level of average hits-per-rally is achieved by the neuronal MCC and HCC cultures while PPO is outperformed by all the opponents. The average $\%$ of aces is lowest for the neuronal cultures compared to all deep RL baseline methods. The average $\%$ of long rallies reaches its highest levels for MCC and HCC.
  {\bf e)} Average performance of groups over time. Only biological cultures have significant within-group improvement and increase in their performance at the second time interval (One-way ANOVA test, p = 5.854e-6, p = 7.936e-17, for MCC and HCC respectively; p = 0.231, p = 0.318, and p = 0.400 for DQN, A2C, and PPO respectively). {\bf f)}  Average \% of aces within groups and over time. Only MCC and HCC (One-way ANOVA test, p = 0.014, p = 2.907e-08, respectively) differed significantly over time. No significant change was detected within the DQN, A2C, or PPO groups (One-way ANOVA test, p = 0.080, p = 0.195, and p = 0.308, respectively). {\bf g)} Average \% of long-rallies ($\geq$~3) performed in a session. All groups showed an increase in the average number of long rallies where this within-group increase was significant only for MCC, HCC, and A2C (One-way ANOVA test, p = 1.172e-7, p = 1.525e-24 for MCC and HCC, respectively and p = 0.605, p = 0.002, and p = 0.684 for DQN, A2C, and PPO, respectively). *$p < 0.05$, **$p<0.01$, and ***$p < 0.001$. {\bf h)} Pairwise Tukey's post-hoc test shows that HCC and MCC groups significantly outperform PPO, A2C, and DQN in the last 15 minutes interval.
  {\bf i)} Using pairwise Tukey's post-hoc test, the HCC group significantly outperforms the PPO in the last 15 minutes interval with a lower average of \% Aces. A2C also outperforms PPO in this time interval. {\bf j)} Pairwise comparison using Tukey's test only shows a significant difference in the percentage of long rallies between HCC and the rest of the groups in the first 5 minutes. However, this is later altered in the direction of all groups having an increased \% of long rallies with MCC outperforming PPO in the last 15 minutes of the game. Box plots show interquartile range, with bars demonstrating 1.5X interquartile range, the line marks the median and the black triangle marks the mean. Error bands = 1 SE}
  \label{fig:Image}
\end{figure*}

For comparison, every 70-episode run of each RL algorithm was mapped to approximately 20 real-time minutes by normalizing the actual total length of each run in minutes and then multiplying by 20. This approximates the number of rallies biological cultures would experience in a 20-minute session. Details of the implemented RL algorithms and information about the selected hyper-parameters are included in \nameref{supp_mat} \ref{RL}.
Figures \ref{fig:Image}, \ref{fig:LocationVec}, and \ref{fig:2dInput} represent the main findings for comparisons between biological cultures and the \Ima{}, \Lo{}, and \twod{} designs of the RL methods. The intent behind different input designs was to determine whether varying the amount of information input into the algorithm altered sample efficiency and learning characteristics of these systems. In particular, the \Lo{}, and \twod{} methods were intended to be more accurate comparisons to the information density presented to the \textit{DishBrain} system. \nameref{ex_data} Tables S3 and S4 present all multivariate statistical tests performed in relation to the following results. All \textit{post-hoc} follow-up tests are presented in \nameref{ex_data} Table S2. 

\subsection{Comparison in performance between \textit{DishBrain} and three RL algorithms with various information densities}
\label{comparisons}
\begin{figure*}[htbp]
  \centering
  \vspace{-1.5cm}
  {\includegraphics[width=0.85\textwidth]{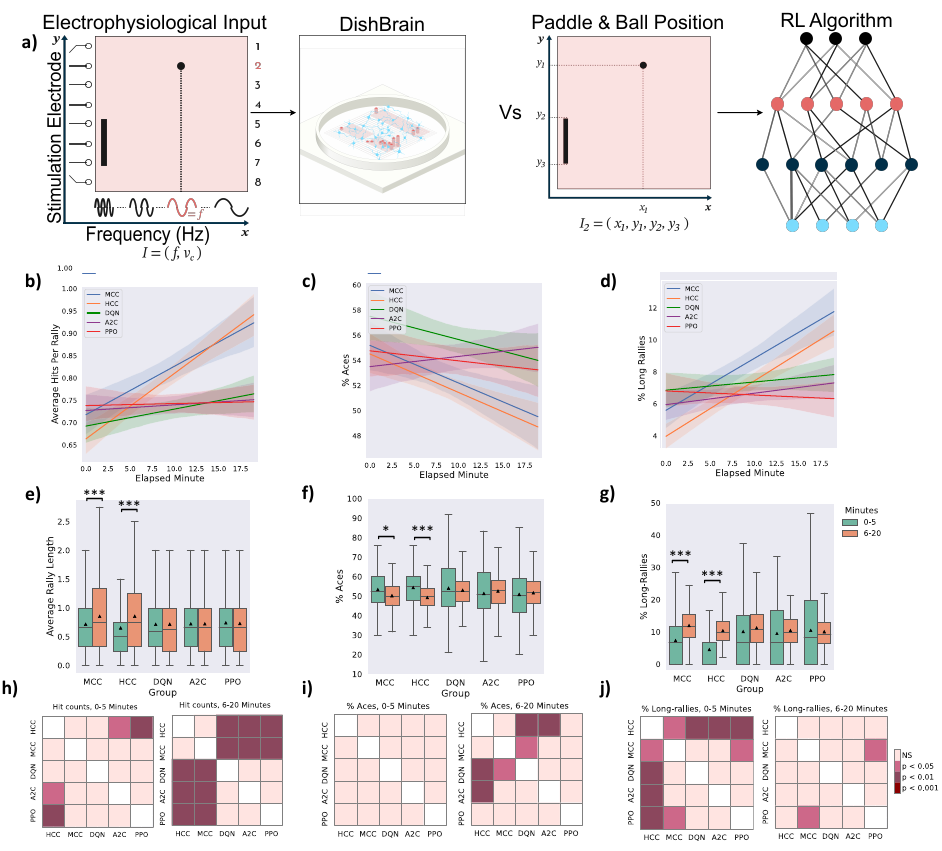}}
  \caption{\textbf{\Lo{} to the deep RL algorithms.} \textbf{a)} Schematic highlighting figure comparisons are between biological DishBrain system and \Lo{} to RL algorithms. Average number of  \textbf {b)} hits-per-rally, \textbf {c)} $\%$ of aces, and \textbf {d)} $\%$ of long rallies over 20 minutes real-time equivalent of training DQN, A2C, PPO, and MCC, HCC cultures. A regressor line on the mean values with a 95\% confidence interval highlights the learning trends. The highest level of average hits-per-rally is achieved by the MCC and HCC cultures. The average $\%$ of aces is lowest for the neuronal cultures compared to all deep RL baseline methods. The average $\%$ of long rallies reaches its highest levels for MCC and HCC. 
  {\bf e)} Average rally length over time only showed a significant increase in the biological cultures between the two time intervals (One-way ANOVA test, p = 0.913, p = 0.958, and p = 0.610 for DQN, A2C, and PPO respectively). {\bf f)}  Average $\%$ of aces within groups and over time only showed a significant difference in the MCC and HCC  groups. No significant change was detected within the DQN, A2C, or PPO groups (One-way ANOVA test, p = 0.463, p = 0.338, and p = 0.544 respectively). {\bf g)} Average $\%$ of long-rallies ($\geq$~3) performed in a session increased in the second time interval in all groups. This within-group difference was only significant for the MCC and HCC groups (One-way ANOVA test, p = 1.172e-7, p = 1.525e-24, p = 0.233, p = 0.320, and p = 0.650 for MCC, HCC, DQN, A2C, and PPO, respectively). *$p < 0.05$, **$p<0.01$, and ***$p < 0.001$. {\bf h)} Pairwise Tukey's post-hoc test shows that the HCC group is significantly outperformed by A2C and PPO in the first 5 minutes in terms of the hit counts. Biological cultures, however, do significantly better compared to all deep RL opponents in the 15 minutes interval.
  {\bf i)} Using pairwise Tukey's post-hoc test, HCC group significantly outperforms the DQN and A2C groups in the last 15 minutes interval with a lower average of \% Aces. DQN is also outperformed by the MCC group in this time interval. {\bf j)} Pairwise comparison using Tukey's test shows a significant difference in the percentage of long rallies between HCC and the rest of the groups in the first 5 minutes all outperforming the HCC. However, this is later altered in the last 15 minutes with only MCC outperforming PPO significantly having an increased \% of long rallies. Box plots show interquartile range, with bars demonstrating 1.5X interquartile range, the line marks the median, and the black triangle marks the mean. Error bands = 1 SE}
  \label{fig:LocationVec}
\end{figure*}

\begin{figure*}[htbp]
  \centering
%   \hspace{-0.6cm}
  {\includegraphics[width=0.89\textwidth]{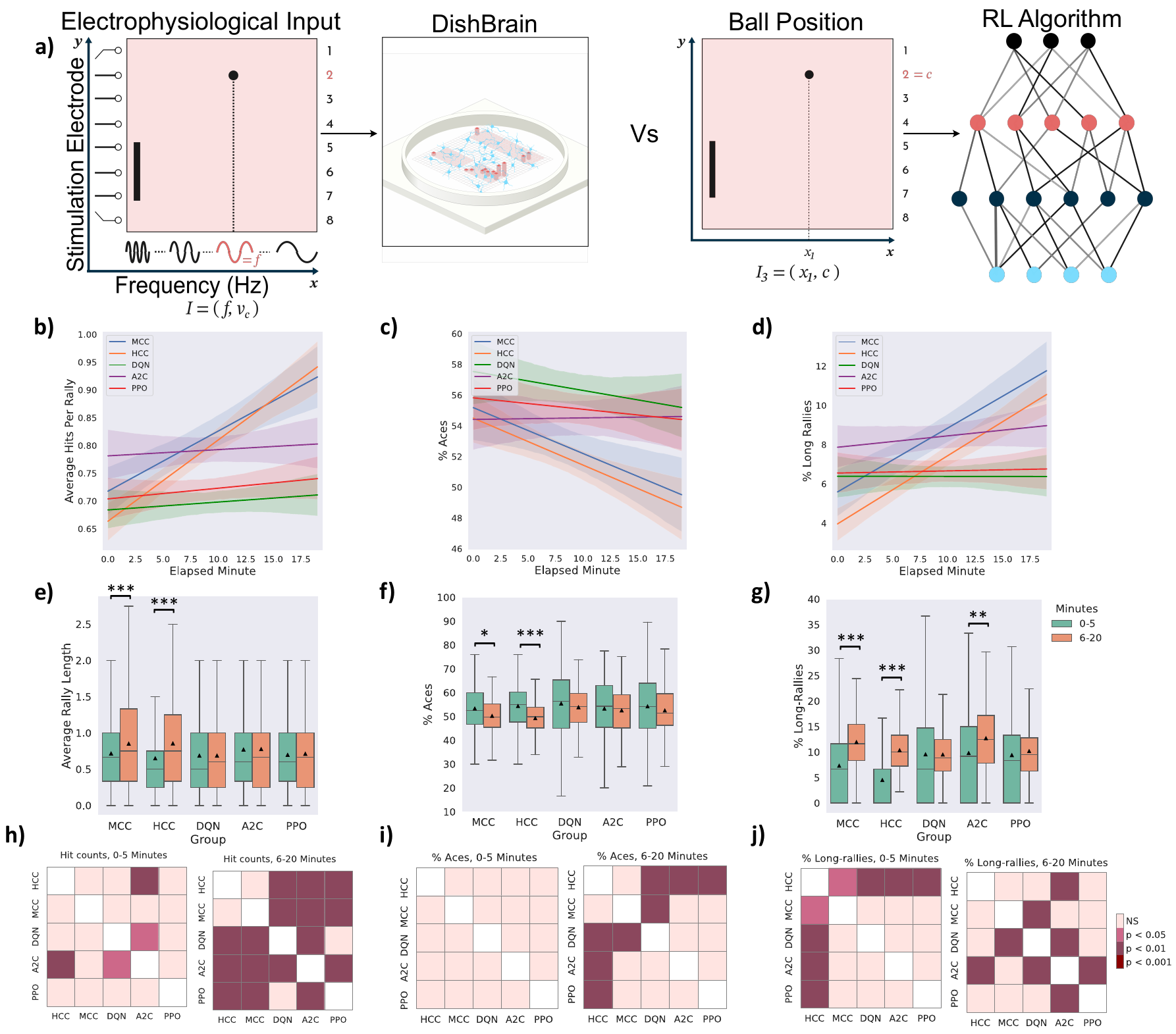}}
  \caption{\textbf{\twod{} to the deep RL algorithms.} \textbf{a)} Schematic highlighting figure comparisons are between biological DishBrain system and \twod{} to RL algorithms. Average number of  \textbf {b)} hits-per-rally, \textbf {c)} $\%$ of aces, and \textbf {d)} $\%$ of long rallies over 20 minutes real-time equivalent of training DQN, A2C, PPO, and MCC, HCC cultures. A regressor line on the mean values with a 95\% confidence interval highlights the learning trends. The highest level of average hits-per-rally is achieved by the neuronal MCC and HCC cultures. The average $\%$ of aces is lowest for the neuronal cultures compared to all deep RL baseline methods. The average $\%$ of long rallies reaches its highest levels for MCC and HCC. Comparing to the same findings for the HCC and MCC groups,
  {\bf e)} average rally length over time only showed a significant increase in the biological cultures between the two time intervals (One-way ANOVA test, p = 0.995, p = 0.812, and p = 0.547 for DQN, A2C, and PPO respectively). {\bf f)}  Average \% of aces within groups and over time only showed a significant difference in the MCC and HCC  groups. No significant change was detected within the DQN, A2C, or PPO groups (One-way ANOVA test, p = 0.241, p = 0.581, and p = 0.216 respectively). {\bf g)} Average \% of long-rallies ($\geq$~3) performed in a session increased in the second time interval in all groups except DQN. This within-group difference was only significant for MCC, HCC, and A2C groups with p = 0.002 for the A2C group. *$p < 0.05$, **$p<0.01$, and ***$p < 0.001$. {\bf h)} Pairwise Tukey's post-hoc test shows that biological cultures significantly outperform all deep RL groups in the last 15 minutes in terms of the hit counts or rally length.
  {\bf i)} Using pairwise Tukey's post-hoc test, the HCC group significantly outperforms all the deep RL groups in the last 15 minutes interval while MCC also outperforms DQN with a lower average of \% Aces. {\bf j)} Pairwise comparison using Tukey's test shows a significant out-performance of all groups over HCC in the percentage of long rallies in the first 5 minutes. In the second time interval, MCC shows a significantly higher $\%$ of long rallies compared to DQN with HCC now being outperformed only by A2C. Box plots show interquartile range, with bars demonstrating 1.5X interquartile range, the line marks the median and the black triangle marks the mean. Error bands = 1 SE}
  \label{fig:2dInput}
\end{figure*}

In all three designs, biological cultures (i.e. HCC and MCC) outperform all RL baseline algorithms (see Subfigures \ref{fig:Image}.b, \ref{fig:LocationVec}.b, and \ref{fig:2dInput}.b) in terms of the highest level of average hits-per-rally achieved. The cultures demonstrate faster learning rates over time. Subfigures \ref{fig:Image}.c, \ref{fig:LocationVec}.c, and \ref{fig:2dInput}.c compare the $\%$ of aces among the biological cultures and the RL groups given the three different designs. HCC and MCC achieve the lowest percentage of aces compared to the deep RL algorithms in Subfigure \ref{fig:Image}.c and the other RL baseline designs in Subfigures \ref{fig:LocationVec}.c, and \ref{fig:2dInput}.c. The increasing trend in $\%$ of long rallies is observed in all groups and among all designs except the DQN and PPO groups in the \Ima{} design and PPO in the \Lo{} design, as illustrated in Subfigures \ref{fig:Image}.d, \ref{fig:LocationVec}.d, and \ref{fig:2dInput}.d. Average $\%$ of long rallies was highest for MCC and HCC compared to RL baselines. 

Key activity metrics in the first 5 minutes versus the last 15 minutes in each session were compared to identify any significant improvement occurring in the learning process within each group. 

Panel (e) in Figures \ref{fig:Image}, \ref{fig:LocationVec}, and \ref{fig:2dInput} compares average rally length between the two defined time intervals within groups. The results imply that the within-group increasing trend in rally length is significant only in the biological groups. 

Panel (f) in Figures \ref{fig:Image}, \ref{fig:LocationVec}, and \ref{fig:2dInput} represents the change in average percentage of aces over time. A significant decrease in number of aces (where the ball was missed immediately in an episode with no accurate hits) implies an improved game performance.  Only MCC and HCC had a significant decrease in average ace percentage as opposed to the rest of RL based algorithms with different input designs. 

Panel (g) in Figures \ref{fig:Image}, \ref{fig:LocationVec}, and \ref{fig:2dInput}, shows that the percentage of long rallies in the first 5 minutes versus the last 15 minutes only significantly increased for biological cultures and A2C with the \Ima{} and \twod{} designs.

% which showed no significant difference in \% of aces between any pairs of groups in the first 5 minutes. HCC and MCC outperformed DQN in the last 15 minutes of the experiments (see 
% Pairwise Tukey's post hoc test was performed between groups for both time intervals in Figure \ref{fig:5vs15}.e which suggests an outperformance (i.e. higher percentage of long rallies) of all groups over HCC in the first 5 minutes. However, the increasing trend of \% long rallies in all the groups alters the results in the last 15 minutes. The results show that in the 6-15 minutes interval, only PPO and could outperform the biological cultures significantly in terms of \% of long rallies.
Inter-group comparison was carried out for both time intervals (0-5 and 6-20 minutes) and all three metrics using Tukey's post-hoc test as represented in panels (h), (i), and (j) in Figures \ref{fig:Image}, \ref{fig:LocationVec}, and \ref{fig:2dInput} for rally length (i.e. hit counts), \% of aces, and \% of long rallies respectively. 
% \nameref{ex_data} Table S3 presents all multivariate statistical tests performed in these figures while all \textit{post-hoc} follow-up tests in these figures are presented in \nameref{ex_data} Table S2.

Note, in the \Ima{} design, where average rally length of deep RL methods comes closest to the biological cultures, the input information density is starkly different between the two groups. While RL agents received pixel data with a density of 40 $\times$ 40 pixels, biological cultures only receive input from 8 stimulation points with a given integer rate code of 4Hz–40Hz, highlighting important efficiency differences in informational input between these learning systems. The possibility of higher input information dimensionality having adverse effects on overall sample efficiency of RL algorithms is further nullified by evaluating two alternative input structures (\Lo{} and \twod{} designs).

\subsection{Examining impact of paddle movement speed on learning rates}
\begin{figure*}[htbp]
  \centering
%   \hspace{-0.6cm}
  {\includegraphics[width=1\textwidth]{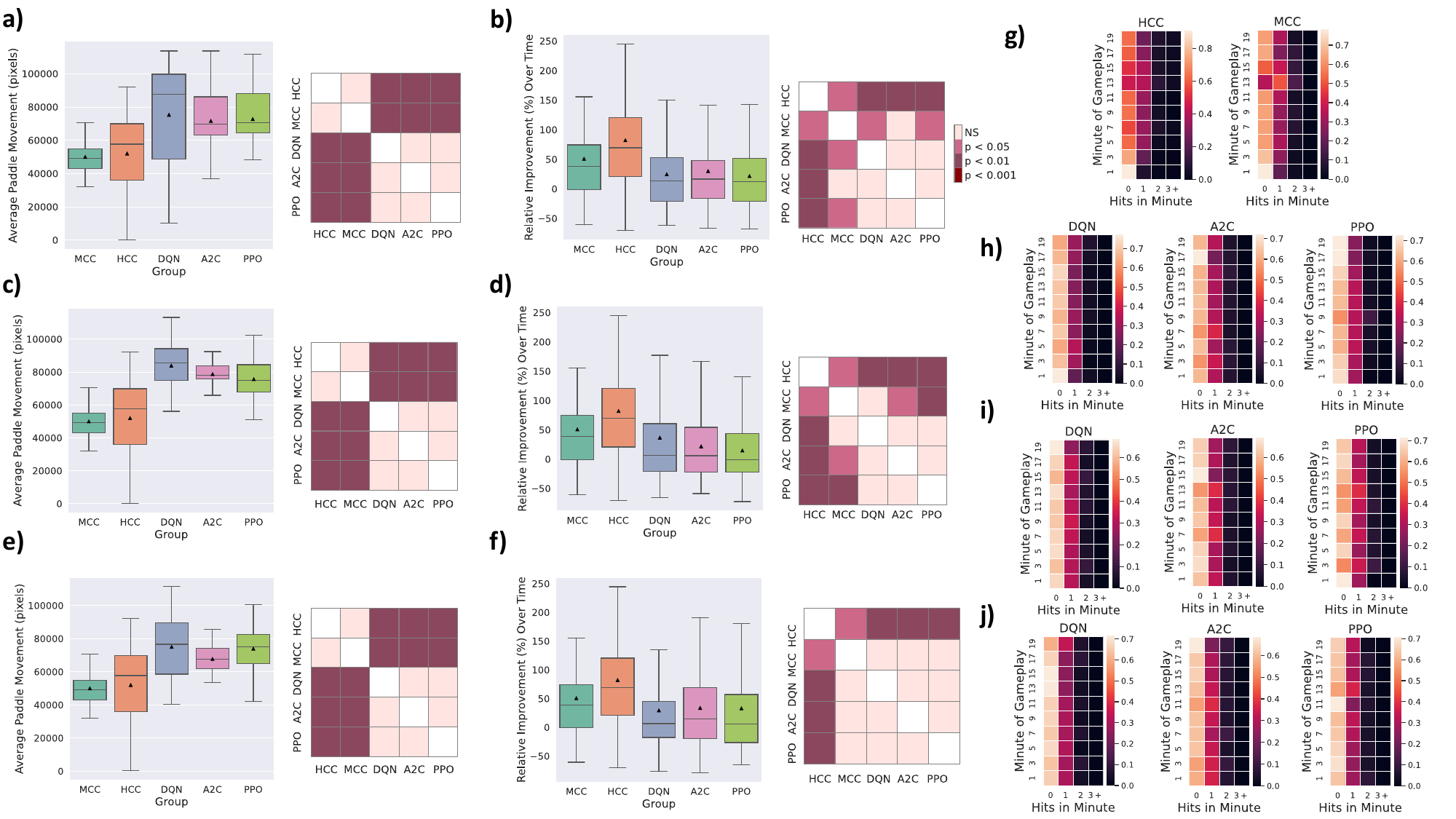}}
  \caption{\textbf{Paddle movement and relative improvement.} The average paddle movement in pixels in all the different groups for the \textbf{a)} \Ima{}, \textbf{c)} \Lo{}, and \textbf{e)} \twod{} to the deep RL algorithms. Tukey's post-hoc test was conducted showing that DQN, PPO, and A2C had a significantly higher average paddle movement compared to HCC and MCC in all scenarios. Relative improvement (\%) in the average hit counts between the first 5 minutes and the last 15 minutes of all sessions in each separate group for the \textbf{b)} \Ima{}, \textbf{d)} \Lo{}, and \textbf{f)} \twod{} to the deep RL algorithms. The biological groups show higher improvements with HCC outperforming all. \textbf{b)} Using Games Howell post-hoc test, the inter-group differences were significant with HCC outperforming all other groups, as well as MCC significantly outperforming PPO. \textbf{d)} HCC showed a significantly higher relative improvement compared to all the other groups while MCC also outperformed A2C and PPO in terms of relative improvement over time. \textbf{f)} Finally, HCC could still perform significantly better than all the deep RL groups with the \twod{} design to the deep RL algorithms with MCC outperforming PPO and DQN in this design. Distribution of frequency of mean summed hits per minute amongst groups for \textbf{g)} biological cultures and  deep RL algorithms with \textbf{h)} \Ima{}, \textbf{i)} \Lo{}, and \textbf{j)} \twod{}. 
%   The performance achieved by both MCC and HCC is of the same level as the deep RL baselines.
}
  \label{fig:paddle_improv}
\end{figure*}

To account for potential effects of paddle movement speed and whether it plays an important role in determining the success rate of paddle control, we derived the average paddle movement (in pixels) for all groups. Subfigures \ref{fig:paddle_improv}.a,c, and e represent these results for the \Ima{}, \Lo{}, and \twod{} designs, respectively. Using Tukey's post-hoc tests, a consistently significant difference between pairs of DQN, PPO or A2C with MCC or HCC was found in terms of average paddle movement, with RL algorithms having the higher average. This occurs when all the RL algorithms with different input designs have significantly higher average paddle movement compared to both groups of biological cultures.
As per previous findings \citep{kagan2022vitro}, increased paddle movement speed in RL algorithms does not translate to improved game performance, likely suggesting a more stochastic paddle control.

Subfigures \ref{fig:paddle_improv}.b, d, and f compare relative improvement in performance between biological cultures and RL algorithms for \Ima{}, \Lo{}, and \twod{}, respectively. This measure identifies the relative increase in average accurate hit counts in the second 15 minutes of the game compared to the first 5 minutes. The HCC group shows the highest improvement in time. Post-hoc tests showed significant differences between HCC and all the RL methods across all of the three different input designs. The MCC group also outperforms PPO in both \Ima{} and \Lo{} designs as well as DQN and A2C in the \Ima{} and \Lo{} designs, respectively. 
% \nameref{ex_data} Table S4 presents full details on all multivariate statistical tests performed in relation to these figures while all details for \textit{post-hoc} follow-up tests related to this figure are presented in \nameref{ex_data} Table S2.

Subfigures \ref{fig:paddle_improv}.g, h, i, and j compare frequency tables for distributions of mean summed hits per minute amongst groups for the \Ima{}, \Lo{}, and \twod{} designs respectively. These tables were not significantly different (Two-sample \textit{t}-test).

\begin{figure*}[!ht]
  \centering
  {\includegraphics[width=0.9\textwidth]{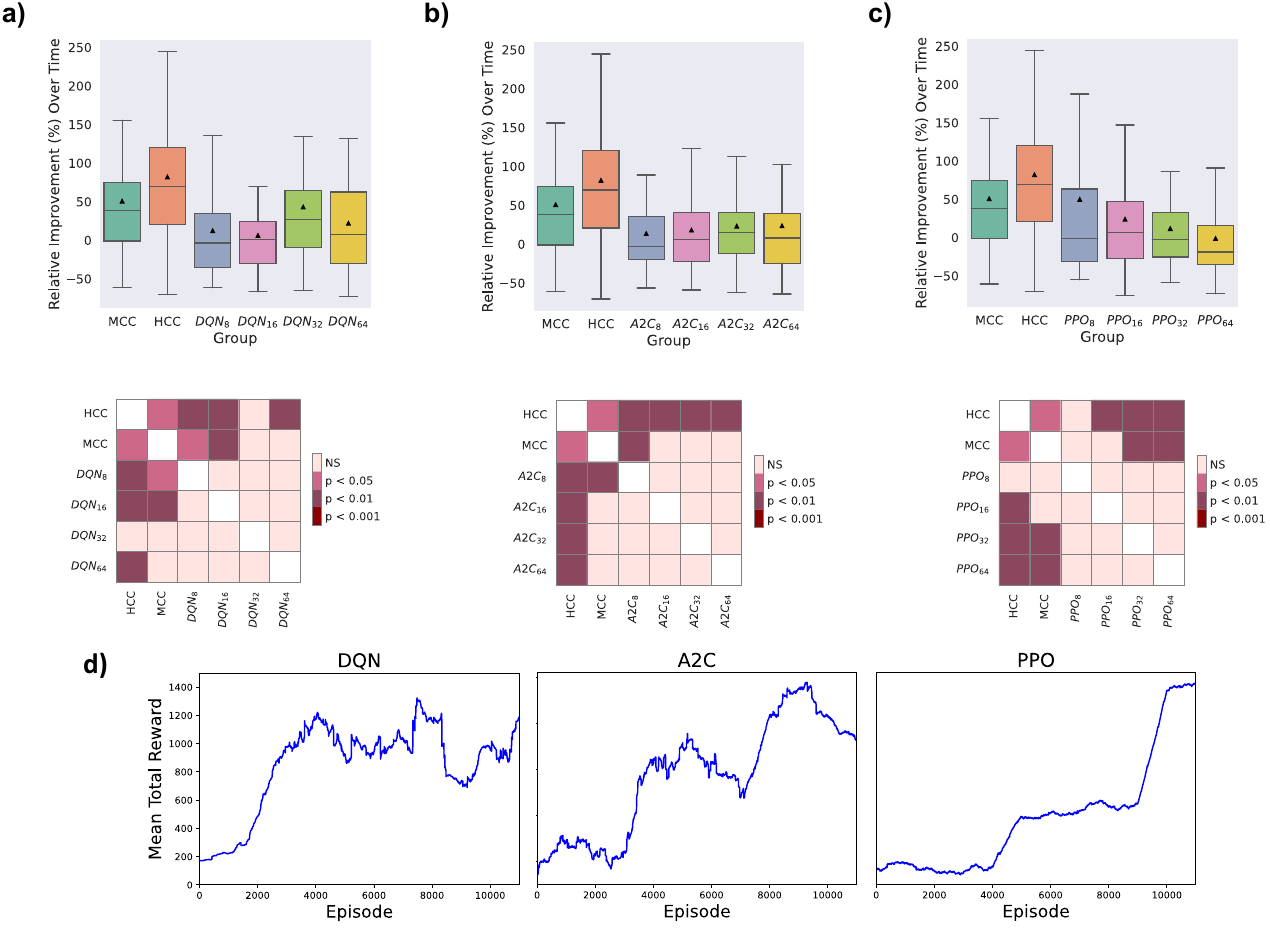}}
  \caption{\textbf{Batch size effect and extended training epochs for the RL algorithms with \Ima{}.} Relative improvement (\%) in the average hit counts between the first 5 minutes and the last 15 minutes of all sessions as well as the post-hoc tests in each separate group for batch sizes of 8, 16, 32, and 64 in the \textbf{a)} DQN, \textbf{b)} A2C, and \textbf{c)} PPO groups compared to biological cultures. Games Howell post-hoc tests show the inter-group differences which were not significant between any pair of different batch sizes for any of the DQN, A2C, or PPO groups. \textbf{d)} Extended training episodes for the deep RL algorithms. The plots show a moving average of the total episode reward with a window size of 100. }
  \label{fig:hyperparams}
  % \label{fig:long_runs}
  % \label{fig:Rel-batchsizes}
\end{figure*}

Details of the implemented RL algorithms and hyper-parameters can be found in the data repository provided in Section \ref{code}. 
In summary, a comprehensive grid search was conducted within the parameter space of \textit{learning rate}, \textit{replay buffer size}, and the training \textit{batch size} aiming to identify the optimal parameter configuration and it was found that similar results were obtained across a variety of hyper-parameters, strongly supporting the initial conclusions of this work.
Here, we include the example of how increasing the \textit{batch size} affects the overall performance of the RL algorithms, while keeping the rest of the parameters set to their initial values in the search space (For further details and exploration of selected hyper-parameters, see \nameref{supp_mat} \ref{hyper-p}, \ref{RL} and \nameref{ex_data} Figures \ref{fig:hp_exp}, \ref{fig:DQN_batchsizes}, \ref{fig:A2C_batchsizes}, \ref{fig:PPO_batchsizes}, \ref{fig:Hidden}, and \ref{fig:Rel-hiddenlayers}). In general, we observed some quantitative changes in outcome metrics when varying the batch size for these algorithms, but these adjustments did not alter the ultimate conclusions of our work. Focusing on the quality of learning in each group and the comparison of sample efficiency, both of these were unaffected or in some cases worsened by increasing the batch size. Specifically, when examining the changes in average accurate hit counts during the first 5 minutes versus the last 15 minutes of training and the overall relative improvement, the increased batch size did not appear to significantly impact the resulting sample efficiency in any of the algorithms as seen in Figure \ref{fig:hyperparams}.a-c.\\
Finally, Figure \ref{fig:hyperparams}.d illustrates the mean total reward of the RL algorithms with the \Ima{} using the same hyper-parameter sets as in Figure \ref{fig:Image} for an extended training period of 11000 game episodes. These results reveal that all three algorithms successfully learned and showed improved performance across an extended number of training episodes. Training the implemented deep RL algorithms for 11,000 game episodes using the same set of hyper-parameters as in Section \ref{comparisons} illustrates an increasing trend in their performance and high levels of total reward, measured as episode duration in terms of the number of frames to complete the episode. However, as anticipated, 70 game episodes—the same number used to train the biological cultures—were not sufficient for any of the algorithms.

\section{Discussion}
The advantages and disadvantages of biological versus machine intelligence are often discussed, yet technical limitations have prevented meaningful comparisons in terms of performance. In this work, we compare performance of biological neuronal networks with that of state-of-the-art deep reinforcement algorithms (deep RL). Using a controllable game environment of a simplified \textit{pong} simulation, it was possible to compare key traits between these different learning systems, with a focus on sample efficiency. Human or mouse cortical cells (HCC or MCC) along with three deep RL algorithms (DQN, A2C, and PPO), were compared in sessions with an average episode number of 70 games played. While direct comparisons between these systems are naturally constrained (even what is referred to as a "neuron" is inconsistent between fields of research), the aim of this work was to determine whether meaningful performance differences would arise between learning systems (as contained systems) that may merit further exploration of BNNs as information processing machines. This approach allowed an examination of the overall performance of each group with respect to various gameplay characteristics and, for the RL methods, in response to varying information input.
% Table \ref{Tab:Tcr} provides a summary of comparisons between different gameplay characteristics amongst groups with the \Ima{} to the RL algorithms and whether the differences were statistically significant. 

Across all types of information input, BNN outperformed all RL baselines in terms of average hit-per-rally (Subfigure \ref{fig:Image}.a),  \% of aces (Subfigure \ref{fig:Image}.b), and \% of long rallies achieved (Subfigure \ref{fig:Image}.c). 
Moreover, the increase in average rally length, decrease in number of aces, and increase in number of long rallies were significant only within the HCC and MCC groups and the A2C algorithm with the \Ima{} and \twod{} designs in terms of the increase in the percentage of long rallies, when comparing the first 5 and the last 15 minutes during gameplay (see Subfigures \ref{fig:Image}.d, e, and f). Additionally, we found that the HCC group had the highest relative improvement in average number of hits between the first 5 minutes and last 15 minutes of the game as depicted in Subfigures \ref{fig:paddle_improv}.b, d, and f. 

Results show that the game performance of deep RL algorithms in terms of relative learning improvement in time and average hits-per-rally is outperformed by biological cultures when number of allowable samples are fixed. This supports the conclusion that RL algorithms showed significantly lower sample efficiency compared to BNN, having lower improvements in learning over an episode-matched training duration provided for all groups. This matches theoretical expectations previously outlined where it was proposed that biological learning is inherently more sample efficient \citep{neftci_reinforcement_2019, whittington_theories_2019}. Given how rapidly synaptic plasticity or behaviour changes have occurred for both \textit{in vitro} and \textit{in vivo} models, this finding is consistent with such observations \citep{tessadori_modular_2012, bakkum_spatio-temporal_2008, hamid_mesolimbic_2016, costa_amygdala_2016, habibollahi2023critical}. Furthermore, although difficult to directly compare energy consumption, it should be noted that biological systems use magnitudes less than traditional computing systems used for ML \citep{jouppi_domain-specific_2020}. 

Moreover, the comparison between the various machine learning algorithms is also consistent with past research. A2C and PPO often achieve better results compared to DQN which is in line with previous studies proposing that algorithms optimizing a stochastic policy generally perform better than DQN \citep{siddique2020learning,el2022learning} which is known to suffer more from low sample efficiency \citep{lee2019sample}.  This can best be seen in the relative performance between different levels of information input. 
When a CNN was integrated into the RL models, some degree of learning (that did not reach statistical significance) was observed for these systems. BNN received only a fraction of the input information density compared to their RL opponents in this condition (8-pixel combination of rate coded and place coded stimulation compared to 40 $\times$ 40 pixels of the input image). Moreover, it was reasonable to consider whether the curse of dimensionality (where higher dimension input can require additional episodes to converge to a minima) may be adversely impacting the RL agents under the \Ima{} condition. To account for potential disadvantages occurring as a result of increased input dimensionality, we also examined two alternative designs for input structure to the RL algorithms (i.e. \Lo{} and \twod{} designs). In-depth comparison between BNN performance and these alternative RL algorithms did not provide any significantly different outcome in favour of the RL baselines' sample efficiency (see Figures \ref{fig:LocationVec} and \ref{fig:2dInput}). 
% However, we observed far poorer performance across these metrics when input information became more sparse. 

That BNN could perform with such sparsely coded informational input conforms to coding mechanisms known to be used in biological intelligence \citep{harrell_elaborate_2020, buchanan2018organoids, bastos_visual_2015}. While RL algorithms use back-propagation, it has been argued that this method is likely too inefficient to function within biological systems \citep{whittington_theories_2019, friston_reinforcement_2009, tsividis2017human, song2022inferring, whittington2017approximation, hinton2022forward}. A more dynamic reconfiguration of network activity has been proposed to be necessary for the learning rates observed in biological cultures \citep{song2022inferring,felleman1991distributed,tsividis2017human,whittington2017approximation}. Theories of how this learning may occur include predictive coding, active inference, prospective configurations, and Hopfield networks, which have been used to describe how neural systems may reorganise activity for learning tasks \citep{hopfield1982neural, song2022inferring, rao1999predictive, friston2005theory, de2010predictive}. While nuances amongst these different theories exist, the general notion supports the idea of a more biological consistent forward-based learning process compared to backpropagation.

To explore this, we explored a biologically inspired algorithm, implementing an active inference agent that uses counterfactual learning and reported the comparison results in \nameref{supp_mat} \ref{active_inf} and \nameref{ex_data} Figure \ref{fig:active_inf}. Improved learning rates observed in the biological inspired learning protocol supports the potential of active inference agents to provide valuable insights into optimized learning strategies, thereby enhancing our understanding of these dynamics. However, these active inference algorithms are still highly dependent on the chosen hyper-parameters and require relatively higher power consumption compared to biological systems. Nonetheless, these results highlight the value of further exploring biologically-inspired systems of learning and support the notion that SBI systems may offer a useful pathway to do this in the future.  

Interplay between individual neuronal activity and population level activity adds further complexity to determining the mechanisms of learning within biological cultures. While limitations in study design (specifically the use of opaque chips) prevent a robust assessment of the specific learning processes within the cultures used in this study beyond that previously reported \citep{kagan2022vitro, habibollahi2023critical}, findings endorse this approach for future exploration of these dynamics with altered study designs. Future work has potential to not only understand how biological intelligence arises, but also how one may implement more advanced biologically inspired learning protocols that may surpass current performance.
    
% \end{comment}

This work acts as the first direct comparison (to our knowledge) between an SBI system and state-of-the-art RL algorithms on a comparable task. A potential limitation of the work results from the fact that the space of hyper-parameters is too large for an exhaustive search in each algorithm. However to explore a significant number of hyper-parameters we used values utilized in the original paper that introduced each algorithm. We tuned the hyper-parameters that were most sensitive by a grid search in a limited space of those parameters. As a result of their sensitivity to hyper-parameter selection, state-of-the-art deep RL algorithms remain challenging to apply. The use of model-based RL is proposed for achieving higher sample efficiencies. Model-free algorithms, however, often perform significantly better asymptotically than these algorithms \citep{chua2018deep}. Recently, different accelerated approaches have also been proposed for deep RL \citep{lee2019sample,chua2018deep,franke2020sample}. Nonetheless, many still lag behind the performance of the original algorithms or require modern computers and a combination of CPUs and GPUs prompting even higher computational costs \citep{stooke2018accelerated}. As a future pathway, these modified algorithms may be utilized for further comparisons.
Arguably, biological cultures operating with the \textit{DishBrain} system do not require such fine-tuning of parameters or manipulation of the architecture. Nonetheless, the results of this work supports that even rudimentary SBI systems with limited informational input are viable learning systems that can compete and even outperform established RL algorithms on sample efficiency. Coupled with the promise of significant gains in power efficiencies, flexibility of tasks, and upcoming improvements in the associated technologies \citep{smirnova2023organoid}, these biological-based intelligence systems present a compelling pathway for realizing real-time learning unachievable by current silicon-based approaches alone. 

\section{Methods}
\subsection{\textit{DishBrain} System}
The initial validation of the \textit{DishBrain} system was previously presented in \citep{kagan2022vitro}. Briefly, cortical cells were either differentiated from human induced pluripotent stem cells (hiPSC) using a modified Dual SMAD inhibition protocol or surgically extracted from E15 mouse embryos. By setting up cultures from multiple cell sources this helped ensure that results would generalize across different species and preparations. Ethical approvals for animal work were obtained (E/1876/2019/M: Alfred Research Alliance Animal Ethics Committee B)  for animal work with all cell culture work according to relevant ethical guidelines. Cell line characterisation and approvals are reported in \citep{kagan2022vitro}.

Approximately 10\textsuperscript{6} cells were plated and integrated onto a high-density multi-electrode array (HD-MEA; Maxwell Biosystems, AG, Switzerland). Cell cultures were maintained in BrainPhys™ Neuronal Medium (Stemcell Technologies Australia, Melbourne, Australia) supplemented with 1\% penicillin-streptomycin during testing.  The \textit{DishBrain} system was developed as a low latency, real-time system which interacts with the HD-MEA software to allow closed-loop stimulation and recording which has previously been described in detail \citep{kagan2022vitro}. Using this method, activity from a neuronal culture can be read, along with providing structured stimulation to the same culture in real-time. \textit{DishBrain} was then utilized to embody neural cultures in a virtual game-world, to simulate the classic arcade game `Pong'. Biphasic electrical stimulation was used to stimulate neurons consistent with previous attempts to elicit action potentials in comparable cultures \citep{ruaro2005toward}. Electrical stimulation was arranged to transmit a variety of task-related information between the cells and the simulated virtual environment using appropriate coding schemes via routed electrodes on the MEA that were divided into discrete regions as in Figure \ref{fig:dish_feed&diagram}.b. 

Specifically, stimulation was applied using a combination of rate coding (4Hz - 40Hz) electrical pulses to communicate the position on the $x$-axis and place coding (on a given electrode that was arranged topographically from an egocentric representation for the culture) to communicate information on the $y$-axis into a predefined bounded two-dimensional sensory area consisting of 8 sensory electrodes to deliver this input information. Three types of input were provided: the sensory stimulation as explained above, or stimulation in response to activity designated as either `Predictable' or `Unpredictable' feedback (see Figure \ref{fig:dish_feed&diagram}.a). Cultures received Unpredictable stimulation when they missed connecting the paddle with the ‘ball’, i.e. when a ‘miss’ occurred. Using a feedback stimulus at a voltage of 150 mV and a frequency of 5 Hz, an unpredictable external stimulus could be added to the system. Random stimulation took place at random sites over the 8 predefined sensory electrodes at random timescales for a period of four seconds, followed by a configurable rest period of four seconds where stimulation paused, then the next rally began. Should no miss occur, the game would continue until either a miss occurred or the timer of 20 minutes expired, which would end the session. In contrast, cultures were exposed to Predictable stimulation when a ‘hit’ was registered - that is, when the ‘paddle’ connected successfully with the ‘ball’. This was delivered across all 8 stimulation electrodes simultaneously at 75mV at 100Hz over 100ms and replaced other sensory information for 100 ms.

The movement of the paddle was controlled by the level of electrophysiological activity measured in a predefined ‘motor area’ of the cultured network as shown in Figure \ref{fig:dish_feed&diagram}.b. , which was collected in real-time. Incoming samples were filtered with a 2nd order high-pass Bessel filter with 100Hz cut-off. The absolute value was smoothed using a 1st order low-pass Bessel filter with a 1 Hz cut-off and the spike threshold is proportional to this smoothed absolute value. A relative activity spike of 6 sigma greater than background noise was then used to define an action potential. Detected action potentials from counterbalanced motor regions were then summed together, where higher activity in a given pair of regions would cause the virtual paddle to move in one direction, while activity in the other regions would result in the inverse movement. Information about ball position relative to the paddle was adjusted in a closed-loop manner with a spike-to-stim latency of approximately 5ms. Figure \ref{fig:dish_feed&diagram}.a,b illustrate the input information, feedback loop setup, and electrode configurations in the \textit{DishBrain} system.

The gameplay performance of cell cultures subjected to the simplified pong environment via the \textit{DishBrain} system was assessed. In each episode of the game, the average number of rallies before the ball was missed for the first time was then compared with different deep RL baseline methods. Each recording session of the cultures during gameplay was 20 minutes. During a gameplay session, the average number of rallies (i.e., episodes) an average biological culture would perform was 69.04 ± 7.95 rallies/episodes. Therefore, to compare sample efficiency in a matched comparison, a total of 70 training episodes were provided to deep reinforcement learning algorithms during training.

More details of this system are introduced in \nameref{supp_mat} \ref{cell}, \ref{MEA}, and \ref{dishbrain} as well as \nameref{ex_data} Figure \ref{DishBrain_config}.

% \begin{figure*}[!ht]
%   \centering
%   \hspace{-1cm}
%   {\includegraphics[width=1\textwidth]{Figures/DishBrain_Feedback.pdf}}
%   \caption{\textbf{DishBrain system.} \textbf{a)} \textit{DishBrain} feedback loop setup. \textbf{b)} Electrode configuration and predefined sensory and motor regions. Figures adapted and modified from \citep{kagan2022vitro}}
%   \label{fig:dish_feed}
% \end{figure*}

\subsection{Deep Reinforcement Learning Algorithms}
In this work, we use three state-of-the-art deep reinforcement learning algorithms: Deep Q Network (DQN) \citep{mnih2015human}, Advantage Actor-Critic (A2C) \citep{arulkumaran2017deep} and Proximal Policy Optimization (PPO) \citep{schulman2017proximal}, established to have good performance in Atari games. Benefiting from deep learning advantages in automated feature extraction, specifically exploiting Convolutional Neural Networks (CNN) in their structures, these methods are robust tools in reinforcement tasks, particularly in games where the system's input is an image. 
In this work, aiming to account for potential detriments to sample efficiency resulting from the increased dimensionality of the image input to the deep RL algorithms \citep{bellman1957dynamic}, we designed two additional types of input information to the RL algorithms. We compare all three different designs with the performance of biological cultures. We attempt to study whether the curse of dimensionality and increased size of the feature vectors when directly utilizing image inputs affect the comparison between biological cultures and RL algorithms in terms of their sample efficiency. All the algorithms follow a common strategy although they are different in structure. The three different input categories and RL algorithm designs are introduced below:

\begin{itemize}
    \item \textbf{\Ima{}:}  The current state is a tensor of the difference of pixel values from the two most recent frames (i.e. another 40 $\times$ 40 grayscale pixel image) \footnote{We also experimented with an alternative design where the input consisted of a stack of the four most recent frames for all algorithms. However, this modification led to a noticeable decline in the performance of all the methods because it failed to capture the sense of motion between frames.}. This current state is then input into the CNN to obtain the selected action. Next, based on the action taken, a reward is received, and a new state is formed. The ultimate goal is to find a policy that indicates the best action in each state to maximize the reward function. 
    
    \item \textbf{\Lo{}:} Instead of the grayscale image, a 4-dimensional vector encoding the $x$ and $y$ coordinates of the ball (distance to the paddle/wall and distance to the floor in pixels) and the $y$ coordinates of the paddle's top and the bottom was obtained. All values are integers between  $[4,40]$. The current state which is the input to each algorithm is then a tensor of the difference of values from the two most recent 4-dimensional location vectors. No additional CNN layer is utilized in this case.
    
    \item \textbf{\twod{}:} A design as similar to the \textit{DishBrain} system's input structure as possible was also examined. For this case, the $y$-axis of the gameplay environment was divided into 8 equal segments each mimicking one of the sensory electrodes in the biological cultures, and place coding the information about the ball's $y$-axis position as an integer in the $[1,8]$ interval. Then, the ball's $x$-axis position is used as the second element of this input vector being an integer value in $[4,40]$ similar to the rate coded component of the stimulation applied to the biological cultures. No additional CNN layer is utilized in this design.
\end{itemize}

The overview of the implemented DQN, A2C, and PPO algorithms are represented in \nameref{supp_mat} \ref{RL} (see Algorithms \ref{DQN}, \ref{A2C}, and \ref{PPO}).
% For a more detailed definition of these methods, see Supplementary Materials \ref{RL}. 

All the deep RL implementations run on a 2.3 GHz Quad-Core Intel Core i5. PyTorch 1.8.1 was used to build neural network blocks and Open AI Gym environment to define our game environment represented by a 40 $\times$ 40 pixel grayscale image.
In the training phase of all RL algorithms, every algorithm was run for 150 random seeds and a total number of 70 episodes for each seed. These seeds imply 150 different neural networks trained separately, resembling 150 different recorded cultures. In this work, we report the average value of each metric among all seeds.

\subsection{Data Availability}
All data generated for or used within this manuscript have been deposited at Open Science Framework (OSF) and are publicly available here: \url{https://osf.io/cnpzf/?view_only=a33b7083f78e4c55a20b6c021a695a4a}.

\subsection{Code Availability}\label{code}
All code for deep reinforcement learning models or used for data analysis to generate the results in this manuscript have been deposited at Open Science Framework (OSF) and are publicly available via \url{https://osf.io/cnpzf/?view_only=a33b7083f78e4c55a20b6c021a695a4a}.

\subsection{Supplementary information}
\nameref{supp_mat}; \nameref{ex_data}; 
Tables S1 - S3

\subsection{Acknowledgments}
The authors thank and acknowledge Dr Haytham Fayek, Dr Hon Weng Chong and Mr Amitesh
Gaurav for their input and advice on the manuscript and experimental design.

\subsection{Competing interests}
B.J.K., F.H., and M.K. were contracted or employed by Cortical Labs during the course of this research. B.J.K. has shares in Cortical Labs and an interest in patents related to this work. There are no other competing interests to declare. 

\subsection{Author contributions}
B.J.K., M.K., and F.H. conceived and designed the work. M.K. developed the models and performed the experiments under the guidance of B.J.K. F.H. and M.K. analysed the data and conducted method comparisons. M.K., F.H., and B.J.K contributed to materials and the analysis tool. A.P. and A.R. conducted additional analysis. M.K., F.H., and B.J.K. drafted the initial manuscript. All authors contributed to reviewing the manuscript.

% \bibliographystyle{unsrtnat}
% \bibliography{references}  %%% Uncomment this line and comment out the ``thebibliography'' section below to use the external .bib file (using bibtex) .

% \thebibliography

\newpage
\begin{appendices}

\renewcommand{\theequation}{S\arabic{equation}}
\renewcommand{\theHequation}{S\arabic{equation}}  % Adjust for hyperref
\setcounter{equation}{0}

\renewcommand{\thefigure}{S\arabic{figure}}
\renewcommand{\theHfigure}{S\arabic{figure}}  % Adjust for hyperref
\setcounter{figure}{0}        
\renewcommand{\thetable}{S\arabic{table}}
\setcounter{table}{0}  
\renewcommand{\theHtable}{S\arabic{table}}  % Adjust for hyperref
\renewcommand{\theequation}{S\arabic{equation}}

\section{Supplementary Materials}\label{supp_mat}

\subsection{Cell Culture}\label{cell}
Neural cells were cultured either from the cortices of E15.5 mouse embryos or differentiated from human induced pluripotent stem cells via a dual SMAD inhibition (DSI) protocol as previously described \citep{kagan2022vitro}. Cells were cultured until plating onto MEA. For primary mouse neurons, this occurred at day-in-vitro (DIV) 0, for DSI cultures this occurred at between DIV 30 - 33 depending on culture development.  

\subsection{MEA Setup and Plating}\label{MEA}
MaxOne Multielectrode Arrays (MEA; Maxwell Biosystems, AG, Switzerland) was used and is a high-resolution electrophysiology platform featuring 26,000 platinum electrodes arranged over an 8 mm2. The MaxOne system is based on complementary meta-oxide-semiconductor (CMOS) technology and allows recording from up to 1024 channels. MEAs were coated with either polyethylenimine (PEI) in borate buffer for primary culture cells or Poly-D-Lysine for cells from an iPSC background before being coated with either 10 µg/ml mouse laminin or 10 µg/ml human 521 Laminin (Stemcell Technologies Australia, Melbourne, Australia) respectively to facilitate cell adhesion. Approximately $10^6$ cells were plated on MEA after preparation as per \citep{kagan2022vitro}. Cells were allowed approximately one hour to adhere to the MEA surface before the well was flooded. The day after plating, cell culture media was changed for all culture types to BrainPhys™ Neuronal Medium (Stemcell Technologies Australia, Melbourne, Australia) supplemented with 1\% penicillin-streptomycin. Cultures were maintained in a low O2 incubator kept at 5\% CO2, 5\% O2, 36°C and 80\% relative humidity. Every two days, half the media from each well was removed and replaced with free media. Media changes always occurred after all recording sessions. 

\subsection{\textit{DishBrain} platform and electrode configuration}\label{dishbrain}
The current \textit{DishBrain} platform is configured as a low-latency, real-time MEA control system with on-line spike detection and recording software. The \textit{DishBrain} platform provides on-line spike detection and recording configured as a low-latency, real-time MEA control. The \textit{DishBrain} software runs at 20 kHz and allows recording at an incredibly fine timescale. There is the option of recording spikes in binary files, and regardless of recording, they are counted over a period of 10 milliseconds (200 samples), at which point the game environment is provided with how many spikes are detected in each electrode in each predefined motor region as described below. Based on which motor region the spikes occurred in, they are interpreted as motor activity, moving the ‘paddle’ up or down in the virtual space. As the ball moves around the play area at a fixed speed and bounces off the edge of the play area and the paddle, the pong game is also updated at every 10ms interval. Once the ball hits the edge of the play area behind the paddle, one rally of pong has come to an end at which point a 'miss' would be recorded and an unpredictable stimulation would be delivered to the cells. Using a feedback stimulus at a voltage of 150 mV and a frequency of 5 Hz, unpredictable external stimulus could be added to the system. Random stimulation took place at random sites over the 8 predefined input electrodes at random timescales for a period of four seconds, followed by a configurable rest period of four seconds where stimulation paused, then the next rally began.\newline

\begin{figure*}[!ht]
  \centering
  {\includegraphics[width=0.9\textwidth]{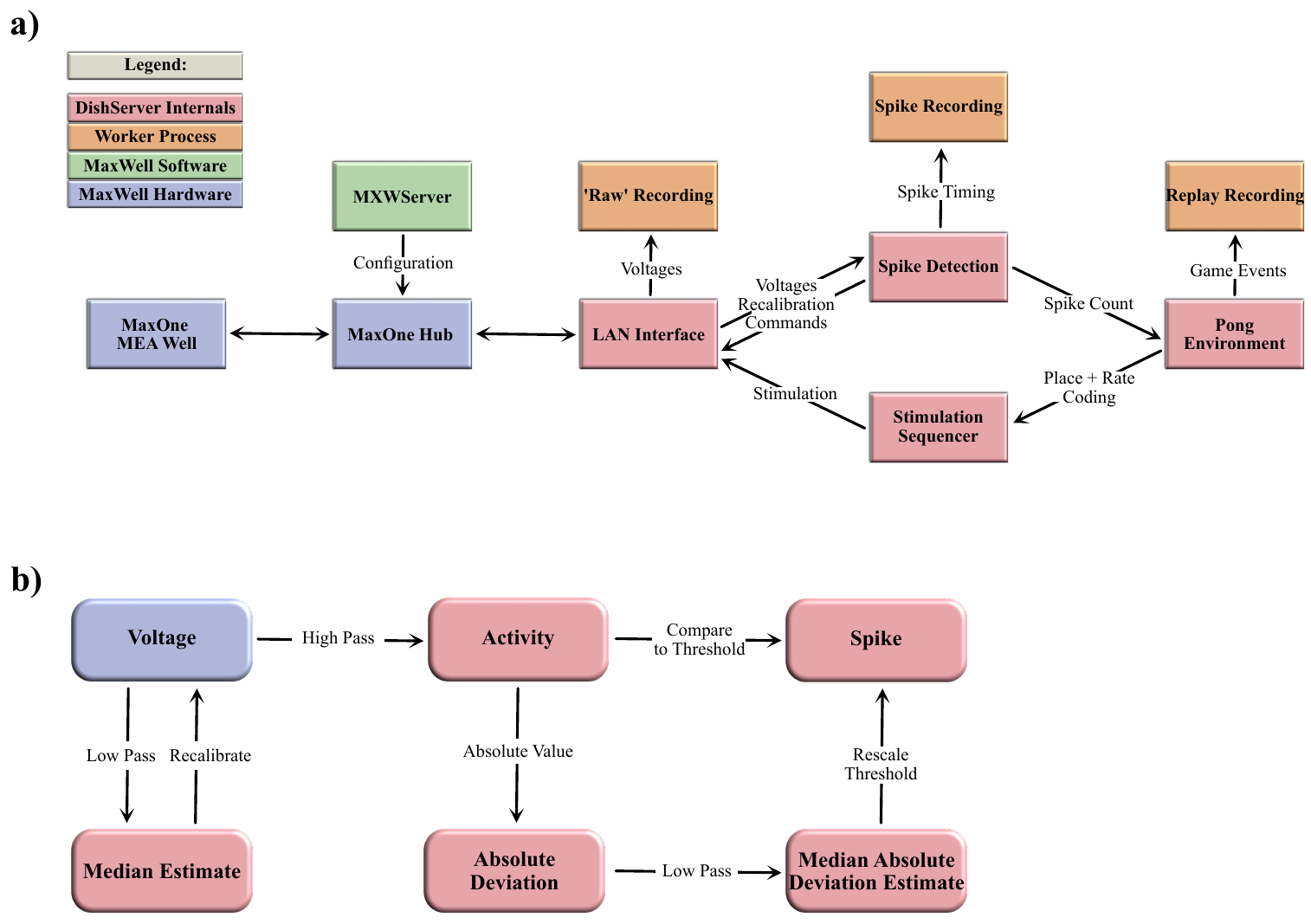}}  
  \caption{ \textbf{DishBrain software schematics.} \textbf{a)} Software components and data flow in the \textit{DishBrain} closed loop system. Voltage samples flow from the MEA to the `Pong' environment, and sensory information flows from the `Pong' environment back to the MEA, forming a closed loop. The blue rectangles mark proprietary pieces of hardware from MaxWell, including the MEA well which may contain a live culture of neurons. The green MXWServer is a piece of software provided by MaxWell which is used to configure the MEA and Hub, using a private API directly over the network. The red rectangles mark components of the `DishServer’ program, a high-performance program consisting of four components designed to run asynchronously, despite being run on a single CPU thread. The `LAN Interface’ component stores the network state, for talking to the Hub, and produces arrays of voltage values for processing. Voltage values are passed to the `Spike Detection’ component, which stores feedback values and spike counts, and passes recalibration commands back to the LAN Interface. When the pong environment is ready to run, it updates the state of the paddle based on the spike counts, updates the state of the ball based on its velocity and collision conditions, and re-configures the stimulation sequencer based on the relative position of the ball and current state of the game. The stimulation sequencer stores and updates indices and countdowns relating to the stimulations it must produce and converts these into commands each time the corresponding countdown reaches zero, which are finally passed back to the LAN Interface, to send to the MEA system, closing the loop. The procedures associated with each component are run one after the other in a simple loop control flow, but the `Pong’ environment only moves forward every 200th update, short-circuiting otherwise. Additionally, up to three worker processes are launched in parallel, depending on which parts of the system need to be recorded. They receive data from the main thread via shared memory and write it to file, allowing the main thread to continue processing data without having to hand control to the operating system and back again. \textbf{b)} Numeric operations in the real-time spike detection component of the \textit{DishBrain} closed loop system, including multiple IIR filters. Running a virtual environment in a closed loop imposes strict performance requirements, and digital signal processing is the main bottleneck of this system, with close to 42 MB of data to process every second. Simple sequences of IIR digital filters are applied to incoming data, storing multiple arrays of 1024 feedback values in between each sample. First, spikes on the incoming data are detected by applying a high pass filter to determine the deviation of the activity and comparing that to the MAD, which is itself calculated with a subsequent low pass filter. Then, a low pass filter is applied to the original data to determine whether the MEA hardware needs to be re-calibrated, affecting future samples. This system was able to keep up with the incoming data on a single thread of an Intel Core i7-8809G. Figures adapted from \citep{kagan2022vitro}. }
  \label{DishBrain_config}
\end{figure*}

In contrast, a predictable stimulus feedback is provided when the ball contacts the paddle under the standard stimulus condition. Predictable stimulus feedback involves 75mV stimulation at 100Hz over 100ms occurring when the simulated ball struck the paddle and replaced other sensory information. All 8 stimulation electrodes simultaneously would receive predictable stimulation at this frequency and period. A ‘stimulation sequencer’ module tracks the location of the ball relative to the paddle during each rally and encodes it as stimulation to one of eight stimulation sites. Each time a sample is received from the MEA, the stimulation sequencer is updated 20,000 times a second, while the game itself runs at 100Hz. After the previous lot of MEA commands has completed, the \textit{DishBrain} system constructs a new sequence of MEA commands based on the information it has been configured to transmit based on both place codes and rate codes. The stimulations take the form of a short square bi-phasic pulse that is a positive voltage, then a negative voltage. This pulse sequence is read and applied to the electrode by a Digital to Analog Converter (or DAC) on the MEA. A real-time interactive version of the game visualizer is available at \url{https://spikestream.corticallabs.com/}. Alternatively, cells could be recorded at ‘rest’ in a gameplay environment where activity was recorded to move the paddle but no stimulation was delivered, with corresponding outcomes still recorded. Using this spontaneous activity alone as a baseline, the gameplay characteristics of a culture were determined. Low level code for interacting with Maxwell API was written in C to minimize processing latencies-so packet processing latency was typically $<$50 $\mu$s. High-level code was written in Python, including configuration setups and general instructions for game settings. A 5 ms spike-to-stim latency was achieved, which was substantially due to MaxOne's inbuilt hardware buffering. Figure \ref{DishBrain_config} illustrates a schematic view of Software components and data flow in the \textit{DishBrain} closed loop system.

\subsection{Deep Reinforcement Learning Algorithms}\label{RL}
\begin{figure*}[!ht]
  \centering
  \hspace{-1cm}
  {\includegraphics[width=1\textwidth]{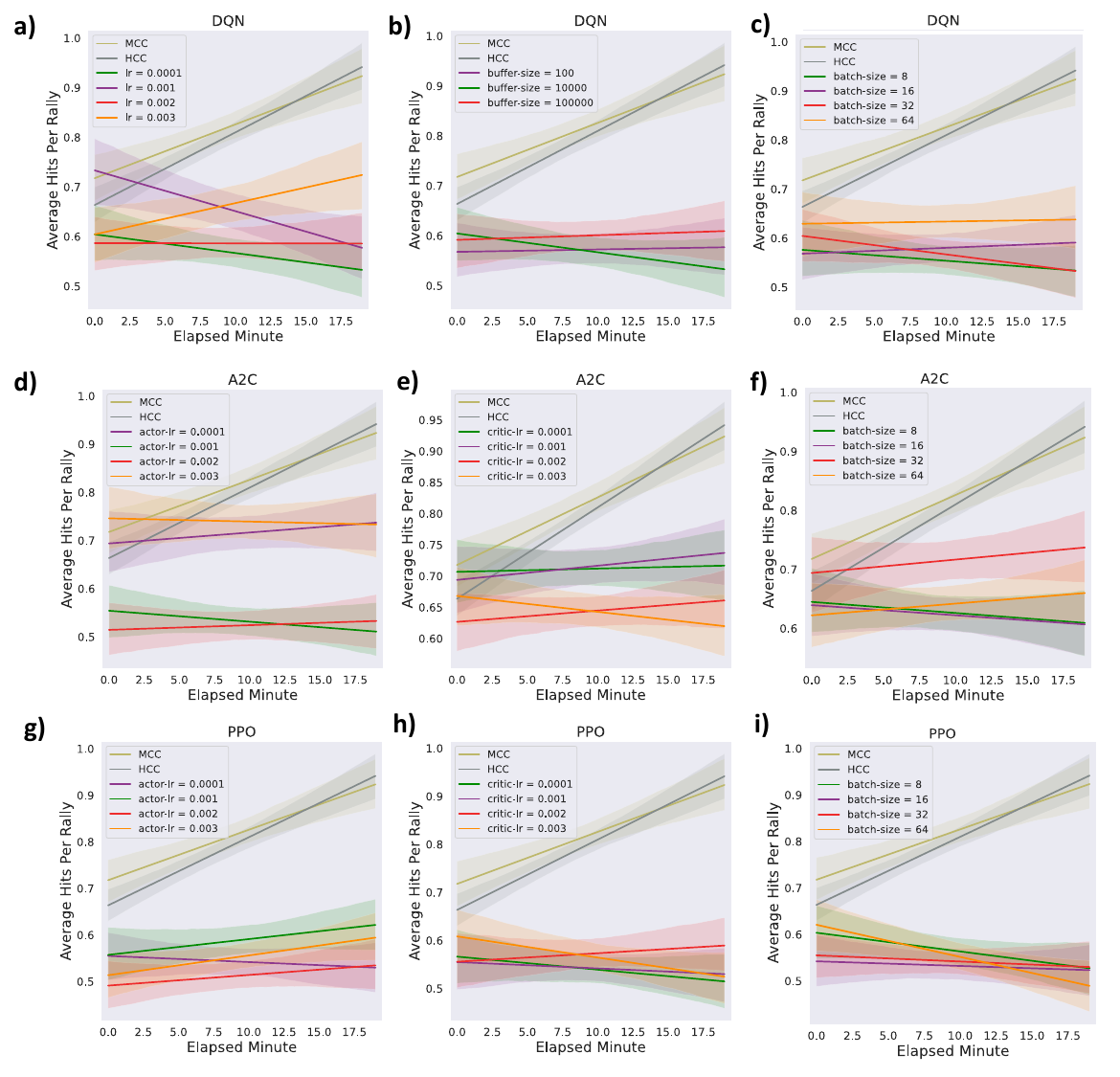}}
  \caption{\textbf{Hyper-parameter exploration of RL algorithms.} The changes in average hits-per-rally for each RL algorithm in several sample points of the grid search space. \textbf{a)} Effects of changing the learning rate on DQN performance. replay buffer size = 10000 and batch size = 32; \textbf{b)} Effects of changing the replay buffer size on DQN performance. learning rate = 0.0001 and batch size = 32;  \textbf{c)} Effects of changing the batch size on DQN performance. learning rate = 0.0001 and replay buffer size = 10000; \textbf{d)} Effects of changing the actor learning rate on A2C performance. critic learning rate = 0.001 and batch size = 32;  \textbf{e)} Effects of changing the critic learning rate on A2C performance. actor learning rate = 0.0001 and batch size = 32;  \textbf{f)} Effects of changing the batch size on A2C performance. actor learning rate = 0.0001 and critic learning rate = 0.001; \textbf{g)} Effects of changing the actor learning rate on PPO performance. critic learning rate = 0.001 and batch size = 32;  \textbf{h)} Effects of changing the critic learning rate on PPO performance. actor learning rate = 0.0001 and batch size = 32;  \textbf{i)} Effects of changing the batch size on PPO performance. actor learning rate = 0.0001 and critic learning rate = 0.001.}
  \label{fig:hp_exp}
\end{figure*}

\textbf{Deep Q Network (DQN):}
The utilized DQN algorithm begins by extracting spatiotemporal features from inputs, such as the movement of the ball in the game of `Pong'. Multiple fully connected layers are used to process the final feature map, which implicitly encodes the effects of actions. As opposed to traditional controllers that use fixed preprocessing steps, this method can adapt the processing of the state based on changes in the learning signal. An epsilon-greedy algorithm was employed in this work to balance the exploration and exploitation capabilities of the DQN algorithm.\\ For the results represented in this manuscript, a comprehensive grid search was conducted within the parameter space of \textit{learning rate} ($[0.0001,0.004]$), \textit{replay buffer size} ($[10,100000]$), and the training \textit{batch size} ($[5,128]$) with starting point of $0.0001,32,10000$, respectively, aiming to identify the optimal parameter configuration. The results presented in this paper are derived from the superior set of hyper-parameters obtained through this search process. As the outcome of this search for the DQN algorithm, we selected \textit{learning rate = 0.002}, \textit{replay buffer size = 10000}, and \textit{batch size = 16} for the results of Figure \ref{fig:Image}, \textit{learning rate = 0.001}, \textit{replay buffer size = 10000}, and \textit{batch size = 16} for the results of Figure \ref{fig:LocationVec}, and \textit{learning rate = 0.001}, \textit{replay buffer size = 10000}, and \textit{batch size = 32} for the results of Figure \ref{fig:2dInput}.
Figure \ref{fig:hp_exp} illustrates the performance of the DQN algorithm with \Ima{} design in terms of average rally length in several sample points of the mentioned search space. While exploring each hyper-parameter in Figure \ref{fig:hp_exp}, the remaining pair are set to the same values as the starting point of the search (i.e. \textit{learning rate = 0.0001}, \textit{batch size = 32}, and \textit{replay buffer size = 10000}).\\
For additional details on the set of explored hyper-parameters and network architectures, see Table \ref{tab:hyp}.

\begin{algorithm}[H]

    \caption{\small{Deep Q Network (DQN) with Experience Replay}}\label{DQN}
    \small
    \begin{algorithmic}[1]
    \Require\\
   $\mathcal{D}$: Replay buffer with size $N$ (Default: 10000)\\
    $\theta$: Initial network parameters \\
    $\tilde{\theta}$: Copy of $\theta$\\
    $\gamma$: Discount factor (Default: 0.95) \\
    $N_b$: Training batch size (Default: 16)\\
    $\tilde{N}$: Target network update frequency (Default: 10)\\
    $x_t$: Input matrix at time $t$\\
    $S$: Number of seeds (Default: 150) \\
    $e_{max}$: Maximum number of episodes (Default: 70)
    \For{seed $ \in \{1,\cdots, S \}$}
        \For{episode $e \in \{1,\cdots,e_{max} \}$}
        \State Set state $s_1 \gets x_1$ and preprocess $\phi_1=\phi(s_1)$
        \State $t=1$
        \While {$\phi_t$ is  non-terminal }
        \State With probability $\epsilon$ select a random action $a_t$
        \State otherwise select $a_t= max_a Q^*(\phi(s_t),a;\theta)$
        \State Execute action $a_t$ and observe reward $r_t$ and input $x_{t+1}$
        \State Set new state $s_{t+1}$ and preprocess $\phi_{t+1}=\phi(s_{t+1})$
        \State Store transition $(\phi_t,a_t,r_t,\phi_{t+1})$ in $\mathcal{D}$
        \State Sample random minibatch of $N_b$ transitions$(\phi_j,a_j,r_j,\phi_{j+1})$ from $\mathcal{D}$
        \State Set $y_j = \begin{cases}r_j & \text{for terminal $\phi_{j+1}$}\\r_j + \gamma max_{a'} Q(\phi_{j+1},a';\theta)  & \text{for non-terminal $\phi_{j+1}$}\end{cases} $ 
        \State Perform a gradient descent step on $(y_j-Q(\phi_j,a_j;\theta))^2$
        \State Replace target parameters $\tilde{\theta} \gets \theta$ every  $\tilde{N}$ steps 
        \State $t=t+1$
        \EndWhile
        \EndFor
    \EndFor
    \end{algorithmic}
\end{algorithm}

\textbf{Advantage Actor-Critic (A2C):}
% Parameterized policies can be improved by using gradients as a learning signal. The policy gradient methods can be made more stable by decreasing the gradient variance which can be done by making the baseline stat-dependant. 
In an A2C model, the total reward itself could be represented as a \textit{value} of the state plus the advantage of the action. The \textit{value} of each policy is learned while following it. The policy gradient can be calculated by knowing the \textit{value} for any state. The policy network is then updated such that the probability of actions with a higher advantage value is increased. Here, the policy network (which returns a probability distribution of actions) is called the \textit{actor}, as it tells the agents what to do. \textit{Critic} is another network that enables the evaluation of the actions to decide whether they were good or not. In this case, policy and value are implemented as separate heads of the network, which transform the output from the common body into either probability distributions or single numbers representing the state's value. Thus, low-level features can be shared between the two networks. \\
For the results represented in the main paper, a comprehensive grid search was conducted within the parameter space of actor learning rate ($[0.0001,0.004]$), critic learning rate ($[0.0001,0.004]$), and the training batch size ($[5,128]$), to identify the optimal parameter configuration. As the outcome of this search for A2C, we selected \textit{actor learning rate = 0.001, 0.0001, 0.003}, \textit{critic learning rate = 0.001, 0.001 , 0.001}, and \textit{batch size = 32, 32,5} for the results of Figure \ref{fig:Image}, \ref{fig:LocationVec}, and \ref{fig:2dInput}, respectively. Figure \ref{fig:hp_exp} contains the results of this hyper-parameter search for the A2C algorithm with the \Ima{} design in terms of average rally length in several sample points of the mentioned search space.  While exploring each hyper-parameter in Figure \ref{fig:hp_exp}, the remaining pair are set to the same values as the starting point of the search (i.e. \textit{actor learning rate = 0.0001}, \textit{batch size = 32}, and \textit{critic learning rate = 0.001}).

\begin{algorithm}[H]
    \caption{\small{Advantage Actor-Critic (A2C)}}\label{A2C}
    \small
    \begin{algorithmic}[1]
       \Require\\
    $\theta_v$: Initial parameter vector for the value net (critic) \\
    $\theta_{\pi}$: Initial parameter vector for the policy net (actor)\\
    $\gamma$: Discount factor (Default: 0.95) \\
    $N$: Number of consecutive steps to play current policy in the environment (Default: 5) \\
    $x_t$: Input matrix at time $t$\\
    $S$: Number of seeds (Default: 150) \\
    $e_{max}$: Maximum number of episodes (Default: 70)
    \For{seed $ \in \{1,\cdots,S \}$}
   \State $t=1$
   \State $e=1$
    \Repeat
    \State $\partial \theta_{\pi} \gets 0$ and $\partial \theta_v \gets 0$
    \State $t_{start}= t$
    \State Set state $s_t \gets x_t$ and preprocess $\phi_t=\phi(s_t)$
    \Repeat
    \State Select $a_t$ according to $\pi(a_t\mid \phi_t;\theta)$
    \State Execute action $a_t$ and observe reward $r_t$ and input $x_{t+1}$
    \State Set new state $s_{t+1}$ and preprocess $\phi_{t+1}=\phi(s_{t+1})$ 
    \State $t \gets t+1$
    \Until{$\phi_t$ is terminal \textbf{or} $t-t_{start}=N$}
    \State $R = \begin{cases} 0 & \text{for terminal $\phi_{t}$}\\V(\phi_t;\theta_v)  & \text{for non-terminal $\phi_{t}$}\end{cases} $ 
    \For{$i \in \{t-1,\cdots,t_{start} \}$}
    \State $R \gets r_i + \gamma R$
    \State Accumulate the policy gradients: $\partial \theta_{\pi} \gets \partial \theta_{\pi} + \nabla_{\theta}\log \pi(a_i\mid\phi_i;\theta)\big(R-V(\phi_i,\theta_v)\big)$
    \State Accumulate the value gradients: $\partial \theta_{v} \gets \partial \theta_{v} + \frac{\partial\big(R-V(\phi_i,\theta_v)\big)^2}{\partial \theta_v}$
    \EndFor
    \State Update $\theta_{\pi}$ and $\theta_{v}$ using $\partial \theta_{\pi}$ and $\partial \theta_{v}$, respectively. 
    \If{$\phi_t$ is terminal}
    \State $e \gets e+1$
    \EndIf
    \Until{$e>e_{max}$}
    \EndFor
    \end{algorithmic}
\end{algorithm}

\textbf{Proximal Policy Optimization (PPO):}
PPO models are a family of policy gradient methods for reinforcement learning. The PPO method uses a slightly different training procedure: An extended set of samples is taken from the environment, and then the advantage is estimated for the whole set or sequence of samples before several epochs of training are performed
To estimate policy gradients, instead of using the gradient of action probabilities, the PPO method uses a different objective: the ratio between the new and the old policy scaled by the advantages.\\
Once more, for the results represented in the main paper, we used the outcome of a grid search for the PPO algorithm in the same space as A2C above and utilized \textit{actor learning rate = 0.003, 0.0001, 0.001}, \textit{critic learning rate = 0.003, 0.001 , 0.001}, and \textit{batch size = 16, 16, 32} to generate the results of Figure \ref{fig:Image}, \ref{fig:LocationVec}, and \ref{fig:2dInput}, respectively.\\
Figure \ref{fig:hp_exp} represents the performance of the PPO algorithm with the \Ima{} design in terms of average rally length in several sample points of the mentioned search space.  While exploring each hyper-parameter in Figure \ref{fig:hp_exp}, the remaining pair are set to the same values as the starting point of the search (i.e. \textit{actor learning rate = 0.0001}, \textit{batch size = 32}, and \textit{critic learning rate = 0.001}).

\begin{algorithm}[H]
    \caption{\small{Proximal Policy Optimization (PPO)}}\label{PPO}
    \small
    \begin{algorithmic}[1]
        \Require\\
    
    $\theta$: Initial policy parameter vector \\
    $\epsilon$: Clipping threshold (Default: 0.2) \\
     $\gamma$: Discount factor (Default: 0.95) \\
     $\lambda$: GAE parameter (Default: 1) \\
    $N$: Number of consecutive steps to play current policy in the environment (Default: 32) \\
    $x_t$: Input matrix at time $t$\\
    $S$: Number of seeds (Default: 150) \\
    $e_{max}$: Maximum number of episodes (Default: 70)
    \For{seed $ \in \{1,\cdots,S \}$}
    \State $t=1$
    \State $e=1$
    \Repeat
    \State $t_{start}= t$
    \State Set state $s_t \gets x_t$ and preprocess $\phi_t=\phi(s_t)$
    \Repeat
    \State Select $a_t$ according to $\pi(a_t\mid\phi_t;\theta)$
    \State Execute action $a_t$ and observe reward $r_t$ and input $x_{t+1}$
    \State Set new state $s_{t+1}$ and preprocess $\phi_{t+1}=\phi(s_{t+1})$ 
    \State $t \gets t+1$
    \Until{$\phi_t$ is terminal \textbf{or} $t-t_{start}=N$}
    \State Collect set of partial trajectories $\mathcal{D}$ on current policy $\pi$
    \State Estimate Advantages $\hat{A}^{\pi}_t = \sigma_t + (\gamma \lambda)\sigma_{t+1} + \cdots + (\gamma \lambda)^{N-t-1}\sigma_{N-1}$, where $\sigma_t = r_t + \gamma V(\phi_{t+1})-V(\phi_{t})$
    \State $\theta \gets \text{argmax}_{\theta} \mathcal{L}^{CLIP}_{\theta}(\theta)$
    \State where $\mathcal{L}^{CLIP}_{\theta}(\theta) =\mathbb{E}_{\tau \sim \pi} \bigg[ \sum^{T}_{t=0}\big[min( r_t(\theta)\hat{A}^{\pi}_t, clip(r_t(\theta),1-\epsilon,1+\epsilon)\hat{A}^{\pi}_t  ) \big] \bigg]$ 
    \If{$\phi_t$ is terminal}
    \State $e \gets e+1$
    \EndIf
    \Until{$e>e_{max}$}
    \EndFor
    \end{algorithmic}
\end{algorithm}
% \end{minipage}

% Figure \ref{fig:long_runs} illustrates the mean total reward of the RL algorithms using the same hyper-parameter sets as in Figure \ref{fig:Image} for an extended training period of 11000 game episodes. These results demonstrate successful learning and improved performance over an extended number of training episodes for all three algorithms. 

\subsection{Additional Hyper-parameter Exploration} \label{hyper-p}

\begin{table}[ht!]
\centering
\caption{Experimented Hyper-parameter and network architecture details}
\vspace{0.6cm}
\begin{threeparttable}
\resizebox{8cm}{!}{%
    \begin{tabular}{ccc}
    \toprule
    \textbf{\small Hyper-parameter} & \textbf{Algorithm} & \textbf{Tested Values} \\ 
    \midrule
    $\text{Conv}_1$ size & DQN, A2C, PPO & \textbf{(16 $\times$ 16)}, (64 $\times$ 64) \\
    $\text{Conv}_2$ size & DQN, A2C, PPO & \textbf{(32 $\times$ 32)}, (64 $\times$ 64) \\
    $\text{Conv}_3$ size & DQN, A2C, PPO & \textbf{(32 $\times$ 32)}, (64 $\times$ 64) \\
    Last hidden layer size & DQN, A2C, PPO & \{100, 256, \textbf{512}\} \\
    Number of seeds & DQN, A2C, PPO & \textbf{150} \\
    Kernel size & DQN, A2C, PPO & \{\textbf{5}, 4\} \\
    Stride & DQN, A2C, PPO & \textbf{2} \\
    Batch size & DQN, A2C, PPO & [5, 128] \\
    Discount factor & DQN, A2C, PPO & \{0.85, 0.95, 0.99, 0.999\} \\
    Learning rate & DQN & [0.0001, 0.004] \\
    Replay buffer size & DQN & [10, 100000] \\
    Actor-learning rate & A2C, PPO & [0.0001, 0.004] \\
    Critic-learning rate & A2C, PPO & [0.0001, 0.004] \\
    Clipping threshold & PPO & \{0.1, \textbf{0.2}, 0.3\} \\
    Num of epochs & PPO & \{\textbf{5}, 8, 10\} \\
    \bottomrule
    \end{tabular}
}
\begin{tablenotes}
    \item[] \tiny{\hspace{.3cm} * The parameter values jointly chosen for all algorithms are highlighted in bold.}
\end{tablenotes}
\end{threeparttable}
\label{tab:hyp}
\end{table}

\textbf{Effect of Batch Size on Deep RL Algorithm Performances:}\\
From a technical standpoint, there exist no foolproof techniques for identifying the ideal hyper-parameter configuration for training deep RL algorithms. In addition, the batch size has an impact on the convergence rate of the prediction network, with smaller batch sizes resulting in faster convergence and well-known degradation in model quality and generalization abilities that can occur with increased batch sizes \citep{keskar2016large}. As such, originally we aimed to select batch sizes that would converge within sample numbers comparable to the training period of biological cultures while attempting to prioritize computational efficiency, which is a significant area of interest in this study. Hence, opting for large batch sizes may significantly slow down the model convergence and would not confer any benefit to the RL algorithms under investigation.

\begin{figure*}[!ht]
  \centering
  \hspace{-1cm}
  {\includegraphics[width=1\textwidth]{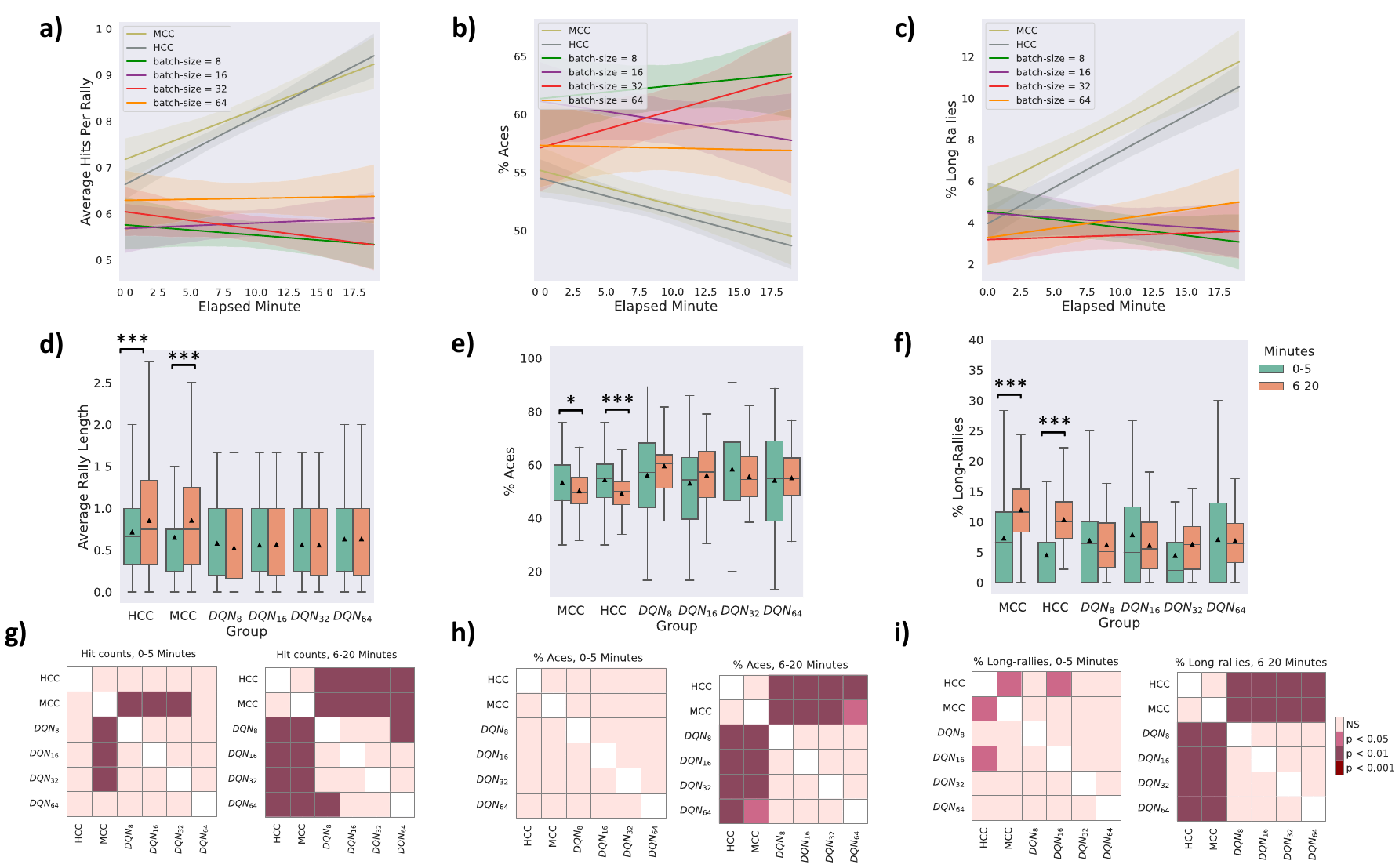}}
  \caption{\textbf{\Ima{} to DQN - Effects of changing the batch size.} The Average number of  \textbf {a)} hits-per-rally, \textbf {b)} $\%$ of aces, and \textbf {c)} $\%$ of long rallies over 20 minutes real-time equivalent of training DQN with batch sizes 8, 16, 32, 64, compared to the MCC and HCC cultures. {\bf d)} average rally length over time, {\bf e)}  Average \% of aces within groups and over time. {\bf f)} Average \% of long-rallies ($\geq$~3) performed in a session. {\bf g,h and i)} Pairwise Tukey's post-hoc test. Box plots show interquartile range, with bars demonstrating 1.5X interquartile range, the line marks the median and the black triangle marks the mean. Error bands = 1 SE.}
  \label{fig:DQN_batchsizes}
\end{figure*}

\begin{figure*}[!ht]
  \centering
  \hspace{-1cm}
  {\includegraphics[width=1\textwidth]{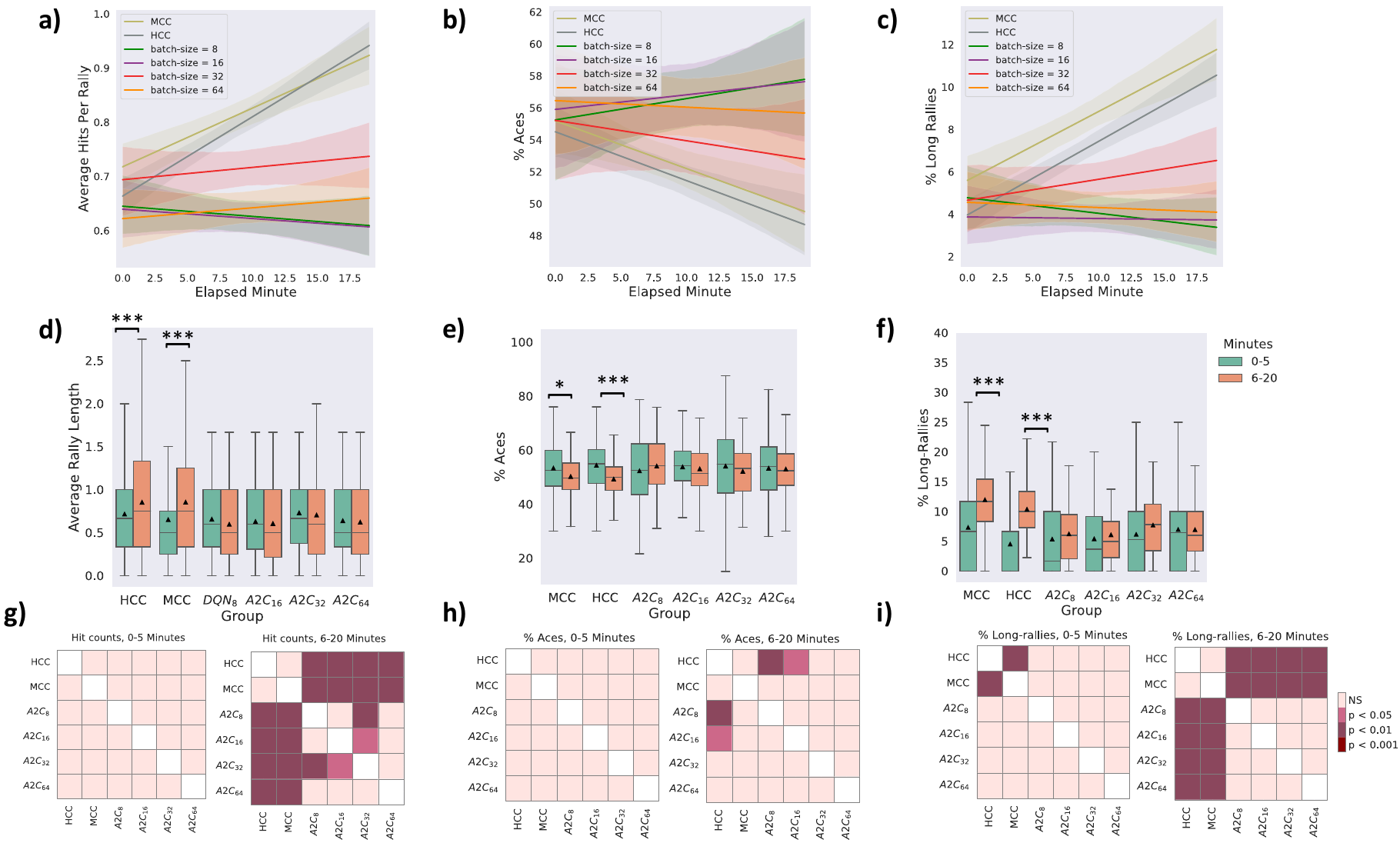}}
  \caption{\textbf{\Ima{} to A2C - Effects of changing the batch size.} The Average number of  \textbf {a)} hits-per-rally, \textbf {b)} $\%$ of aces, and \textbf {c)} $\%$ of long rallies over 20 minutes real-time equivalent of training A2C with batch sizes 8, 16, 32, 64, compared to the MCC and HCC cultures. {\bf d)} average rally length over time, {\bf e)}  Average \% of aces within groups and over time. {\bf f)} Average \% of long-rallies ($\geq$~3) performed in a session. {\bf g,h and i)} Pairwise Tukey's post-hoc test. Box plots show interquartile range, with bars demonstrating 1.5X interquartile range, the line marks the median and the black triangle marks the mean. Error bands = 1 SE.}
  \label{fig:A2C_batchsizes}
\end{figure*}

\begin{figure*}[!ht]
  \centering
  \hspace{-1cm}
  {\includegraphics[width=1\textwidth]{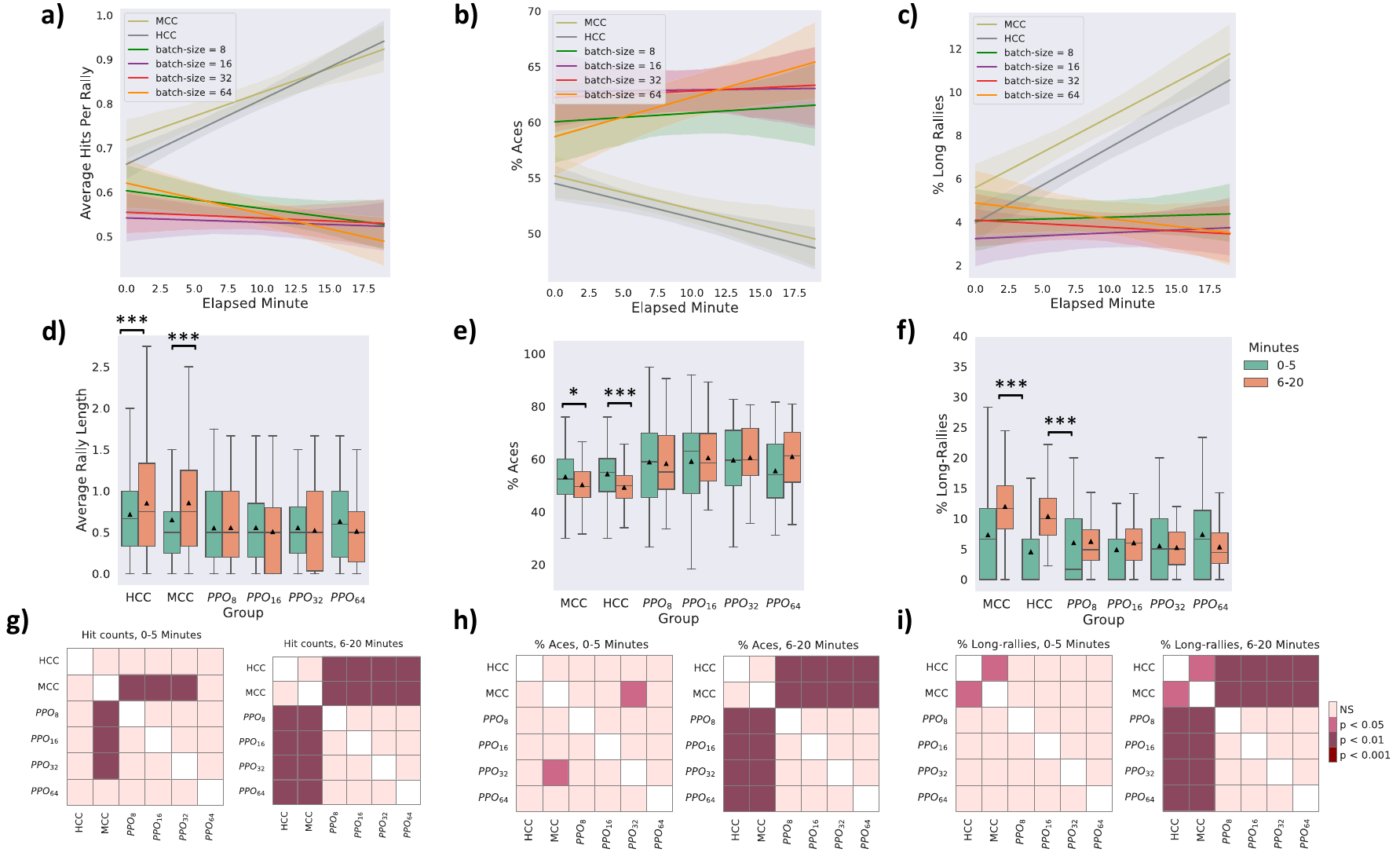}}
  \caption{\textbf{\Ima{} to PPO - Effects of changing the batch size.} The Average number of  \textbf {a)} hits-per-rally, \textbf {b)} $\%$ of aces, and \textbf {c)} $\%$ of long rallies over 20 minutes real-time equivalent of training PPO with batch sizes 8, 16, 32, 64, compared to the MCC and HCC cultures. {\bf d)} average rally length over time, {\bf e)}  Average \% of aces within groups and over time. {\bf f)} Average \% of long-rallies ($\geq$~3) performed in a session. {\bf g,h and i)} Pairwise Tukey's post-hoc test. Box plots show interquartile range, with bars demonstrating 1.5X interquartile range, the line marks the median and the black triangle marks the mean. Error bands = 1 SE.}
  \label{fig:PPO_batchsizes}
\end{figure*}

Figures \ref{fig:DQN_batchsizes}, \ref{fig:A2C_batchsizes}, and \ref{fig:PPO_batchsizes} investigate the impact of changing batch sizes utilizing the \Ima{} design by incorporating batch sizes of 8, 16,  32, and 64 while keeping the rest of the hyper-parameters in each algorithm fixed at default levels similar to Figure \ref{fig:hp_exp} (i.e. \textit{learning rate = 0.0001}, \textit{batch size = 32}, and \textit{replay buffer size = 10000} for DQN and \textit{actor learning rate = 0.0001}, \textit{batch size = 32}, and \textit{critic learning rate = 0.001} for A2C and PPO).

% In general, we observed some quantitative changes in outcome metrics when varying the batch size for these algorithms, but these adjustments did not alter the ultimate conclusions of our work. Focusing on the quality of learning in each group and the comparison of sample efficiency, both of these were unaffected or in some cases worsened by increasing the batch size. Specifically, when examining the statistical significance of metric changes during the first 5 minutes versus the last 15 minutes of training and overall relative improvement, increased batch size did not appear to significantly impact the resulting sample efficiency in any of the algorithms as seen in \nameref{ex_data} Figure \ref{fig:Rel-batchsizes}.\\

In some cases, these results illustrate an unwanted trend in the main metrics of interest when increasing the batch size above certain levels. For instance, an increasing \% of aces in the DQN and PPO algorithms, decreasing average rally length in PPO, and decreasing \% of long rallies in both A2C and PPO algorithms are observed
which may eventually prevent the model from converging to the optima. This suggests that if the comparison were to be extended to a larger number of episodes for all groups, the increase in batch size would not necessarily yield improved performances, as evidenced by the undesirable trend observed in the aforementioned metrics (\nameref{ex_data} Figures \ref{fig:DQN_batchsizes}, \ref{fig:A2C_batchsizes}, and \ref{fig:PPO_batchsizes}). 
Notably, this may occur due to the fact that larger batch sizes make larger gradient steps than smaller batch sizes for the same number of samples seen and the update is heavily dependent on the specific samples drawn from the dataset. Conversely, a small batch size leads to updates that are more consistent in size, with the size of the update being only weakly dependent on which particular samples are selected from the dataset. In conclusion, it is possible that in deep neural networks, optimal weight configurations are located far from the initial weights. Hence, averaging the loss function over large batch sizes may not allow the model to explore a large enough space to reach the optimal weight configurations within the same number of training epochs.\\

\begin{figure*}[!ht]
  \centering
  \vspace{-0.5cm}
  % \hspace{-1cm}
  {\includegraphics[width=1\textwidth]{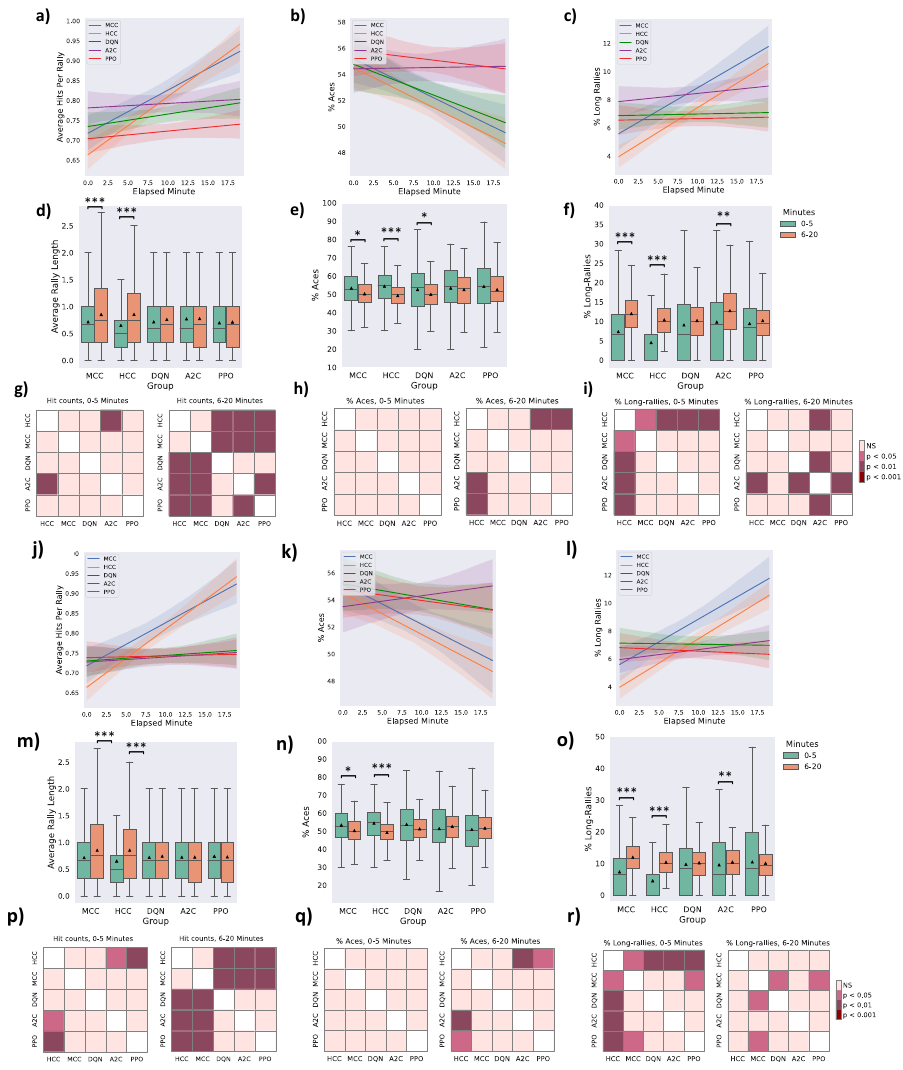}}
  \caption{\textbf{Additional hidden layers in the DQN algorithm.} \twod{} to the RL Algorithms: The average number of \textbf {a)} hits-per-rally, \textbf {b)} $\%$ of aces, and \textbf {c)} $\%$ of long rallies over 20 minutes real-time equivalent of training DQN (2 additional hidden layers, batch size = 32), A2C, PPO, and MCC, HCC cultures. {\bf d)} average rally length over time, {\bf e)}  Average \% of aces within groups and over time. {\bf f)} Average \% of long-rallies ($\geq$~3) performed in a session. {\bf g,h and i)} Pairwise Tukey's post-hoc test.
  \Lo{} to the RL Algorithms: The average number of  \textbf {j)} hits-per-rally, \textbf {k)} $\%$ of aces, and \textbf {l)} $\%$ of long rallies over 20 minutes real-time equivalent of training DQN (2 additional hidden layers, batch size = 32), A2C, PPO, and MCC, HCC cultures. {\bf m)} average rally length over time, {\bf n)}  Average \% of aces within groups and over time. {\bf o)} Average \% of long-rallies ($\geq$~3) performed in a session. {\bf p,q and r)} Pairwise Tukey's post hoc test. 
  Box plots show interquartile range, with bars demonstrating 1.5X interquartile range, the line marks the median and the black triangle marks the mean. Error bands = 1 SE.}
  \label{fig:Hidden}
\end{figure*}

\noindent \textbf{Effects of Adding Hidden Layers on DQN Performance:}\\
To evaluate the effect of adding extra hidden layers on the performance of \twod{} and \Lo{} designs to the DQN algorithm, we implemented them by adding 2 additional hidden layers before the output layer and incorporating a batch size = 32. \nameref{ex_data} Figure \ref{fig:Hidden} shows the outcomes of these adjustments.

\begin{figure*}[!ht]
  \centering
  \hspace{-1cm}
  {\includegraphics[width=1\textwidth]{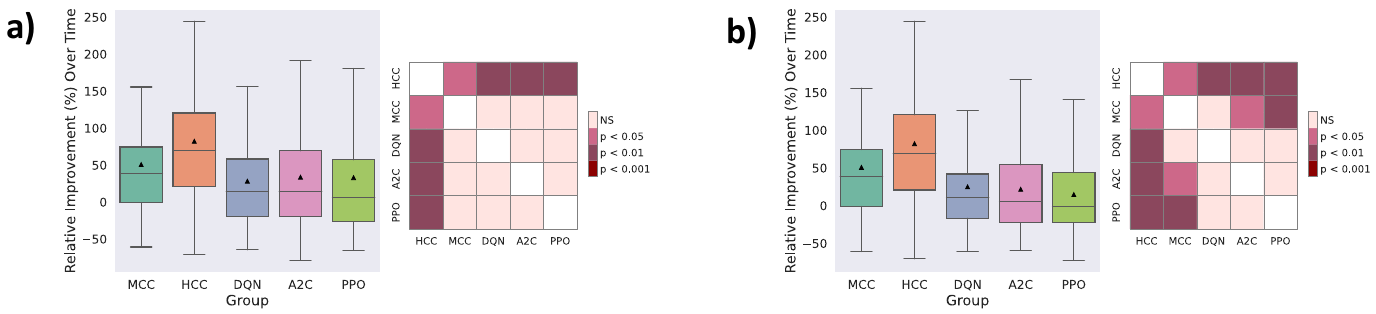}}
  \caption{\textbf{Relative improvement over time with additional hidden layers in DQN algorithm.} Relative improvement (\%) in the average hit counts between the first 5 minutes and the last 15 minutes of all sessions in each separate group for \textbf{a)} \twod{} design for DQN with 2 additional hidden layers, \textbf{b)} \Lo{} design for DQN with 2 additional hidden layers.}
  \label{fig:Rel-hiddenlayers}
\end{figure*}

This further analysis revealed that although certain metrics exhibited qualitative and quantitative changes in their trends, the overall sample efficiency performance remained unaffected and even worsened with the inclusion of additional hidden layers. 
For example, we noted a degradation and an unwanted decreasing trend in the DQN's performance in the \% of long rallies for the \Lo{} design. This resulted in the MCC group significantly outperforming DQN \Lo{} design in terms of \% of long Rallies during the second 15 minutes. The performance of DQN in terms of average rally length was also deteriorated by the addition of these layers. On the other hand, MCC no longer demonstrated a significantly superior performance in terms of \% of aces by the addition of hidden layers to the \Lo{} design. While some level of improvement was detected in the DQN group with the \twod{} design (specifically in the \% of aces achieved) by the addition of the extra layers, overall performance in all 3 metrics was still inferior to those of the biological cultures in all the metrics.
Specifically as illustrated in \nameref{ex_data} Figure \ref{fig:Rel-hiddenlayers}, the HCC group still demonstrated significant outperformance compared to the DQN \Lo{} and \twod{} designs in terms of relative improvement. The relative improvement in both of the \Lo{} and \twod{} designs showed a decay compared to the results reported in the main text, where this level of outperformance of MCC over DQN was not observed in the absence of hidden layers.
 
The observed deteriorated performance in terms of relative improvement in the \Lo{} design may be attributed to decreased generalization capabilities and higher variance resulting from the introduction of additional hidden layers. Because, for simpler tasks, a smaller network with fewer hidden layers might be sufficient to achieve good performance, and adding more layers could lead to overfitting. Thereby, this declined performance in the relative improvement as well as the low dimensionality of the input information in these designs combined with the faster computational performance of the algorithm with fewer hidden layers can justify the use of the shallower design for comparison reasons.\\

\subsection{Active Inference Agent}\label{active_inf}
While RL algorithms use back-propagation, it has been argued that this method is likely too inefficient to function within biological systems. Therefore, we attempted to evaluate the sample efficiency of more biologically inspired algorithms, by implementing a counterfactual learning active inference agent \citep{isomura2020reverse, isomura2022canonical}. Our preliminary findings show that one can use a generic active inference agent which can then mimic the performance of the DishBrain system depending on additional parameters such as memory.

\begin{figure*}[!ht]
  \centering
  % \hspace{-1cm}
  {\includegraphics[width=1\textwidth]{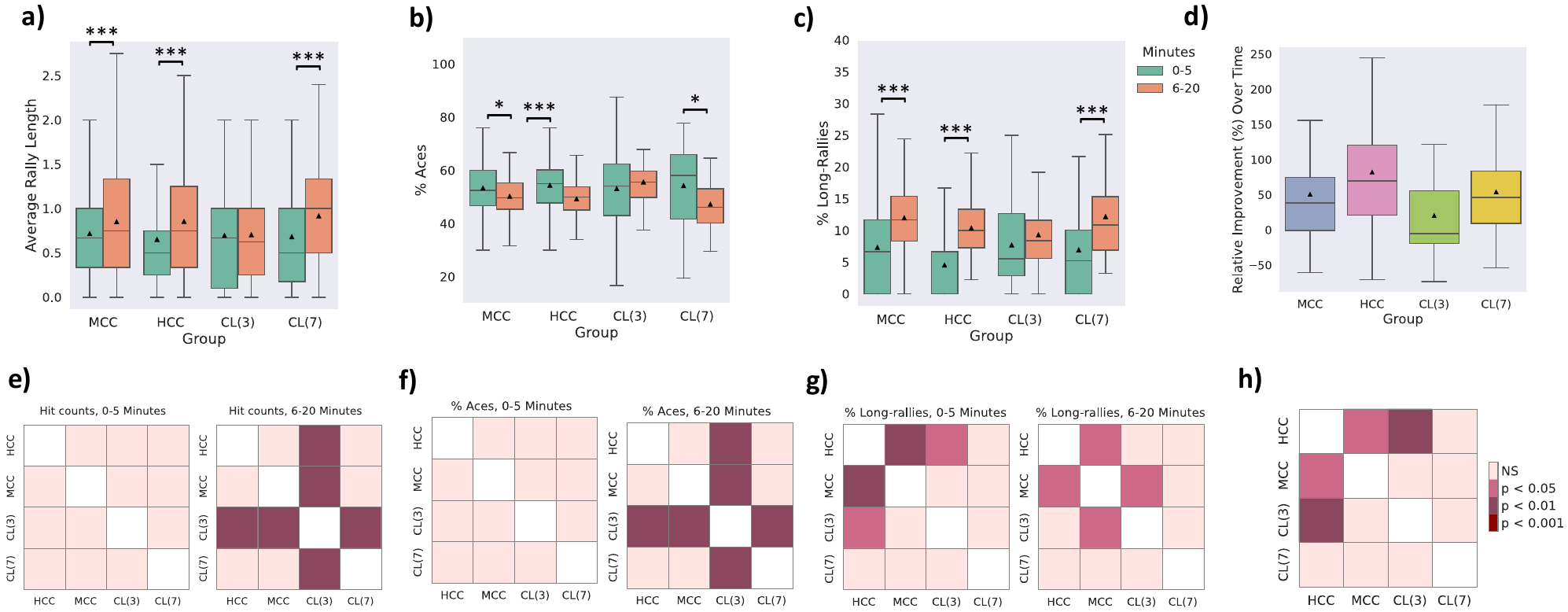}}
  \caption{\textbf{Comparing Active Inference agent with biological neurons.} {\bf a)} Average rally length over time where this within-group increase was significant for all groups except CL(3) (One-way ANOVA test, p = 5.854e-6, p = 7.936-17, p = 0.873, and p = 2.254e-6, for MCC, HCC, CL(3), and CL(7) respectively). {\bf b)}  Average \% of aces within groups and over time where this within-group increase was significant only for HCC, MCC, and CL(7) groups (One-way ANOVA test, p = 0.014, p = 2.907e-08, p = 0.380, and p = 0.016, for MCC, HCC, CL(3), and CL(7) respectively). {\bf c)} Average \% of long-rallies ($\geq$~3) performed in a session where the increase over time was significant for all groups except CL(3) (One-way ANOVA test, p = 1.172e-7, p = 1.525e-24, p = 0.253, and p = 8.944e-4 for MCC, HCC, CL(3), and CL(7), respectively). 
  {\bf d} Relative improvement (\%) in the average hit counts between the first 5 minutes and the last 15 minutes of all sessions in each separate group. {\bf e,f,g and h)} Pairwise post hoc tests. Box plots show interquartile range, with bars demonstrating 1.5X interquartile range, the line marks the median and the black triangle marks the mean. Error bands = 1 SE.}
  \label{fig:active_inf}
\end{figure*}

The active inference framework is a formal way of modelling the behaviour of self-organising systems that interface with the external world and maintain a consistent form over time \citep{Friston2021,Kaplan2018,Kuchling2020}. The framework assumes that agents embody generative models of the environment they interact with, on which they base their behaviour \citep{Tschantz2020, Parr2018}. A recent active inference scheme is shown to be mathematically equivalent to a particular class of neural networks accompanied by some neuromodulations of synaptic plasticity \citep{isomura2020reverse, isomura2022canonical}. It uses counterfactual learning (CL) to accumulate a measure of risk over time based on feedback from the environment. Subsequent work that validates this scheme experimentally using \textit{in vitro} neural networks has also appeared recently \citep{Isomura2022}. Of particular note, the training schematic for the DishBrain system was inspired by implications from theory on active inference via the Free Energy Principle, making it the most suitable algorithm to compare here \citep{kagan2022vitro}. 
Here, we focus on generative models in the form of Partially Observable Markov Decision Processes (POMDPs) for their simplicity and ubiquitous use in the optimal control literature \citep{Lovejoy1991,Shani2013,Kaelbling1998}.\\
Gameplay performance of these agents with two different memory horizons of 3 (CL(3)) and 7 (CL(7)) is summarised in Figure \ref{fig:active_inf}. We see that the CL(7) agents perform at par and in some cases better than the HCC group and are the only group where the HCC has no significant outperformance over them in terms of the relative improvement in time (see Figure \ref{fig:active_inf}.h). However, this is not the case for CL(3) agents which have a smaller memory horizon. While further exploring this active inference framework is out of scope for this paper, it does highlight the value of using biologically inspired algorithms in terms of sample efficiency.

\noindent \textbf{Generative model of the pong game environment:}\\
Assuming agents have a discrete representation of their surrounding environment, we turn to the POMDP framework \citep{Kaelbling1998}. POMDPs offer a fairly expressive structure to model discrete state-space environments where parameters can be expressed as tractable categorical distributions. The POMDP-based generative model can be formally defined as a tuple of finite sets $(S, O, U, \mathbb{B},\mathbb{A})$:

\begin{itemize}
    \item [$\circ$] $s \in S:$ $S$ is a set of hidden states ($s$) causing observations $o$.
    \item [$\circ$] $o \in O:$ $O$ is a set of observations, where $o=s$, in the fully observable setting. In a partially observable setting,  $o=f(s)$.
    \item [$\circ$] $u \in U:$ $U$ is a set of actions ($u$). E.g., $U = $ 
    $ \{ Up, Stay, Down \}$. 
    \item [$\circ$]$\mathbb{B}:$ encodes the one-step transition dynamics, $P(s_{t} \vert s_{t-1}, u_{t-1})$ i.e., the probability that when action $u_{t-1}$ is taken while being in state $s_{t-1}$ (at time $t-1$) results in $s_{t}$ at time $t$.
    \item [$\circ$] $\mathbb{A}:$ encodes the likelihood mapping, $P(o_{\tau} \vert s_{\tau})$ for  the partially observable setting.
    \item [$\circ$] $\mathbb{D}:$ Encodes the prior of the agent about the hidden state factor $s$.
    \item [$\circ$] $\mathbb{E}:$ Encodes the prior of the agent about actions $u$.
    
\end{itemize}

In a POMDP, the hidden states ($s$) generate observations ($o$) through the likelihood mapping ($\mathbb{A}$) in the form of a categorical distribution, $P(o_{\tau} \vert s_{\tau}) = \mathrm{Cat}(\mathbb{A} \times s_\tau)$. %where $\mathbb{A}[:,i]$ is the i-th column in $\mathbb{A}$.
$\mathbb{B}$ is a collection of square matrices $\mathbb{B}_{u}$, where $\mathbb{B}_{u}$ represents transition dynamics $P(s_{t} \vert s_{t-1}, u_{t-1} = u$): The transition matrix ($\mathbb{B}$) determines the dynamics of $s$  given the agent's action $u$ as $P(s_{t} \vert s_{t-1}, u_{t-1}) = \mathrm{Cat}(\mathbb{B}_{u_{t-1}} \times s_{t-1})$. In $\left[\mathbb{A} \times s_\tau \right]$ and $\left[ \mathbb{B}_{u_\tau} \times s_\tau \right]$, $s_\tau$ is represented as a one-hot vector that is multiplied through regular matrix multiplication \footnote{One-hot is a group of bits among which the legal combinations of values are only those with a single high (1) bit and all the others low (0). Here, the bit (1) is allocated to the state $s= s_\tau$ }. The \textit{Markovianity} of POMDPs means that state transitions are independent of history (i.e. state $s_{t}$ only depends upon the state-action pair $(s_{t-1}, u_{t-1})$ and not $s_{t-2}, ~u_{t-2}$ etc.).

The generative model can be summarised as follows,
\begin{equation}
P(o_{1:t},s_{1:t},u_{1:t}) = P(\mathbb{A}) P(\mathbb{B}) P(\mathbb{D}) P(\mathbb{E}) \prod_{\tau=1}^t P(o_{\tau} \vert s_{\tau}, \mathbb{A}) \prod_{\tau=2}^t P(s_{\tau} \vert s_{\tau-1}, u_{\tau-1},\mathbb{B}).
\end{equation}
So, from the agent's perspective, when encountering a stream of observations in time, such as $(o_{1}, o_{2}, o_{3}, ..., o_{t})$, as a consequence of performing a stream of actions $(u_{1}, u_{2}, u_{3}, ..., u_{t-1})$, the generative model quantitatively couples and quantifies the causal relationship from action to observation through some assumed hidden states of the environment. These are called `hidden' states because, in POMDPs, the agent cannot observe them directly. Based on this representation, an agent can now attempt to optimise its actions to keep receiving preferred observations.\\

The generative model structure used explicitly for the pong game environment is summarised below:

\begin{itemize}
    \item \textbf{$x-$axis location of the ball:} Communicated to DishBrain using a stimulation between 4-40 HZ, i.e. 37 states. 
    \item \textbf{$y-$axis location of the ball:} Communicated to DishBrain through 8 sensory electrodes, i.e. 8 states.
    \item \textbf{$y-$axis location of the paddle:} Assumed to be part of DishBrain's generative model as control is exerted, i.e. 8 states.    
    \item \textbf{Structure:}
    State Space =
    $37 * 8 * 8$ states,
    Action Space =
    $\{$Up, Down, Stay$\}$
\end{itemize}

\noindent \textbf{Counterfactual learning algorithm:}\\
In the counterfactual variant of active inference, the agent learns a state-action mapping $\mathbb{C_{P}}$. For the exact form of the generative model and free energy, refer to \citep{isomura2020reverse}. This state-action mapping is learned using a 'Risk' parameter $\Gamma(t)$ using the update equation as given in \citep{isomura2020reverse} as:

\begin{equation}
    \mathbb{C}_{P} \leftarrow \mathbb{C}_{P} + t ~ \langle (1 - 2 ~\Gamma(t)) \langle u_{t} \otimes s_{t-1} \rangle \rangle.
    \label{cp-clmodeleqn}
\end{equation}

Here, $\langle \cdot \rangle$ refers to the average over time, and $\otimes$ is the Kronecker-product operator.
Given the state-action mapping $\mathbb{C}_{P}$, agent samples actions from the distribution, 
\begin{equation}
    P(u \vert s)_{CL} = \sigma \left( \ln~ \mathbb{C}_{P} \cdot s_{t-1} \right).
    \label{cl-action-selection-eqn}
\end{equation}

For the complete model, refer to \citep{isomura2020reverse}. The free parameter in our model is the number of past instances (of state-action pairs) the agent stores in memory use in every time-step to learn $\mathbb{C}_{P}$ in Eq.\ref{cp-clmodeleqn}. In the article, we use active inference agents with memory horizons of 3 and 7.

The functional form of $\Gamma(t)$ used in the simulations of this work is:
\begin{equation}
    \Gamma(t)_{prior} = 0.55 
    \label{gamma-prior-eqn}
\end{equation}

The value of 0.55 corresponds to a bias of ``higher risk'' in the CL method. An initial value greater than 0.5 is necessary to enable learning.

For updating $\Gamma$, we use the equation,
\begin{equation}
    \Gamma(t) \leftarrow \Gamma(t) - \frac{1}{T_{goal} - t}.
    \label{gamma-update-eqn}
\end{equation}

Here, $T_{goal}$ is when the agent reaches the goal state (received a positive reward from the environment). So, the sooner the agent reaches the goal state, the quicker the $\Gamma(t)$, i.e., risk converges to zero. All the update rules defined in the paper can be derived from the postulate that the agent tries to minimise the (variational) free energy w.r.t the generative model \citep{paul2021active,isomura2020reverse}.\\

% \end{comment}

\newpage
\section{Extended Data}\label{ex_data}

\clearpage
% \begin{figure*}[!ht]
% \vspace{10cm}
  % \centering
  \hspace{-3cm}
\AtEndDocument{

\section*{Supplementary material}

**Table S2.** Follow up main and supplementary text post-hoc tests for multivariate tests, including means, standard error (SE), t-scores, degree of freedom and exact p-values with hedges.

| Figure | Panel | Parameters | Source | A | B | Mean (A) | Mean (B) | diff | se | T | p-val | hedges | Method |
|--------|-------|-----------|--------|---|---|----------|----------|------|-----|---|-------|--------|--------|
| 2 | h | Hit Counts | 0-5 Minutes | A2C | DQN | 0.705 | 0.709 | -0.004 | 0.027 | -0.147 | 0.900 | -0.008 | Tukey's |
| | | | | A2C | HCC | 0.705 | 0.651 | 0.055 | 0.025 | 2.151 | 0.199 | 0.110 | |
| | | | | A2C | MCC | 0.705 | 0.716 | -0.011 | 0.029 | -0.373 | 0.900 | -0.021 | |
| | | | | A2C | PPO | 0.705 | 0.716 | -0.011 | 0.029 | -0.388 | 0.900 | -0.022 | |
| | | | | DQN | HCC | 0.709 | 0.651 | 0.059 | 0.025 | 2.310 | 0.142 | 0.117 | |
| | | | | DQN | MCC | 0.709 | 0.716 | -0.007 | 0.029 | -0.237 | 0.900 | -0.014 | |
| | | | | DQN | PPO | 0.709 | 0.716 | -0.007 | 0.029 | -0.251 | 0.900 | -0.014 | |
| | | | | HCC | MCC | 0.651 | 0.716 | -0.065 | 0.027 | -2.386 | 0.120 | -0.131 | |
| | | | | HCC | PPO | 0.651 | 0.716 | -0.066 | 0.027 | -2.410 | 0.113 | -0.132 | |
| | | | | MCC | PPO | 0.716 | 0.716 | -0.000 | 0.030 | -0.013 | 0.900 | -0.001 | |
| | | | 6-20 Minutes | A2C | DQN | 0.738 | 0.738 | 0.000 | 0.018 | 0.004 | 0.900 | 0.000 | |
| | | | | A2C | HCC | 0.738 | 0.854 | -0.117 | 0.017 | -6.726 | 0.001 | -0.198 | |
| | | | | A2C | MCC | 0.738 | 0.852 | -0.115 | 0.020 | -5.715 | 0.001 | -0.194 | |
| | | | | A2C | PPO | 0.738 | 0.709 | 0.029 | 0.019 | 1.506 | 0.551 | 0.049 | |
| | | | | DQN | HCC | 0.738 | 0.854 | -0.117 | 0.017 | -6.737 | 0.001 | -0.198 | |
| | | | | DQN | MCC | 0.738 | 0.852 | -0.115 | 0.020 | -5.723 | 0.001 | -0.194 | |
| | | | | DQN | PPO | 0.738 | 0.709 | 0.029 | 0.019 | 1.503 | 0.552 | 0.049 | |
| | | | | HCC | MCC | 0.854 | 0.852 | 0.002 | 0.020 | 0.101 | 0.900 | 0.003 | |
| | | | | HCC | PPO | 0.854 | 0.709 | 0.146 | 0.019 | 7.818 | 0.001 | 0.246 | |
| | | | | MCC | PPO | 0.852 | 0.709 | 0.144 | 0.021 | 6.778 | 0.001 | 0.243 | |
| | i | % Aces | 0-5 Minutes | A2C | DQN | 51.842 | 52.190 | -0.347 | 1.425 | -0.244 | 0.900 | -0.028 | Tukey's |
| | | | | A2C | HCC | 51.842 | 54.382 | -2.539 | 1.375 | -1.847 | 0.348 | -0.205 | |
| | | | | A2C | MCC | 51.842 | 53.333 | -1.490 | 1.549 | -0.962 | 0.859 | -0.120 | |
| | | | | A2C | PPO | 51.842 | 54.731 | -2.889 | 1.425 | -2.028 | 0.254 | -0.234 | |
| | | | | DQN | HCC | 52.190 | 54.382 | -2.192 | 1.375 | -1.595 | 0.501 | -0.177 | |
| | | | | DQN | MCC | 52.190 | 53.333 | -1.143 | 1.549 | -0.738 | 0.900 | -0.092 | |
| | | | | DQN | PPO | 52.190 | 54.731 | -2.542 | 1.425 | -1.784 | 0.385 | -0.205 | |
| | | | | HCC | MCC | 54.382 | 53.333 | 1.049 | 1.503 | 0.698 | 0.900 | 0.085 | |
| | | | | HCC | PPO | 54.382 | 54.731 | -0.350 | 1.375 | -0.254 | 0.900 | -0.028 | |
| | | | | MCC | PPO | 53.333 | 54.731 | -1.399 | 1.549 | -0.903 | 0.893 | -0.113 | |
| | | | 6-20 Minutes | A2C | DQN | 50.284 | 50.136 | 0.148 | 0.946 | 0.157 | 0.900 | 0.018 | |
| | | | | A2C | HCC | 50.284 | 49.259 | 1.025 | 0.912 | 1.123 | 0.768 | 0.125 | |
| | | | | A2C | MCC | 50.284 | 50.232 | 0.052 | 1.028 | 0.051 | 0.900 | 0.006 | |
| | | | | A2C | PPO | 50.284 | 53.254 | -2.970 | 0.946 | -3.141 | 0.015 | -0.362 | |
| | | | | DQN | HCC | 50.136 | 49.259 | 0.877 | 0.912 | 0.961 | 0.860 | 0.107 | |
| | | | | DQN | MCC | 50.136 | 50.232 | -0.096 | 1.028 | -0.093 | 0.900 | -0.012 | |
| | | | | DQN | PPO | 50.136 | 53.254 | -3.118 | 0.946 | -3.298 | 0.009 | -0.380 | |
| | | | | HCC | MCC | 49.259 | 50.232 | -0.973 | 0.998 | -0.975 | 0.852 | -0.118 | |
| | | | | HCC | PPO | 49.259 | 53.254 | -3.995 | 0.912 | -4.378 | 0.001 | -0.487 | |
| | | | | MCC | PPO | 50.232 | 53.254 | -3.022 | 1.028 | -2.940 | 0.028 | -0.368 | |

| | | | | | | | | | | | | | |
|---|---|---|---|---|---|---|---|---|---|---|---|---|---|
| | j | % Long Rally | 0-5 Minutes | A2C | DQN | 7.421 | 9.789 | -2.368 | 0.913 | -2.594 | 0.073 | -0.299 | Tukey's |
| | | | | A2C | HCC | 7.421 | 4.523 | 2.898 | 0.881 | 3.290 | 0.009 | 0.366 | |
| | | | | A2C | MCC | 7.421 | 7.318 | 0.103 | 0.993 | 0.103 | 0.900 | 0.013 | |
| | | | | A2C | PPO | 7.421 | 8.122 | -0.701 | 0.913 | -0.767 | 0.900 | -0.088 | |
| | | | | DQN | HCC | 9.789 | 4.523 | 5.267 | 0.881 | 5.978 | 0.001 | 0.665 | |
| | | | | DQN | MCC | 9.789 | 7.318 | 2.471 | 0.993 | 2.489 | 0.094 | 0.312 | |
| | | | | DQN | PPO | 9.789 | 8.122 | 1.667 | 0.913 | 1.826 | 0.360 | 0.210 | |
| | | | | HCC | MCC | 4.523 | 7.318 | -2.796 | 0.963 | -2.903 | 0.031 | -0.353 | |
| | | | | HCC | PPO | 4.523 | 8.122 | -3.599 | 0.881 | -4.085 | 0.001 | -0.454 | |
| | | | | MCC | PPO | 7.318 | 8.122 | -0.803 | 0.993 | -0.809 | 0.900 | -0.101 | |
| | | | 6-20 Minutes | A2C | DQN | 10.034 | 10.248 | -0.214 | 0.623 | -0.344 | 0.900 | -0.040 | |
| | | | | A2C | HCC | 10.034 | 10.365 | -0.331 | 0.601 | -0.550 | 0.900 | -0.061 | |
| | | | | A2C | MCC | 10.034 | 11.972 | -1.938 | 0.677 | -2.863 | 0.035 | -0.358 | |
| | | | | A2C | PPO | 10.034 | 8.506 | 1.528 | 0.623 | 2.454 | 0.102 | 0.283 | |
| | | | | DQN | HCC | 10.248 | 10.365 | -0.116 | 0.601 | -0.194 | 0.900 | -0.022 | |
| | | | | DQN | MCC | 10.248 | 11.972 | -1.724 | 0.677 | -2.547 | 0.082 | -0.319 | |
| | | | | DQN | PPO | 10.248 | 8.506 | 1.743 | 0.623 | 2.798 | 0.042 | 0.322 | |
| | | | | HCC | MCC | 10.365 | 11.972 | -1.608 | 0.657 | -2.447 | 0.104 | -0.297 | |
| | | | | HCC | PPO | 10.365 | 8.506 | 1.859 | 0.601 | 3.094 | 0.017 | 0.344 | |
| | | | | MCC | PPO | 11.972 | 8.506 | 3.467 | 0.677 | 5.121 | 0.001 | 0.641 | |
| 3 | h | Hit Counts | 0-5 Minutes | A2C | DQN | 0.722 | 0.713 | 0.009 | 0.027 | 0.325 | 0.900 | 0.017 | Tukey's |
| | | | | A2C | HCC | 0.722 | 0.651 | 0.072 | 0.026 | 2.761 | 0.046 | 0.141 | |
| | | | | A2C | MCC | 0.722 | 0.716 | 0.006 | 0.029 | 0.216 | 0.900 | 0.012 | |
| | | | | A2C | PPO | 0.722 | 0.740 | -0.018 | 0.027 | -0.641 | 0.900 | -0.035 | |
| | | | | DQN | HCC | 0.713 | 0.651 | 0.063 | 0.026 | 2.428 | 0.108 | 0.123 | |
| | | | | DQN | MCC | 0.713 | 0.716 | -0.003 | 0.029 | -0.087 | 0.900 | -0.005 | |
| | | | | DQN | PPO | 0.713 | 0.740 | -0.026 | 0.027 | -0.968 | 0.856 | -0.052 | |
| | | | | HCC | MCC | 0.651 | 0.716 | -0.065 | 0.028 | -2.335 | 0.134 | -0.128 | |
| | | | | HCC | PPO | 0.651 | 0.740 | -0.089 | 0.026 | -3.428 | 0.006 | -0.175 | |
| | | | | MCC | PPO | 0.716 | 0.740 | -0.024 | 0.029 | -0.815 | 0.900 | -0.047 | |
| | | | 6-20 Minutes | A2C | DQN | 0.724 | 0.716 | 0.008 | 0.018 | 0.415 | 0.900 | 0.013 | |
| | | | | A2C | HCC | 0.724 | 0.854 | -0.131 | 0.018 | -7.461 | 0.001 | -0.220 | |
| | | | | A2C | MCC | 0.724 | 0.852 | -0.129 | 0.020 | -6.354 | 0.001 | -0.216 | |
| | | | | A2C | PPO | 0.724 | 0.727 | -0.004 | 0.018 | -0.217 | 0.900 | -0.007 | |
| | | | | DQN | HCC | 0.716 | 0.854 | -0.138 | 0.017 | -7.918 | 0.001 | -0.232 | |
| | | | | DQN | MCC | 0.716 | 0.852 | -0.136 | 0.020 | -6.743 | 0.001 | -0.229 | |
| | | | | DQN | PPO | 0.716 | 0.727 | -0.011 | 0.018 | -0.633 | 0.900 | -0.019 | |
| | | | | HCC | MCC | 0.854 | 0.852 | 0.002 | 0.020 | 0.100 | 0.900 | 0.003 | |
| | | | | HCC | PPO | 0.854 | 0.727 | 0.127 | 0.018 | 7.233 | 0.001 | 0.213 | |
| | | | | MCC | PPO | 0.852 | 0.727 | 0.125 | 0.020 | 6.158 | 0.001 | 0.210 | |
| | i | % Aces | 0-5 Minutes | A2C | DQN | 51.318 | 54.016 | -2.698 | 1.469 | -1.837 | 0.354 | -0.212 | Tukey's |
| | | | | A2C | HCC | 51.318 | 54.382 | -3.064 | 1.417 | -2.162 | 0.196 | -0.240 | |
| | | | | A2C | MCC | 51.318 | 53.333 | -2.014 | 1.597 | -1.262 | 0.690 | -0.158 | |
| | | | | A2C | PPO | 51.318 | 50.866 | 0.453 | 1.469 | 0.308 | 0.900 | 0.035 | |
| | | | | DQN | HCC | 54.016 | 54.382 | -0.366 | 1.417 | -0.258 | 0.900 | -0.029 | |
| | | | | DQN | MCC | 54.016 | 53.333 | 0.683 | 1.597 | 0.428 | 0.900 | 0.054 | |

| | | | | | | | | | | | | |
|---|---|---|---|---|---|---|---|---|---|---|---|---|
| | | | | DQN | PPO | 54.016 | 50.866 | 3.150 | 1.469 | 2.145 | 0.202 | 0.247 | |
| | | | | HCC | MCC | 54.382 | 53.333 | 1.049 | 1.550 | 0.677 | 0.900 | 0.082 | |
| | | | | HCC | PPO | 54.382 | 50.866 | 3.516 | 1.417 | 2.481 | 0.096 | 0.276 | |
| | | | | MCC | PPO | 53.333 | 50.866 | 2.467 | 1.597 | 1.545 | 0.529 | 0.193 | |
| | | | 6-20 Minutes | A2C | DQN | 52.596 | 53.001 | -0.404 | 0.919 | -0.440 | 0.900 | -0.051 | |
| | | | | A2C | HCC | 52.596 | 49.259 | 3.337 | 0.887 | 3.762 | 0.002 | 0.418 | |
| | | | | A2C | MCC | 52.596 | 50.232 | 2.364 | 0.999 | 2.366 | 0.126 | 0.296 | |
| | | | | A2C | PPO | 52.596 | 51.658 | 0.938 | 0.919 | 1.020 | 0.826 | 0.118 | |
| | | | | DQN | HCC | 53.001 | 49.259 | 3.741 | 0.887 | 4.218 | 0.001 | 0.469 | |
| | | | | DQN | MCC | 53.001 | 50.232 | 2.769 | 0.999 | 2.771 | 0.045 | 0.347 | |
| | | | | DQN | PPO | 53.001 | 51.658 | 1.342 | 0.919 | 1.460 | 0.577 | 0.168 | |
| | | | | HCC | MCC | 49.259 | 50.232 | -0.973 | 0.970 | -1.003 | 0.836 | -0.122 | |
| | | | | HCC | PPO | 49.259 | 51.658 | -2.399 | 0.887 | -2.705 | 0.054 | -0.301 | |
| | | | | MCC | PPO | 50.232 | 51.658 | -1.427 | 0.999 | -1.428 | 0.595 | -0.179 | |
| | j | % Long Rally | 0-5 Minutes | A2C | DQN | 9.519 | 10.105 | -0.586 | 0.990 | -0.591 | 0.900 | -0.068 | Tukey's |
| | | | | A2C | HCC | 9.519 | 4.523 | 4.997 | 0.955 | 5.230 | 0.001 | 0.581 | |
| | | | | A2C | MCC | 9.519 | 7.318 | 2.201 | 1.076 | 2.045 | 0.246 | 0.256 | |
| | | | | A2C | PPO | 9.519 | 10.462 | -0.942 | 0.990 | -0.952 | 0.865 | -0.110 | |
| | | | | DQN | HCC | 10.105 | 4.523 | 5.582 | 0.955 | 5.843 | 0.001 | 0.650 | |
| | | | | DQN | MCC | 10.105 | 7.318 | 2.787 | 1.076 | 2.589 | 0.074 | 0.324 | |
| | | | | DQN | PPO | 10.105 | 10.462 | -0.357 | 0.990 | -0.360 | 0.900 | -0.042 | |
| | | | | HCC | MCC | 4.523 | 7.318 | -2.796 | 1.044 | -2.677 | 0.059 | -0.325 | |
| | | | | HCC | PPO | 4.523 | 10.462 | -5.939 | 0.955 | -6.217 | 0.001 | -0.691 | |
| | | | | MCC | PPO | 7.318 | 10.462 | -3.144 | 1.076 | -2.921 | 0.030 | -0.366 | |
| | | | 6-20 Minutes | A2C | DQN | 10.431 | 11.238 | -0.807 | 0.616 | -1.311 | 0.661 | -0.151 | |
| | | | | A2C | HCC | 10.431 | 10.365 | 0.066 | 0.594 | 0.111 | 0.900 | 0.012 | |
| | | | | A2C | MCC | 10.431 | 11.972 | -1.541 | 0.669 | -2.303 | 0.145 | -0.288 | |
| | | | | A2C | PPO | 10.431 | 10.049 | 0.382 | 0.616 | 0.620 | 0.900 | 0.071 | |
| | | | | DQN | HCC | 11.238 | 10.365 | 0.873 | 0.594 | 1.470 | 0.571 | 0.163 | |
| | | | | DQN | MCC | 11.238 | 11.972 | -0.734 | 0.669 | -1.097 | 0.783 | -0.137 | |
| | | | | DQN | PPO | 11.238 | 10.049 | 1.189 | 0.616 | 1.931 | 0.302 | 0.222 | |
| | | | | HCC | MCC | 10.365 | 11.972 | -1.608 | 0.649 | -2.475 | 0.097 | -0.301 | |
| | | | | HCC | PPO | 10.365 | 10.049 | 0.316 | 0.594 | 0.531 | 0.900 | 0.059 | |
| | | | | MCC | PPO | 11.972 | 10.049 | 1.923 | 0.669 | 2.873 | 0.034 | 0.360 | |
| 4 | h | Hit Counts | 0-5 Minutes | A2C | DQN | 0.771 | 0.687 | 0.084 | 0.028 | 2.980 | 0.024 | 0.159 | Tukey's |
| | | | | A2C | HCC | 0.771 | 0.651 | 0.121 | 0.027 | 4.507 | 0.001 | 0.229 | |
| | | | | A2C | MCC | 0.771 | 0.716 | 0.055 | 0.030 | 1.826 | 0.359 | 0.105 | |
| | | | | A2C | PPO | 0.771 | 0.698 | 0.073 | 0.028 | 2.593 | 0.072 | 0.139 | |
| | | | | DQN | HCC | 0.687 | 0.651 | 0.037 | 0.027 | 1.371 | 0.628 | 0.070 | |
| | | | | DQN | MCC | 0.687 | 0.716 | -0.029 | 0.030 | -0.951 | 0.866 | -0.055 | |
| | | | | DQN | PPO | 0.687 | 0.698 | -0.011 | 0.028 | -0.375 | 0.900 | -0.020 | |
| | | | | HCC | MCC | 0.651 | 0.716 | -0.065 | 0.029 | -2.262 | 0.158 | -0.124 | |
| | | | | HCC | PPO | 0.651 | 0.698 | -0.047 | 0.027 | -1.759 | 0.399 | -0.090 | |
| | | | | MCC | PPO | 0.716 | 0.698 | 0.018 | 0.030 | 0.598 | 0.900 | 0.034 | |
| | | | 6-20 Minutes | A2C | DQN | 0.777 | 0.687 | 0.090 | 0.018 | 4.982 | 0.001 | 0.150 | |
| | | | | A2C | HCC | 0.777 | 0.854 | -0.077 | 0.018 | -4.348 | 0.001 | -0.128 | |

| | | | | | | | | | | | |
|---|---|---|---|---|---|---|---|---|---|---|---|
| | | | A2C | MCC | 0.777 | 0.852 | -0.075 | 0.020 | -3.662 | 0.002 | -0.125 | |
| | | | A2C | PPO | 0.777 | 0.712 | 0.065 | 0.018 | 3.576 | 0.003 | 0.108 | |
| | | | DQN | HCC | 0.687 | 0.854 | -0.167 | 0.018 | -9.521 | 0.001 | -0.278 | |
| | | | DQN | MCC | 0.687 | 0.852 | -0.165 | 0.020 | -8.119 | 0.001 | -0.275 | |
| | | | DQN | PPO | 0.687 | 0.712 | -0.025 | 0.018 | -1.389 | 0.617 | -0.042 | |
| | | | HCC | MCC | 0.854 | 0.852 | 0.002 | 0.020 | 0.099 | 0.900 | 0.003 | |
| | | | HCC | PPO | 0.854 | 0.712 | 0.142 | 0.018 | 8.044 | 0.001 | 0.236 | |
| | | | MCC | PPO | 0.852 | 0.712 | 0.140 | 0.020 | 6.854 | 0.001 | 0.233 | |
| i | % Aces | 0-5 Minutes | A2C | DQN | 53.293 | 55.443 | -2.150 | 1.473 | -1.459 | 0.577 | -0.168 | Tukey's |
| | | | A2C | HCC | 53.293 | 54.382 | -1.089 | 1.422 | -0.766 | 0.900 | -0.085 | |
| | | | A2C | MCC | 53.293 | 53.333 | -0.040 | 1.602 | -0.025 | 0.900 | -0.003 | |
| | | | A2C | PPO | 53.293 | 54.248 | -0.956 | 1.473 | -0.649 | 0.900 | -0.075 | |
| | | | DQN | HCC | 55.443 | 54.382 | 1.061 | 1.422 | 0.746 | 0.900 | 0.083 | |
| | | | DQN | MCC | 55.443 | 53.333 | 2.110 | 1.602 | 1.317 | 0.658 | 0.165 | |
| | | | DQN | PPO | 55.443 | 54.248 | 1.194 | 1.473 | 0.811 | 0.900 | 0.093 | |
| | | | HCC | MCC | 54.382 | 53.333 | 1.049 | 1.554 | 0.675 | 0.900 | 0.082 | |
| | | | HCC | PPO | 54.382 | 54.248 | 0.133 | 1.422 | 0.094 | 0.900 | 0.010 | |
| | | | MCC | PPO | 53.333 | 54.248 | -0.916 | 1.602 | -0.572 | 0.900 | -0.072 | |
| | | 6-20 Minutes | A2C | DQN | 52.530 | 53.879 | -1.349 | 0.966 | -1.397 | 0.613 | -0.161 | |
| | | | A2C | HCC | 52.530 | 49.259 | 3.270 | 0.932 | 3.508 | 0.004 | 0.390 | |
| | | | A2C | MCC | 52.530 | 50.232 | 2.298 | 1.050 | 2.188 | 0.185 | 0.274 | |
| | | | A2C | PPO | 52.530 | 52.511 | 0.018 | 0.966 | 0.019 | 0.900 | 0.002 | |
| | | | DQN | HCC | 53.879 | 49.259 | 4.620 | 0.932 | 4.955 | 0.001 | 0.551 | |
| | | | DQN | MCC | 53.879 | 50.232 | 3.647 | 1.050 | 3.472 | 0.005 | 0.435 | |
| | | | DQN | PPO | 53.879 | 52.511 | 1.368 | 0.966 | 1.415 | 0.602 | 0.163 | |
| | | | HCC | MCC | 49.259 | 50.232 | -0.973 | 1.019 | -0.954 | 0.864 | -0.116 | |
| | | | HCC | PPO | 49.259 | 52.511 | -3.252 | 0.932 | -3.488 | 0.005 | -0.388 | |
| | | | MCC | PPO | 50.232 | 52.511 | -2.280 | 1.050 | -2.170 | 0.192 | -0.272 | |
| j | % Long Rally | 0-5 Minutes | A2C | DQN | 9.810 | 9.554 | 0.256 | 0.935 | 0.274 | 0.900 | 0.032 | Tukey's |
| | | | A2C | HCC | 9.810 | 4.523 | 5.288 | 0.902 | 5.861 | 0.001 | 0.652 | |
| | | | A2C | MCC | 9.810 | 7.318 | 2.492 | 1.016 | 2.452 | 0.103 | 0.307 | |
| | | | A2C | PPO | 9.810 | 9.403 | 0.408 | 0.935 | 0.436 | 0.900 | 0.050 | |
| | | | DQN | HCC | 9.554 | 4.523 | 5.032 | 0.902 | 5.577 | 0.001 | 0.620 | |
| | | | DQN | MCC | 9.554 | 7.318 | 2.236 | 1.016 | 2.200 | 0.181 | 0.275 | |
| | | | DQN | PPO | 9.554 | 9.403 | 0.151 | 0.935 | 0.162 | 0.900 | 0.019 | |
| | | | HCC | MCC | 4.523 | 7.318 | -2.796 | 0.986 | -2.834 | 0.038 | -0.344 | |
| | | | HCC | PPO | 4.523 | 9.403 | -4.880 | 0.902 | -5.410 | 0.001 | -0.601 | |
| | | | MCC | PPO | 7.318 | 9.403 | -2.085 | 1.016 | -2.051 | 0.243 | -0.257 | |
| | | 6-20 Minutes | A2C | DQN | 12.722 | 9.511 | 3.211 | 0.632 | 5.083 | 0.001 | 0.585 | |
| | | | A2C | HCC | 12.722 | 10.365 | 2.357 | 0.610 | 3.868 | 0.001 | 0.430 | |
| | | | A2C | MCC | 12.722 | 11.972 | 0.750 | 0.687 | 1.092 | 0.786 | 0.137 | |
| | | | A2C | PPO | 12.722 | 10.183 | 2.540 | 0.632 | 4.020 | 0.001 | 0.463 | |
| | | | DQN | HCC | 9.511 | 10.365 | -0.854 | 0.610 | -1.401 | 0.611 | -0.156 | |
| | | | DQN | MCC | 9.511 | 11.972 | -2.461 | 0.687 | -3.584 | 0.003 | -0.449 | |
| | | | DQN | PPO | 9.511 | 10.183 | -0.672 | 0.632 | -1.063 | 0.802 | -0.122 | |
| | | | HCC | MCC | 10.365 | 11.972 | -1.608 | 0.666 | -2.412 | 0.113 | -0.293 | |

| | | | | | | | | | | | | | |
|---|---|---|---|---|---|---|---|---|---|---|---|---|---|
| | | | | HCC | PPO | 10.365 | 10.183 | 0.182 | 0.610 | 0.299 | 0.900 | 0.033 | |
| | | | | MCC | PPO | 11.972 | 10.183 | 1.790 | 0.687 | 2.606 | 0.070 | 0.326 | |
| **5** | a | Average Paddle Movement | | A2C | DQN | 71606.154 | 75257.436 | -3651.282 | 4997.725 | -0.731 | 0.900 | -0.164 | Tukey's |
| | | | | A2C | HCC | 71606.154 | 52000.427 | 19605.727 | 3783.228 | 5.182 | 0.001 | 0.886 | |
| | | | | A2C | MCC | 71606.154 | 50007.504 | 21598.650 | 4190.720 | 5.154 | 0.001 | 0.973 | |
| | | | | A2C | PPO | 71606.154 | 72712.500 | -1106.346 | 4966.391 | -0.223 | 0.900 | -0.050 | |
| | | | | DQN | HCC | 75257.436 | 52000.427 | 23257.009 | 3783.228 | 6.147 | 0.001 | 1.051 | |
| | | | | DQN | MCC | 75257.436 | 50007.504 | 25249.932 | 4190.720 | 6.025 | 0.001 | 1.138 | |
| | | | | DQN | PPO | 75257.436 | 72712.500 | 2544.936 | 4966.391 | 0.512 | 0.900 | 0.114 | |
| | | | | HCC | MCC | 52000.427 | 50007.504 | 1992.923 | 2626.345 | 0.759 | 0.900 | 0.090 | |
| | | | | HCC | PPO | 52000.427 | 72712.500 | -20712.073 | 3741.737 | -5.535 | 0.001 | -0.936 | |
| | | | | MCC | PPO | 50007.504 | 72712.500 | -22704.996 | 4153.302 | -5.467 | 0.001 | -1.023 | |
| | b | Relative improvement (%) in the average hit counts | | A2C | DQN | 29.919 | 24.634 | 5.285 | 7.934 | 288.957 | 0.900 | 0.077 | Games Howell |
| | | | | A2C | HCC | 29.919 | 82.147 | -52.227 | 9.623 | 316.974 | 0.001 | -0.603 | |
| | | | | A2C | MCC | 29.919 | 50.755 | -20.836 | 9.830 | 223.464 | 0.215 | -0.265 | |
| | | | | A2C | PPO | 29.919 | 21.602 | 8.318 | 7.665 | 279.006 | 0.789 | 0.125 | |
| | | | | DQN | HCC | 24.634 | 82.147 | -57.512 | 9.026 | 296.959 | 0.001 | -0.708 | |
| | | | | DQN | MCC | 24.634 | 50.755 | -26.121 | 9.246 | 197.121 | 0.041 | -0.354 | |
| | | | | DQN | PPO | 24.634 | 21.602 | 3.033 | 6.900 | 295.706 | 0.900 | 0.051 | |
| | | | | HCC | MCC | 82.147 | 50.755 | 31.391 | 10.731 | 262.994 | 0.030 | 0.355 | |
| | | | | HCC | PPO | 82.147 | 21.602 | 60.545 | 8.791 | 284.214 | 0.001 | 0.766 | |
| | | | | MCC | PPO | 50.755 | 21.602 | 29.154 | 9.016 | 184.940 | 0.012 | 0.405 | |
| | c | Average Paddle Movement | | A2C | DQN | 78719.250 | 83859.000 | -5139.750 | 4264.838 | -1.205 | 0.722 | -0.267 | Tukey's |
| | | | | A2C | HCC | 78719.250 | 52000.427 | 26718.823 | 3233.710 | 8.263 | 0.001 | 1.397 | |
| | | | | A2C | MCC | 78719.250 | 50007.504 | 28711.746 | 3589.396 | 7.999 | 0.001 | 1.497 | |
| | | | | A2C | PPO | 78719.250 | 75665.500 | 3053.750 | 4264.838 | 0.716 | 0.900 | 0.159 | |
| | | | | DQN | HCC | 83859.000 | 52000.427 | 31858.573 | 3233.710 | 9.852 | 0.001 | 1.666 | |
| | | | | DQN | MCC | 83859.000 | 50007.504 | 33851.496 | 3589.396 | 9.431 | 0.001 | 1.765 | |
| | | | | DQN | PPO | 83859.000 | 75665.500 | 8193.500 | 4264.838 | 1.921 | 0.307 | 0.425 | |
| | | | | HCC | MCC | 52000.427 | 50007.504 | 1992.923 | 2269.758 | 0.878 | 0.900 | 0.104 | |

| | | | | | | | | | | | |
|---|---|---|---|---|---|---|---|---|---|---|---|
| | | | HCC | PPO | 52000.427 | 75665.500 | -23665.073 | 3233.710 | -7.318 | 0.001 | -1.238 | |
| | | | MCC | PPO | 50007.504 | 75665.500 | -25657.996 | 3589.396 | -7.148 | 0.001 | -1.338 | |
| d | Relative improvement (%) in the average hit counts-Paddle&Ball Position Input | | A2C | DQN | 21.717 | 36.623 | -14.906 | 10.286 | 245.447 | 0.584 | -0.167 | Games Howell |
| | | | A2C | HCC | 21.717 | 82.147 | -60.429 | 9.165 | 303.151 | 0.001 | -0.733 | |
| | | | A2C | MCC | 21.717 | 50.755 | -29.038 | 9.381 | 203.860 | 0.019 | -0.387 | |
| | | | A2C | PPO | 21.717 | 14.690 | 7.027 | 7.082 | 292.773 | 0.842 | 0.114 | |
| | | | DQN | HCC | 36.623 | 82.147 | -45.523 | 11.531 | 304.565 | 0.001 | -0.439 | |
| | | | DQN | MCC | 36.623 | 50.755 | -14.132 | 11.703 | 257.834 | 0.720 | -0.151 | |
| | | | DQN | PPO | 36.623 | 14.690 | 21.933 | 9.955 | 226.546 | 0.182 | 0.254 | |
| | | | HCC | MCC | 82.147 | 50.755 | 31.391 | 10.731 | 262.994 | 0.030 | 0.355 | |
| | | | HCC | PPO | 82.147 | 14.690 | 67.456 | 8.792 | 284.259 | 0.001 | 0.853 | |
| | | | MCC | PPO | 50.755 | 14.690 | 36.065 | 9.017 | 184.981 | 0.001 | 0.501 | |
| e | Average Paddle Movement | | A2C | DQN | 67718.750 | 75019.250 | -7300.500 | 4333.263 | -1.685 | 0.446 | -0.373 | Tukey's |
| | | | A2C | HCC | 67718.750 | 52000.427 | 15718.323 | 3285.592 | 4.784 | 0.001 | 0.809 | |
| | | | A2C | MCC | 67718.750 | 50007.504 | 17711.246 | 3646.984 | 4.856 | 0.001 | 0.909 | |
| | | | A2C | PPO | 67718.750 | 73952.250 | -6233.500 | 4333.263 | -1.439 | 0.589 | -0.319 | |
| | | | DQN | HCC | 75019.250 | 52000.427 | 23018.823 | 3285.592 | 7.006 | 0.001 | 1.185 | |
| | | | DQN | MCC | 75019.250 | 50007.504 | 25011.746 | 3646.984 | 6.858 | 0.001 | 1.283 | |
| | | | DQN | PPO | 75019.250 | 73952.250 | 1067.000 | 4333.263 | 0.246 | 0.900 | 0.055 | |
| | | | HCC | MCC | 52000.427 | 50007.504 | 1992.923 | 2306.174 | 0.864 | 0.900 | 0.103 | |
| | | | HCC | PPO | 52000.427 | 73952.250 | -21951.823 | 3285.592 | -6.681 | 0.001 | -1.130 | |
| | | | MCC | PPO | 50007.504 | 73952.250 | -23944.746 | 3646.984 | -6.566 | 0.001 | -1.229 | |
| f | Relative improvement (%) in the average hit counts- Ball Poistion Input | | A2C | DQN | 33.724 | 29.397 | 4.327 | 9.789 | 297.513 | 0.900 | 0.051 | Games Howell |
| | | | A2C | HCC | 33.724 | 82.147 | -48.423 | 10.077 | 321.871 | 0.001 | -0.534 | |
| | | | A2C | MCC | 33.724 | 50.755 | -17.031 | 10.274 | 238.311 | 0.464 | -0.207 | |
| | | | A2C | PPO | 33.724 | 33.016 | 0.709 | 10.301 | 292.792 | 0.900 | 0.008 | |
| | | | DQN | HCC | 29.397 | 82.147 | -52.749 | 10.268 | 321.866 | 0.001 | -0.571 | |

| | | | | | | | | | | | | |
|---|---|---|---|---|---|---|---|---|---|---|---|---|
| | | | DQN | MCC | 29.397 | 50.755 | -21.358 | 10.461 | 243.172 | 0.249 | -0.256 | |
| | | | DQN | PPO | 29.397 | 33.016 | -3.618 | 10.487 | 295.423 | 0.900 | -0.040 | |
| | | | HCC | MCC | 82.147 | 50.755 | 31.391 | 10.731 | 262.994 | 0.030 | 0.355 | |
| | | | HCC | PPO | 82.147 | 33.016 | 49.131 | 10.756 | 317.852 | 0.001 | 0.508 | |
| | | | MCC | PPO | 50.755 | 33.016 | 17.740 | 10.941 | 252.147 | 0.486 | 0.203 | |
| 6 | a | Relative improvement (%) in the average hit counts – DQN | DQN_16 | DQN_32 | 6.400 | 43.207 | -36.806 | 17.554 | 75.733 | 0.300 | -0.416 | Games Howell |
| | | | DQN_16 | DQN_64 | 6.400 | 22.119 | -15.719 | 13.194 | 94.597 | 0.820 | -0.236 | |
| | | | DQN_16 | DQN_8 | 6.400 | 12.525 | -6.124 | 12.502 | 97.075 | 0.900 | -0.097 | |
| | | | DQN_16 | HCC | 6.400 | 82.147 | -75.746 | 11.229 | 133.216 | 0.001 | -1.079 | |
| | | | DQN_16 | MCC | 6.400 | 50.755 | -44.355 | 11.407 | 126.260 | 0.002 | -0.660 | |
| | | | DQN_32 | DQN_64 | 43.207 | 22.119 | 21.088 | 18.470 | 84.890 | 0.848 | 0.227 | |
| | | | DQN_32 | DQN_8 | 43.207 | 12.525 | 30.682 | 17.983 | 80.300 | 0.527 | 0.339 | |
| | | | DQN_32 | HCC | 43.207 | 82.147 | -38.940 | 17.122 | 73.462 | 0.218 | -0.364 | |
| | | | DQN_32 | MCC | 43.207 | 50.755 | -7.549 | 17.239 | 74.550 | 0.900 | -0.074 | |
| | | | DQN_64 | DQN_8 | 22.119 | 12.525 | 9.594 | 13.759 | 97.145 | 0.900 | 0.138 | |
| | | | DQN_64 | HCC | 22.119 | 82.147 | -60.027 | 12.614 | 106.983 | 0.001 | -0.761 | |
| | | | DQN_64 | MCC | 22.119 | 50.755 | -28.636 | 12.772 | 105.863 | 0.228 | -0.381 | |
| | | | DQN_8 | HCC | 12.525 | 82.147 | -69.622 | 11.889 | 118.917 | 0.001 | -0.936 | |
| | | | DQN_8 | MCC | 12.525 | 50.755 | -38.231 | 12.057 | 115.635 | 0.023 | -0.538 | |
| | | | HCC | MCC | 82.147 | 50.755 | 31.391 | 10.731 | 262.994 | 0.043 | 0.355 | |
| | b | Relative improvement (%) in the average hit counts – A2C | A2C_16 | A2C_32 | 18.203 | 23.304 | -5.101 | 11.700 | 97.985 | 0.900 | -0.087 | Games Howell |
| | | | A2C_16 | A2C_64 | 18.203 | 23.700 | -5.497 | 13.380 | 92.457 | 0.900 | -0.082 | |
| | | | A2C_16 | A2C_8 | 18.203 | 13.710 | 4.493 | 10.929 | 96.325 | 0.900 | 0.082 | |
| | | | A2C_16 | HCC | 18.203 | 82.147 | -63.944 | 11.098 | 136.551 | 0.001 | -0.921 | |
| | | | A2C_16 | MCC | 18.203 | 50.755 | -32.552 | 11.277 | 128.559 | 0.051 | -0.490 | |
| | | | A2C_32 | A2C_64 | 23.304 | 23.700 | -0.396 | 13.444 | 92.951 | 0.900 | -0.006 | |
| | | | A2C_32 | A2C_8 | 23.304 | 13.710 | 9.594 | 11.007 | 96.006 | 0.900 | 0.173 | |

| | | | | | | | | | | | | | |
|---|---|---|---|---|---|---|---|---|---|---|---|---|---|
| | | | | A2C_32 | HCC | 23.304 | 82.147 | -58.843 | 11.175 | 134.580 | 0.001 | -0.842 | |
| | | | | A2C_32 | MCC | 23.304 | 50.755 | -27.452 | 11.353 | 127.209 | 0.158 | -0.410 | |
| | | | | A2C_64 | A2C_8 | 23.700 | 13.710 | 9.990 | 12.778 | 86.482 | 0.900 | 0.155 | |
| | | | | A2C_64 | HCC | 23.700 | 82.147 | -58.446 | 12.923 | 102.820 | 0.001 | -0.723 | |
| | | | | A2C_64 | MCC | 23.700 | 50.755 | -27.055 | 13.077 | 102.271 | 0.312 | -0.351 | |
| | | | | A2C_8 | HCC | 13.710 | 82.147 | -68.436 | 10.364 | 158.702 | 0.001 | -1.056 | |
| | | | | A2C_8 | MCC | 13.710 | 50.755 | -37.045 | 10.556 | 142.025 | 0.008 | -0.596 | |
| | | | | HCC | MCC | 82.147 | 50.755 | 31.391 | 10.731 | 262.994 | 0.043 | 0.355 | |
| | c | Relative improvement (%) in the average hit counts – PPO | | PPO_16 | PPO_32 | 24.036 | 11.686 | 12.356 | 14.077 | 81.194 | 0.900 | 0.174 | Games Howell |
| | | | | PPO_16 | PPO_64 | 24.036 | -1.291 | 25.326 | 14.037 | 80.683 | 0.470 | 0.358 | |
| | | | | PPO_16 | PPO_8 | 24.036 | 49.866 | -25.830 | 25.190 | 75.516 | 0.900 | -0.204 | |
| | | | | PPO_16 | HCC | 24.036 | 82.147 | -58.111 | 14.132 | 90.262 | 0.001 | -0.658 | |
| | | | | PPO_16 | MCC | 24.036 | 50.755 | -26.720 | 14.274 | 90.898 | 0.428 | -0.318 | |
| | | | | PPO_32 | PPO_64 | 11.686 | -1.291 | 12.977 | 10.338 | 97.989 | 0.783 | 0.249 | |
| | | | | PPO_32 | PPO_64 | 11.686 | 49.866 | -38.180 | 23.333 | 59.662 | 0.568 | -0.325 | |
| | | | | PPO_32 | HCC | 11.686 | 82.147 | -70.461 | 10.468 | 155.189 | 0.001 | -1.076 | |
| | | | | PPO_32 | MCC | 11.686 | 50.755 | -39.070 | 10.658 | 140.107 | 0.005 | -0.622 | |
| | | | | PPO_64 | PPO_8 | -1.291 | 49.866 | -51.157 | 23.308 | 59.445 | 0.256 | -0.436 | |
| | | | | PPO_64 | HCC | -1.291 | 82.147 | -83.437 | 10.414 | 157.004 | 0.001 | -1.281 | |
| | | | | PPO_64 | MCC | -1.291 | 50.755 | -52.046 | 10.605 | 141.109 | 0.001 | -0.833 | |
| | | | | PPO_8 | HCC | 49.866 | 82.147 | -32.280 | 23.366 | 60.514 | 0.712 | -0.221 | |
| | | | | PPO_8 | MCC | 49.866 | 50.755 | -0.889 | 23.454 | 61.224 | 0.900 | -0.006 | |
| | | | | HCC | MCC | 82.147 | 50.755 | 31.391 | 10.731 | 262.994 | 0.043 | 0.355 | |
| **S3** | g | Hit Counts | 0-5 Minutes | DQN_16 | DQN_32 | 0.560 | 0.562 | -0.002 | 0.045 | -0.054 | 0.900 | -0.005 | Tukey's |
| | | | | DQN_16 | DQN_64 | 0.560 | 0.632 | -0.072 | 0.046 | -1.567 | 0.604 | -0.150 | |
| | | | | DQN_16 | DQN_8 | 0.560 | 0.582 | -0.022 | 0.046 | -0.472 | 0.900 | -0.045 | |
| | | | | DQN_16 | HCC | 0.560 | 0.651 | -0.091 | 0.037 | -2.457 | 0.137 | -0.190 | |

| | | | | | | | | | | | | | |
|---|---|---|---|---|---|---|---|---|---|---|---|---|---|
| | | | | DQN_16 | MCC | 0.560 | 0.716 | -0.156 | 0.039 | -3.998 | 0.001 | -0.326 | |
| | | | | DQN_32 | DQN_64 | 0.562 | 0.632 | -0.069 | 0.044 | -1.566 | 0.605 | -0.145 | |
| | | | | DQN_32 | DQN_8 | 0.562 | 0.582 | -0.019 | 0.045 | -0.433 | 0.900 | -0.040 | |
| | | | | DQN_32 | HCC | 0.562 | 0.651 | -0.088 | 0.035 | -2.522 | 0.118 | -0.184 | |
| | | | | DQN_32 | MCC | 0.562 | 0.716 | -0.154 | 0.037 | -4.127 | 0.001 | -0.321 | |
| | | | | DQN_64 | DQN_8 | 0.632 | 0.582 | 0.050 | 0.045 | 1.111 | 0.867 | 0.105 | |
| | | | | DQN_64 | HCC | 0.632 | 0.651 | -0.019 | 0.036 | -0.525 | 0.900 | -0.039 | |
| | | | | DQN_64 | MCC | 0.632 | 0.716 | -0.084 | 0.038 | -2.220 | 0.229 | -0.176 | |
| | | | | DQN_8 | HCC | 0.582 | 0.651 | -0.069 | 0.036 | -1.916 | 0.394 | -0.144 | |
| | | | | DQN_8 | MCC | 0.582 | 0.716 | -0.134 | 0.038 | -3.521 | 0.006 | -0.281 | |
| | | | | HCC | MCC | 0.651 | 0.716 | -0.065 | 0.026 | -2.490 | 0.127 | -0.137 | |
| | | | 6-20 Minutes | DQN_16 | DQN_32 | 0.567 | 0.560 | 0.007 | 0.031 | 0.232 | 0.900 | 0.012 | |
| | | | | DQN_16 | DQN_64 | 0.567 | 0.634 | -0.067 | 0.030 | -2.207 | 0.235 | -0.116 | |
| | | | | DQN_16 | DQN_8 | 0.567 | 0.526 | 0.041 | 0.031 | 1.326 | 0.743 | 0.070 | |
| | | | | DQN_16 | HCC | 0.567 | 0.854 | -0.288 | 0.025 | -11.616 | 0.001 | -0.495 | |
| | | | | DQN_16 | MCC | 0.567 | 0.852 | -0.286 | 0.027 | -10.708 | 0.001 | -0.492 | |
| | | | | DQN_32 | DQN_64 | 0.560 | 0.634 | -0.074 | 0.030 | -2.442 | 0.142 | -0.128 | |
| | | | | DQN_32 | DQN_8 | 0.560 | 0.526 | 0.034 | 0.031 | 1.093 | 0.878 | 0.058 | |
| | | | | DQN_32 | HCC | 0.560 | 0.854 | -0.295 | 0.025 | -11.906 | 0.001 | -0.508 | |
| | | | | DQN_32 | MCC | 0.560 | 0.852 | -0.293 | 0.027 | -10.977 | 0.001 | -0.504 | |
| | | | | DQN_64 | DQN_8 | 0.634 | 0.526 | 0.108 | 0.030 | 3.555 | 0.005 | 0.186 | |
| | | | | DQN_64 | HCC | 0.634 | 0.854 | -0.220 | 0.024 | -9.087 | 0.001 | -0.380 | |
| | | | | DQN_64 | MCC | 0.634 | 0.852 | -0.218 | 0.026 | -8.335 | 0.001 | -0.376 | |
| | | | | DQN_8 | HCC | 0.526 | 0.854 | -0.328 | 0.025 | -13.309 | 0.001 | -0.566 | |
| | | | | DQN_8 | MCC | 0.526 | 0.852 | -0.326 | 0.027 | -12.274 | 0.001 | -0.562 | |
| | | | | HCC | MCC | 0.854 | 0.852 | 0.002 | 0.019 | 0.103 | 0.900 | 0.003 | |
| | h | % Aces | 0-5 Minutes | DQN_16 | DQN_32 | 53.103 | 58.358 | -5.255 | 2.638 | -1.992 | 0.349 | -0.395 | Tukey's |
| | | | | DQN_16 | DQN_64 | 53.103 | 54.163 | -1.061 | 2.638 | -0.402 | 0.900 | -0.080 | |
| | | | | DQN_16 | DQN_8 | 53.103 | 56.084 | -2.981 | 2.638 | -1.130 | 0.856 | -0.224 | |
| | | | | DQN_16 | HCC | 53.103 | 54.382 | -1.279 | 2.117 | -0.604 | 0.900 | -0.097 | |

| | | | | | | | | | | | | |
|---|---|---|---|---|---|---|---|---|---|---|---|---|
| | | | | DQN_16 | MCC | 53.103 | 53.333 | -0.230 | 2.250 | -0.102 | 0.900 | -0.017 | |
| | | | | DQN_32 | DQN_64 | 58.358 | 54.163 | 4.194 | 2.638 | 1.590 | 0.591 | 0.316 | |
| | | | | DQN_32 | DQN_8 | 58.358 | 56.084 | 2.274 | 2.638 | 0.862 | 0.900 | 0.171 | |
| | | | | DQN_32 | HCC | 58.358 | 54.382 | 3.976 | 2.117 | 1.879 | 0.419 | 0.300 | |
| | | | | DQN_32 | MCC | 58.358 | 53.333 | 5.025 | 2.250 | 2.234 | 0.224 | 0.379 | |
| | | | | DQN_64 | DQN_8 | 54.163 | 56.084 | -1.921 | 2.638 | -0.728 | 0.900 | -0.145 | |
| | | | | DQN_64 | HCC | 54.163 | 54.382 | -0.219 | 2.117 | -0.103 | 0.900 | -0.017 | |
| | | | | DQN_64 | MCC | 54.163 | 53.333 | 0.831 | 2.250 | 0.369 | 0.900 | 0.063 | |
| | | | | DQN_8 | HCC | 56.084 | 54.382 | 1.702 | 2.117 | 0.804 | 0.900 | 0.129 | |
| | | | | DQN_8 | MCC | 56.084 | 53.333 | 2.751 | 2.250 | 1.223 | 0.802 | 0.208 | |
| | | | | HCC | MCC | 54.382 | 53.333 | 1.049 | 1.607 | 0.653 | 0.900 | 0.079 | |
| | | | 6-20 Minutes | DQN_16 | DQN_32 | 56.069 | 55.545 | 0.524 | 1.744 | 0.300 | 0.900 | 0.060 | |
| | | | | DQN_16 | DQN_64 | 56.069 | 55.105 | 0.964 | 1.744 | 0.553 | 0.900 | 0.110 | |
| | | | | DQN_16 | DQN_8 | 56.069 | 59.565 | -3.496 | 1.744 | -2.004 | 0.341 | -0.398 | |
| | | | | DQN_16 | HCC | 56.069 | 49.259 | 6.810 | 1.399 | 4.866 | 0.001 | 0.778 | |
| | | | | DQN_16 | MCC | 56.069 | 50.232 | 5.837 | 1.488 | 3.924 | 0.001 | 0.666 | |
| | | | | DQN_32 | DQN_64 | 55.545 | 55.105 | 0.440 | 1.744 | 0.252 | 0.900 | 0.050 | |
| | | | | DQN_32 | DQN_8 | 55.545 | 59.565 | -4.020 | 1.744 | -2.305 | 0.194 | -0.457 | |
| | | | | DQN_32 | HCC | 55.545 | 49.259 | 6.286 | 1.399 | 4.491 | 0.001 | 0.718 | |
| | | | | DQN_32 | MCC | 55.545 | 50.232 | 5.313 | 1.488 | 3.572 | 0.005 | 0.606 | |
| | | | | DQN_64 | DQN_8 | 55.105 | 59.565 | -4.460 | 1.744 | -2.557 | 0.110 | -0.507 | |
| | | | | DQN_64 | HCC | 55.105 | 49.259 | 5.846 | 1.399 | 4.177 | 0.001 | 0.668 | |
| | | | | DQN_64 | MCC | 55.105 | 50.232 | 4.873 | 1.488 | 3.276 | 0.014 | 0.556 | |
| | | | | DQN_8 | HCC | 59.565 | 49.259 | 10.306 | 1.399 | 7.364 | 0.001 | 1.178 | |
| | | | | DQN_8 | MCC | 59.565 | 50.232 | 9.334 | 1.488 | 6.274 | 0.001 | 1.065 | |
| | | | | HCC | MCC | 49.259 | 50.232 | -0.973 | 1.062 | -0.915 | 0.900 | -0.111 | |
| | i | % Long Rally | 0-5 Minutes | DQN_16 | DQN_32 | 7.882 | 4.460 | 3.422 | 1.409 | 2.428 | 0.148 | 0.482 | Tukey's |
| | | | | DQN_16 | DQN_64 | 7.882 | 7.083 | 0.798 | 1.409 | 0.567 | 0.900 | 0.112 | |
| | | | | DQN_16 | DQN_8 | 7.882 | 6.948 | 0.933 | 1.409 | 0.662 | 0.900 | 0.131 | |
| | | | | DQN_16 | HCC | 7.882 | 4.523 | 3.359 | 1.130 | 2.972 | 0.037 | 0.475 | |
| | | | | DQN_16 | MCC | 7.882 | 7.318 | 0.563 | 1.202 | 0.469 | 0.900 | 0.080 | |
| | | | | DQN_32 | DQN_64 | 4.460 | 7.083 | -2.623 | 1.409 | -1.862 | 0.429 | -0.370 | |
| | | | | DQN_32 | DQN_8 | 4.460 | 6.948 | -2.488 | 1.409 | -1.766 | 0.489 | -0.350 | |
| | | | | DQN_32 | HCC | 4.460 | 4.523 | -0.063 | 1.130 | -0.055 | 0.900 | -0.009 | |
| | | | | DQN_32 | MCC | 4.460 | 7.318 | -2.858 | 1.202 | -2.379 | 0.166 | -0.404 | |
| | | | | DQN_64 | DQN_8 | 7.083 | 6.948 | 0.135 | 1.409 | 0.096 | 0.900 | 0.019 | |

| | | | | | | | | | | | | |
|---|---|---|---|---|---|---|---|---|---|---|---|---|
| | | | | DQN_64 | HCC | 7.083 | 4.523 | 2.561 | 1.130 | 2.265 | 0.210 | 0.362 | |
| | | | | DQN_64 | MCC | 7.083 | 7.318 | -0.235 | 1.202 | -0.195 | 0.900 | -0.033 | |
| | | | | DQN_8 | HCC | 6.948 | 4.523 | 2.426 | 1.130 | 2.146 | 0.266 | 0.343 | |
| | | | | DQN_8 | MCC | 6.948 | 7.318 | -0.370 | 1.202 | -0.308 | 0.900 | -0.052 | |
| | | | | HCC | MCC | 4.523 | 7.318 | -2.796 | 0.858 | -3.258 | 0.015 | -0.396 | |
| | | | 6-20 Minutes | DQN_16 | DQN_32 | 6.138 | 6.310 | -0.172 | 0.929 | -0.185 | 0.900 | -0.037 | |
| | | | | DQN_16 | DQN_64 | 6.138 | 6.916 | -0.777 | 0.929 | -0.837 | 0.900 | -0.166 | |
| | | | | DQN_16 | DQN_8 | 6.138 | 6.221 | -0.083 | 0.929 | -0.089 | 0.900 | -0.018 | |
| | | | | DQN_16 | HCC | 6.138 | 10.365 | -4.226 | 0.746 | -5.668 | 0.001 | -0.906 | |
| | | | | DQN_16 | MCC | 6.138 | 11.972 | -5.834 | 0.793 | -7.361 | 0.001 | -1.250 | |
| | | | | DQN_32 | DQN_64 | 6.310 | 6.916 | -0.605 | 0.929 | -0.651 | 0.900 | -0.129 | |
| | | | | DQN_32 | DQN_8 | 6.310 | 6.221 | 0.090 | 0.929 | 0.096 | 0.900 | 0.019 | |
| | | | | DQN_32 | HCC | 6.310 | 10.365 | -4.054 | 0.746 | -5.438 | 0.001 | -0.870 | |
| | | | | DQN_32 | MCC | 6.310 | 11.972 | -5.662 | 0.793 | -7.144 | 0.001 | -1.213 | |
| | | | | DQN_64 | DQN_8 | 6.916 | 6.221 | 0.695 | 0.929 | 0.748 | 0.900 | 0.148 | |
| | | | | DQN_64 | HCC | 6.916 | 10.365 | -3.449 | 0.746 | -4.626 | 0.001 | -0.740 | |
| | | | | DQN_64 | MCC | 6.916 | 11.972 | -5.057 | 0.793 | -6.380 | 0.001 | -1.083 | |
| | | | | DQN_8 | HCC | 6.221 | 10.365 | -4.144 | 0.746 | -5.558 | 0.001 | -0.889 | |
| | | | | DQN_8 | MCC | 6.221 | 11.972 | -5.751 | 0.793 | -7.257 | 0.001 | -1.232 | |
| S4 | g | Hit Counts | 0-5 Minutes | A2C_16 | A2C_32 | 0.629 | 0.730 | -0.101 | 0.044 | -2.308 | 0.191 | -0.214 | Tukey's |
| | | | | A2C_16 | A2C_64 | 0.629 | 0.638 | -0.009 | 0.044 | -0.208 | 0.900 | -0.019 | |
| | | | | A2C_16 | A2C_8 | 0.629 | 0.655 | -0.026 | 0.044 | -0.594 | 0.900 | -0.056 | |
| | | | | A2C_16 | HCC | 0.629 | 0.651 | -0.021 | 0.035 | -0.611 | 0.900 | -0.045 | |
| | | | | A2C_16 | MCC | 0.629 | 0.716 | -0.087 | 0.037 | -2.326 | 0.184 | -0.184 | |
| | | | | A2C_32 | A2C_64 | 0.730 | 0.638 | 0.092 | 0.043 | 2.115 | 0.280 | 0.195 | |
| | | | | A2C_32 | A2C_8 | 0.730 | 0.655 | 0.075 | 0.044 | 1.710 | 0.521 | 0.158 | |
| | | | | A2C_32 | HCC | 0.730 | 0.651 | 0.080 | 0.035 | 2.302 | 0.193 | 0.169 | |
| | | | | A2C_32 | MCC | 0.730 | 0.716 | 0.014 | 0.037 | 0.386 | 0.900 | 0.030 | |
| | | | | A2C_64 | A2C_8 | 0.638 | 0.655 | -0.017 | 0.044 | -0.390 | 0.900 | -0.036 | |
| | | | | A2C_64 | HCC | 0.638 | 0.651 | -0.012 | 0.035 | -0.355 | 0.900 | -0.026 | |
| | | | | A2C_64 | MCC | 0.638 | 0.716 | -0.078 | 0.037 | -2.104 | 0.286 | -0.165 | |
| | | | | A2C_8 | HCC | 0.655 | 0.651 | 0.005 | 0.035 | 0.136 | 0.900 | 0.010 | |
| | | | | A2C_8 | MCC | 0.655 | 0.716 | -0.061 | 0.037 | -1.626 | 0.570 | -0.128 | |
| | | | | HCC | MCC | 0.651 | 0.716 | -0.065 | 0.026 | -2.523 | 0.118 | -0.139 | |
| | | | | A2C_16 | A2C_32 | 0.605 | 0.705 | -0.100 | 0.031 | -3.254 | 0.015 | -0.173 | |

| | | | | | | | | | | | | | |
|---|---|---|---|---|---|---|---|---|---|---|---|---|---|
| | | | 6-20 Minutes | A2C_16 | A2C_64 | 0.605 | 0.622 | -0.017 | 0.031 | -0.541 | 0.900 | -0.029 | |
| | | | | A2C_16 | A2C_8 | 0.605 | 0.597 | 0.008 | 0.031 | 0.256 | 0.900 | 0.014 | |
| | | | | A2C_16 | HCC | 0.605 | 0.854 | -0.250 | 0.025 | -9.962 | 0.001 | -0.430 | |
| | | | | A2C_16 | MCC | 0.605 | 0.852 | -0.248 | 0.027 | -9.188 | 0.001 | -0.426 | |
| | | | | A2C_32 | A2C_64 | 0.705 | 0.622 | 0.083 | 0.031 | 2.735 | 0.069 | 0.144 | |
| | | | | A2C_32 | A2C_8 | 0.705 | 0.597 | 0.108 | 0.031 | 3.526 | 0.006 | 0.186 | |
| | | | | A2C_32 | HCC | 0.705 | 0.854 | -0.149 | 0.024 | -6.122 | 0.001 | -0.257 | |
| | | | | A2C_32 | MCC | 0.705 | 0.852 | -0.147 | 0.026 | -5.595 | 0.001 | -0.254 | |
| | | | | A2C_64 | A2C_8 | 0.622 | 0.597 | 0.025 | 0.031 | 0.801 | 0.900 | 0.043 | |
| | | | | A2C_64 | HCC | 0.622 | 0.854 | -0.233 | 0.025 | -9.427 | 0.001 | -0.401 | |
| | | | | A2C_64 | MCC | 0.622 | 0.852 | -0.231 | 0.027 | -8.673 | 0.001 | -0.397 | |
| | | | | A2C_8 | HCC | 0.597 | 0.854 | -0.258 | 0.025 | -10.339 | 0.001 | -0.444 | |
| | | | | A2C_8 | MCC | 0.597 | 0.852 | -0.256 | 0.027 | -9.531 | 0.001 | -0.440 | |
| | | | | HCC | MCC | 0.854 | 0.852 | 0.002 | 0.019 | 0.103 | 0.900 | 0.003 | |
| | h | % Aces | 0-5 Minutes | A2C_16 | A2C_32 | 53.810 | 54.083 | -0.272 | 2.370 | -0.115 | 0.900 | -0.023 | Tukey's |
| | | | | A2C_16 | A2C_64 | 53.810 | 53.299 | 0.511 | 2.370 | 0.216 | 0.900 | 0.043 | |
| | | | | A2C_16 | A2C_8 | 53.810 | 52.332 | 1.478 | 2.370 | 0.624 | 0.900 | 0.124 | |
| | | | | A2C_16 | HCC | 53.810 | 54.382 | -0.571 | 1.901 | -0.301 | 0.900 | -0.048 | |
| | | | | A2C_16 | MCC | 53.810 | 53.333 | 0.478 | 2.021 | 0.236 | 0.900 | 0.040 | |
| | | | | A2C_32 | A2C_64 | 54.083 | 53.299 | 0.783 | 2.370 | 0.331 | 0.900 | 0.066 | |
| | | | | A2C_32 | A2C_8 | 54.083 | 52.332 | 1.750 | 2.370 | 0.738 | 0.900 | 0.147 | |
| | | | | A2C_32 | HCC | 54.083 | 54.382 | -0.299 | 1.901 | -0.157 | 0.900 | -0.025 | |
| | | | | A2C_32 | MCC | 54.083 | 53.333 | 0.750 | 2.021 | 0.371 | 0.900 | 0.063 | |
| | | | | A2C_64 | A2C_8 | 53.299 | 52.332 | 0.967 | 2.370 | 0.408 | 0.900 | 0.081 | |
| | | | | A2C_64 | HCC | 53.299 | 54.382 | -1.083 | 1.901 | -0.569 | 0.900 | -0.091 | |
| | | | | A2C_64 | MCC | 53.299 | 53.333 | -0.034 | 2.021 | -0.017 | 0.900 | -0.003 | |
| | | | | A2C_8 | HCC | 52.332 | 54.382 | -2.049 | 1.901 | -1.078 | 0.886 | -0.172 | |
| | | | | A2C_8 | MCC | 52.332 | 53.333 | -1.000 | 2.021 | -0.495 | 0.900 | -0.084 | |
| | | | | HCC | MCC | 54.382 | 53.333 | 1.049 | 1.443 | 0.727 | 0.900 | 0.088 | |
| | | | 6-20 Minutes | A2C_16 | A2C_32 | 53.076 | 52.170 | 0.906 | 1.627 | 0.557 | 0.900 | 0.110 | |
| | | | | A2C_16 | A2C_64 | 53.076 | 52.981 | 0.094 | 1.627 | 0.058 | 0.900 | 0.012 | |
| | | | | A2C_16 | A2C_8 | 53.076 | 54.117 | -1.042 | 1.627 | -0.640 | 0.900 | -0.127 | |
| | | | | A2C_16 | HCC | 53.076 | 49.259 | 3.816 | 1.305 | 2.924 | 0.042 | 0.468 | |
| | | | | A2C_16 | MCC | 53.076 | 50.232 | 2.844 | 1.388 | 2.049 | 0.316 | 0.348 | |
| | | | | A2C_32 | A2C_64 | 52.170 | 52.981 | -0.811 | 1.627 | -0.499 | 0.900 | -0.099 | |
| | | | | A2C_32 | A2C_8 | 52.170 | 54.117 | -1.947 | 1.627 | -1.197 | 0.817 | -0.238 | |
| | | | | A2C_32 | HCC | 52.170 | 49.259 | 2.911 | 1.305 | 2.230 | 0.226 | 0.357 | |
| | | | | A2C_32 | MCC | 52.170 | 50.232 | 1.938 | 1.388 | 1.397 | 0.702 | 0.237 | |
| | | | | A2C_64 | A2C_8 | 52.981 | 54.117 | -1.136 | 1.627 | -0.698 | 0.900 | -0.139 | |

| | | | | A2C_64 | HCC | 52.981 | 49.259 | 3.722 | 1.305 | 2.851 | 0.051 | 0.456 | |
| | | | | A2C_64 | MCC | 52.981 | 50.232 | 2.749 | 1.388 | 1.981 | 0.355 | 0.336 | |
| | | | | A2C_8 | HCC | 54.117 | 49.259 | 4.858 | 1.305 | 3.721 | 0.003 | 0.595 | |
| | | | | A2C_8 | MCC | 54.117 | 50.232 | 3.885 | 1.388 | 2.800 | 0.059 | 0.475 | |
| | | | | HCC | MCC | 49.259 | 50.232 | -0.973 | 0.991 | -0.981 | 0.900 | -0.119 | |
| | i | % Long Rally | 0-5 Minutes | A2C_16 | A2C_32 | 5.395 | 6.147 | -0.752 | 1.314 | -0.572 | 0.900 | -0.114 | Tukey's |
| | | | | A2C_16 | A2C_64 | 5.395 | 6.973 | -1.578 | 1.314 | -1.201 | 0.815 | -0.238 | |
| | | | | A2C_16 | A2C_8 | 5.395 | 5.357 | 0.038 | 1.314 | 0.029 | 0.900 | 0.006 | |
| | | | | A2C_16 | HCC | 5.395 | 4.523 | 0.872 | 1.054 | 0.828 | 0.900 | 0.132 | |
| | | | | A2C_16 | MCC | 5.395 | 7.318 | -1.923 | 1.120 | -1.717 | 0.518 | -0.291 | |
| | | | | A2C_32 | A2C_64 | 6.147 | 6.973 | -0.826 | 1.314 | -0.629 | 0.900 | -0.125 | |
| | | | | A2C_32 | A2C_8 | 6.147 | 5.357 | 0.790 | 1.314 | 0.602 | 0.900 | 0.119 | |
| | | | | A2C_32 | HCC | 6.147 | 4.523 | 1.624 | 1.054 | 1.541 | 0.619 | 0.246 | |
| | | | | A2C_32 | MCC | 6.147 | 7.318 | -1.171 | 1.120 | -1.045 | 0.900 | -0.177 | |
| | | | | A2C_64 | A2C_8 | 6.973 | 5.357 | 1.617 | 1.314 | 1.231 | 0.798 | 0.244 | |
| | | | | A2C_64 | HCC | 6.973 | 4.523 | 2.451 | 1.054 | 2.325 | 0.186 | 0.372 | |
| | | | | A2C_64 | MCC | 6.973 | 7.318 | -0.345 | 1.120 | -0.308 | 0.900 | -0.052 | |
| | | | | A2C_8 | HCC | 5.357 | 4.523 | 0.834 | 1.054 | 0.791 | 0.900 | 0.127 | |
| | | | | A2C_8 | MCC | 5.357 | 7.318 | -1.962 | 1.120 | -1.751 | 0.498 | -0.297 | |
| | | | | HCC | MCC | 4.523 | 7.318 | -2.796 | 0.800 | -3.494 | 0.007 | -0.424 | |
| | | | 6-20 Minutes | A2C_16 | A2C_32 | 6.077 | 7.705 | -1.628 | 0.942 | -1.727 | 0.512 | -0.343 | |
| | | | | A2C_16 | A2C_64 | 6.077 | 6.942 | -0.865 | 0.942 | -0.917 | 0.900 | -0.182 | |
| | | | | A2C_16 | A2C_8 | 6.077 | 6.231 | -0.154 | 0.942 | -0.163 | 0.900 | -0.032 | |
| | | | | A2C_16 | HCC | 6.077 | 10.365 | -4.287 | 0.756 | -5.671 | 0.001 | -0.907 | |
| | | | | A2C_16 | MCC | 6.077 | 11.972 | -5.895 | 0.804 | -7.336 | 0.001 | -1.245 | |
| | | | | A2C_32 | A2C_64 | 7.705 | 6.942 | 0.763 | 0.942 | 0.810 | 0.900 | 0.161 | |
| | | | | A2C_32 | A2C_8 | 7.705 | 6.231 | 1.474 | 0.942 | 1.564 | 0.606 | 0.310 | |
| | | | | A2C_32 | HCC | 7.705 | 10.365 | -2.660 | 0.756 | -3.518 | 0.006 | -0.563 | |
| | | | | A2C_32 | MCC | 7.705 | 11.972 | -4.267 | 0.804 | -5.310 | 0.001 | -0.901 | |
| | | | | A2C_64 | A2C_8 | 6.942 | 6.231 | 0.711 | 0.942 | 0.754 | 0.900 | 0.150 | |
| | | | | A2C_64 | HCC | 6.942 | 10.365 | -3.423 | 0.756 | -4.528 | 0.001 | -0.724 | |
| | | | | A2C_64 | MCC | 6.942 | 11.972 | -5.030 | 0.804 | -6.260 | 0.001 | -1.063 | |
| | | | | A2C_8 | HCC | 6.231 | 10.365 | -4.134 | 0.756 | -5.468 | 0.001 | -0.874 | |
| | | | | A2C_8 | MCC | 6.231 | 11.972 | -5.741 | 0.804 | -7.144 | 0.001 | -1.213 | |
| | | | | HCC | MCC | 10.365 | 11.972 | -1.608 | 0.574 | -2.801 | 0.059 | -0.340 | |

| | | | | | | | | | | | | | |
|---|---|---|---|---|---|---|---|---|---|---|---|---|---|
| S5 | g | Hit Counts | 0-5 Minutes | PPO_16 | PPO_32 | 0.557 | 0.555 | 0.002 | 0.044 | 0.035 | 0.900 | 0.003 | Tukey's |
| | | | | PPO_16 | PPO_64 | 0.557 | 0.632 | -0.075 | 0.044 | -1.684 | 0.537 | -0.159 | |
| | | | | PPO_16 | PPO_8 | 0.557 | 0.552 | 0.005 | 0.045 | 0.121 | 0.900 | 0.011 | |
| | | | | PPO_16 | HCC | 0.557 | 0.651 | -0.094 | 0.035 | -2.639 | 0.088 | -0.199 | |
| | | | | PPO_16 | MCC | 0.557 | 0.716 | -0.159 | 0.038 | -4.227 | 0.001 | -0.338 | |
| | | | | PPO_32 | PPO_64 | 0.555 | 0.632 | -0.076 | 0.044 | -1.734 | 0.508 | -0.162 | |
| | | | | PPO_32 | PPO_64 | 0.555 | 0.552 | 0.004 | 0.044 | 0.088 | 0.900 | 0.008 | |
| | | | | PPO_32 | HCC | 0.555 | 0.651 | -0.095 | 0.035 | -2.720 | 0.072 | -0.202 | |
| | | | | PPO_32 | MCC | 0.555 | 0.716 | -0.161 | 0.037 | -4.322 | 0.001 | -0.341 | |
| | | | | PPO_64 | PPO_8 | 0.632 | 0.552 | 0.080 | 0.044 | 1.816 | 0.457 | 0.171 | |
| | | | | PPO_64 | HCC | 0.632 | 0.651 | -0.019 | 0.035 | -0.536 | 0.900 | -0.040 | |
| | | | | PPO_64 | MCC | 0.632 | 0.716 | -0.084 | 0.037 | -2.262 | 0.210 | -0.179 | |
| | | | | PPO_8 | HCC | 0.552 | 0.651 | -0.099 | 0.035 | -2.816 | 0.055 | -0.211 | |
| | | | | PPO_8 | MCC | 0.552 | 0.716 | -0.164 | 0.037 | -4.405 | 0.001 | -0.350 | |
| | | | | HCC | MCC | 0.651 | 0.716 | -0.065 | 0.026 | -2.532 | 0.115 | -0.139 | |
| | | | 6-20 Minutes | PPO_16 | PPO_32 | 0.508 | 0.523 | -0.015 | 0.030 | -0.486 | 0.900 | -0.026 | |
| | | | | PPO_16 | PPO_64 | 0.508 | 0.513 | -0.005 | 0.030 | -0.160 | 0.900 | -0.008 | |
| | | | | PPO_16 | PPO_8 | 0.508 | 0.556 | -0.048 | 0.030 | -1.579 | 0.597 | -0.083 | |
| | | | | PPO_16 | HCC | 0.508 | 0.854 | -0.347 | 0.024 | -14.206 | 0.001 | -0.607 | |
| | | | | PPO_16 | MCC | 0.508 | 0.852 | -0.345 | 0.026 | -13.114 | 0.001 | -0.603 | |
| | | | | PPO_32 | PPO_64 | 0.523 | 0.513 | 0.010 | 0.030 | 0.328 | 0.900 | 0.017 | |
| | | | | PPO_32 | PPO_64 | 0.523 | 0.556 | -0.033 | 0.030 | -1.090 | 0.880 | -0.058 | |
| | | | | PPO_32 | HCC | 0.523 | 0.854 | -0.332 | 0.024 | -13.615 | 0.001 | -0.581 | |
| | | | | PPO_32 | MCC | 0.523 | 0.852 | -0.330 | 0.026 | -12.563 | 0.001 | -0.577 | |
| | | | | PPO_64 | PPO_8 | 0.513 | 0.556 | -0.043 | 0.030 | -1.424 | 0.687 | -0.075 | |
| | | | | PPO_64 | HCC | 0.513 | 0.854 | -0.342 | 0.024 | -14.092 | 0.001 | -0.598 | |
| | | | | PPO_64 | MCC | 0.513 | 0.852 | -0.340 | 0.026 | -12.997 | 0.001 | -0.595 | |
| | | | | PPO_8 | HCC | 0.556 | 0.854 | -0.299 | 0.024 | -12.406 | 0.001 | -0.523 | |
| | | | | PPO_8 | MCC | 0.556 | 0.852 | -0.297 | 0.026 | -11.422 | 0.001 | -0.520 | |

| | | | | | | | | | | | | |
|---|---|---|---|---|---|---|---|---|---|---|---|---|
| | | | HCC | MCC | 0.854 | 0.852 | 0.002 | 0.019 | 0.104 | 0.900 | 0.003 | |
| h | % Aces | 0-5 Minutes | PPO_16 | PPO_32 | 59.143 | 59.654 | -0.511 | 2.553 | -0.200 | 0.900 | -0.040 | Tukey's |
| | | | PPO_16 | PPO_64 | 59.143 | 55.534 | 3.610 | 2.553 | 1.414 | 0.692 | 0.281 | |
| | | | PPO_16 | PPO_8 | 59.143 | 58.904 | 0.239 | 2.553 | 0.094 | 0.900 | 0.019 | |
| | | | PPO_16 | HCC | 59.143 | 54.382 | 4.762 | 2.048 | 2.325 | 0.186 | 0.372 | |
| | | | PPO_16 | MCC | 59.143 | 53.333 | 5.811 | 2.177 | 2.669 | 0.084 | 0.453 | |
| | | | PPO_32 | PPO_64 | 59.654 | 55.534 | 4.120 | 2.553 | 1.614 | 0.577 | 0.320 | |
| | | | PPO_32 | PPO_64 | 59.654 | 58.904 | 0.750 | 2.553 | 0.294 | 0.900 | 0.058 | |
| | | | PPO_32 | HCC | 59.654 | 54.382 | 5.273 | 2.048 | 2.575 | 0.106 | 0.412 | |
| | | | PPO_32 | MCC | 59.654 | 53.333 | 6.322 | 2.177 | 2.904 | 0.044 | 0.493 | |
| | | | PPO_64 | PPO_8 | 55.534 | 58.904 | -3.370 | 2.553 | -1.320 | 0.746 | -0.262 | |
| | | | PPO_64 | HCC | 55.534 | 54.382 | 1.152 | 2.048 | 0.563 | 0.900 | 0.090 | |
| | | | PPO_64 | MCC | 55.534 | 53.333 | 2.201 | 2.177 | 1.011 | 0.900 | 0.172 | |
| | | | PPO_8 | HCC | 58.904 | 54.382 | 4.522 | 2.048 | 2.208 | 0.236 | 0.353 | |
| | | | PPO_8 | MCC | 58.904 | 53.333 | 5.571 | 2.177 | 2.559 | 0.110 | 0.434 | |
| | | | HCC | MCC | 54.382 | 53.333 | 1.049 | 1.555 | 0.675 | 0.900 | 0.082 | |
| | | 6-20 Minutes | PPO_16 | PPO_32 | 60.504 | 60.595 | -0.091 | 1.900 | -0.048 | 0.900 | -0.010 | |
| | | | PPO_16 | PPO_64 | 60.504 | 60.941 | -0.438 | 1.900 | -0.230 | 0.900 | -0.046 | |
| | | | PPO_16 | PPO_8 | 60.504 | 58.316 | 2.187 | 1.900 | 1.151 | 0.844 | 0.228 | |
| | | | PPO_16 | HCC | 60.504 | 49.259 | 11.244 | 1.524 | 7.376 | 0.001 | 1.180 | |
| | | | PPO_16 | MCC | 60.504 | 50.232 | 10.272 | 1.620 | 6.339 | 0.001 | 1.076 | |
| | | | PPO_32 | PPO_64 | 60.595 | 60.941 | -0.347 | 1.900 | -0.182 | 0.900 | -0.036 | |
| | | | PPO_32 | PPO_64 | 60.595 | 58.316 | 2.279 | 1.900 | 1.199 | 0.816 | 0.238 | |
| | | | PPO_32 | HCC | 60.595 | 49.259 | 11.335 | 1.524 | 7.436 | 0.001 | 1.189 | |
| | | | PPO_32 | MCC | 60.595 | 50.232 | 10.363 | 1.620 | 6.395 | 0.001 | 1.086 | |
| | | | PPO_64 | PPO_8 | 60.941 | 58.316 | 2.625 | 1.900 | 1.382 | 0.711 | 0.274 | |
| | | | PPO_64 | HCC | 60.941 | 49.259 | 11.682 | 1.524 | 7.663 | 0.001 | 1.225 | |
| | | | PPO_64 | MCC | 60.941 | 50.232 | 10.709 | 1.620 | 6.609 | 0.001 | 1.122 | |
| | | | PPO_8 | HCC | 58.316 | 49.259 | 9.057 | 1.524 | 5.941 | 0.001 | 0.950 | |
| | | | PPO_8 | MCC | 58.316 | 50.232 | 8.084 | 1.620 | 4.989 | 0.001 | 0.847 | |
| | | | HCC | MCC | 49.259 | 50.232 | -0.973 | 1.157 | -0.840 | 0.900 | -0.102 | |
| i | % Long Rally | 0-5 Minutes | PPO_16 | PPO_32 | 4.885 | 5.501 | -0.616 | 1.359 | -0.453 | 0.900 | -0.090 | Tukey's |
| | | | PPO_16 | PPO_64 | 4.885 | 7.393 | -2.508 | 1.359 | -1.845 | 0.439 | -0.366 | |
| | | | PPO_16 | PPO_8 | 4.885 | 6.020 | -1.135 | 1.359 | -0.835 | 0.900 | -0.166 | |
| | | | PPO_16 | HCC | 4.885 | 4.523 | 0.362 | 1.090 | 0.332 | 0.900 | 0.053 | |

| | | | | | | | | | | | | | |
|---|---|---|---|---|---|---|---|---|---|---|---|---|---|
| | | | | PPO_16 | MCC | 4.885 | 7.318 | -2.433 | 1.159 | -2.099 | 0.290 | -0.356 | |
| | | | | PPO_32 | PPO_64 | 5.501 | 7.393 | -1.892 | 1.359 | -1.392 | 0.705 | -0.276 | |
| | | | | PPO_32 | PPO_64 | 5.501 | 6.020 | -0.519 | 1.359 | -0.382 | 0.900 | -0.076 | |
| | | | | PPO_32 | HCC | 5.501 | 4.523 | 0.979 | 1.090 | 0.897 | 0.900 | 0.144 | |
| | | | | PPO_32 | MCC | 5.501 | 7.318 | -1.817 | 1.159 | -1.568 | 0.604 | -0.266 | |
| | | | | PPO_64 | PPO_8 | 7.393 | 6.020 | 1.373 | 1.359 | 1.010 | 0.900 | 0.201 | |
| | | | | PPO_64 | HCC | 7.393 | 4.523 | 2.871 | 1.090 | 2.633 | 0.091 | 0.421 | |
| | | | | PPO_64 | MCC | 7.393 | 7.318 | 0.075 | 1.159 | 0.065 | 0.900 | 0.011 | |
| | | | | PPO_8 | HCC | 6.020 | 4.523 | 1.497 | 1.090 | 1.373 | 0.716 | 0.220 | |
| | | | | PPO_8 | MCC | 6.020 | 7.318 | -1.298 | 1.159 | -1.120 | 0.862 | -0.190 | |
| | | | | HCC | MCC | 4.523 | 7.318 | -2.796 | 0.828 | -3.377 | 0.010 | -0.410 | |
| | | | 6-20 Minutes | PPO_16 | PPO_32 | 6.008 | 5.224 | 0.784 | 0.871 | 0.900 | 0.900 | 0.179 | |
| | | | | PPO_16 | PPO_64 | 6.008 | 5.339 | 0.669 | 0.871 | 0.769 | 0.900 | 0.153 | |
| | | | | PPO_16 | PPO_8 | 6.008 | 6.234 | -0.226 | 0.871 | -0.260 | 0.900 | -0.052 | |
| | | | | PPO_16 | HCC | 6.008 | 10.365 | -4.357 | 0.699 | -6.237 | 0.001 | -0.997 | |
| | | | | PPO_16 | MCC | 6.008 | 11.972 | -5.964 | 0.743 | -8.032 | 0.001 | -1.363 | |
| | | | | PPO_32 | PPO_64 | 5.224 | 5.339 | -0.114 | 0.871 | -0.131 | 0.900 | -0.026 | |
| | | | | PPO_32 | PPO_64 | 5.224 | 6.234 | -1.010 | 0.871 | -1.160 | 0.839 | -0.230 | |
| | | | | PPO_32 | HCC | 5.224 | 10.365 | -5.140 | 0.699 | -7.358 | 0.001 | -1.177 | |
| | | | | PPO_32 | MCC | 5.224 | 11.972 | -6.748 | 0.743 | -9.088 | 0.001 | -1.543 | |
| | | | | PPO_64 | PPO_8 | 5.339 | 6.234 | -0.896 | 0.871 | -1.029 | 0.900 | -0.204 | |
| | | | | PPO_64 | HCC | 5.339 | 10.365 | -5.026 | 0.699 | -7.195 | 0.001 | -1.151 | |
| | | | | PPO_64 | MCC | 5.339 | 11.972 | -6.634 | 0.743 | -8.934 | 0.001 | -1.517 | |
| | | | | PPO_8 | HCC | 6.234 | 10.365 | -4.130 | 0.699 | -5.913 | 0.001 | -0.946 | |
| | | | | PPO_8 | MCC | 6.234 | 11.972 | -5.738 | 0.743 | -7.728 | 0.001 | -1.312 | |
| | | | | HCC | MCC | 10.365 | 11.972 | -1.608 | 0.530 | -3.031 | 0.031 | -0.368 | |
| S6 | g | Hit Counts | 0-5 Minutes | A2C | DQN | 0.771 | 0.717 | 0.054 | 0.028 | 1.937 | 0.298 | 0.103 | Tukey's |
| | | | | A2C | HCC | 0.771 | 0.651 | 0.121 | 0.027 | 4.508 | 0.001 | 0.229 | |
| | | | | A2C | MCC | 0.771 | 0.716 | 0.055 | 0.030 | 1.827 | 0.359 | 0.105 | |
| | | | | A2C | PPO | 0.771 | 0.698 | 0.073 | 0.028 | 2.594 | 0.072 | 0.139 | |
| | | | | DQN | HCC | 0.717 | 0.651 | 0.066 | 0.027 | 2.476 | 0.096 | 0.126 | |
| | | | | DQN | MCC | 0.717 | 0.716 | 0.001 | 0.030 | 0.025 | 0.900 | 0.001 | |
| | | | | DQN | PPO | 0.717 | 0.698 | 0.019 | 0.028 | 0.668 | 0.900 | 0.036 | |
| | | | | HCC | MCC | 0.651 | 0.716 | -0.065 | 0.029 | -2.263 | 0.157 | -0.124 | |
| | | | | HCC | PPO | 0.651 | 0.698 | -0.047 | 0.027 | -1.760 | 0.399 | -0.090 | |
| | | | | MCC | PPO | 0.716 | 0.698 | 0.018 | 0.030 | 0.598 | 0.900 | 0.034 | |
| | | | 6-20 Minutes | A2C | DQN | 0.777 | 0.762 | 0.016 | 0.018 | 0.866 | 0.900 | 0.026 | |
| | | | | A2C | HCC | 0.777 | 0.854 | -0.077 | 0.018 | -4.329 | 0.001 | -0.127 | |
| | | | | A2C | MCC | 0.777 | 0.852 | -0.075 | 0.021 | -3.647 | 0.002 | -0.124 | |
| | | | | A2C | PPO | 0.777 | 0.712 | 0.065 | 0.018 | 3.561 | 0.003 | 0.108 | |

| | | | | | | | | | | | |
|---|---|---|---|---|---|---|---|---|---|---|---|
| | | | DQN | HCC | 0.762 | 0.854 | -0.093 | 0.018 | -5.262 | 0.001 | -0.153 | |
| | | | DQN | MCC | 0.762 | 0.852 | -0.091 | 0.020 | -4.442 | 0.001 | -0.150 | |
| | | | DQN | PPO | 0.762 | 0.712 | 0.049 | 0.018 | 2.723 | 0.051 | 0.082 | |
| | | | HCC | MCC | 0.854 | 0.852 | 0.002 | 0.020 | 0.099 | 0.900 | 0.003 | |
| | | | HCC | PPO | 0.854 | 0.712 | 0.142 | 0.018 | 8.009 | 0.001 | 0.235 | |
| | | | MCC | PPO | 0.852 | 0.712 | 0.140 | 0.021 | 6.825 | 0.001 | 0.232 | |
| h | % Aces | 0-5 Minutes | A2C | DQN | 53.293 | 52.579 | 0.714 | 1.455 | 0.491 | 0.900 | 0.057 | Tukey's |
| | | | A2C | HCC | 53.293 | 54.382 | -1.089 | 1.404 | -0.776 | 0.900 | -0.086 | |
| | | | A2C | MCC | 53.293 | 53.333 | -0.040 | 1.582 | -0.025 | 0.900 | -0.003 | |
| | | | A2C | PPO | 53.293 | 54.248 | -0.956 | 1.455 | -0.657 | 0.900 | -0.076 | |
| | | | DQN | HCC | 52.579 | 54.382 | -1.803 | 1.404 | -1.284 | 0.677 | -0.143 | |
| | | | DQN | MCC | 52.579 | 53.333 | -0.754 | 1.582 | -0.477 | 0.900 | -0.060 | |
| | | | DQN | PPO | 52.579 | 54.248 | -1.670 | 1.455 | -1.147 | 0.754 | -0.132 | |
| | | | HCC | MCC | 54.382 | 53.333 | 1.049 | 1.535 | 0.683 | 0.900 | 0.083 | |
| | | | HCC | PPO | 54.382 | 54.248 | 0.133 | 1.404 | 0.095 | 0.900 | 0.011 | |
| | | | MCC | PPO | 53.333 | 54.248 | -0.916 | 1.582 | -0.579 | 0.900 | -0.072 | |
| | | 6-20 Minutes | A2C | DQN | 52.530 | 49.935 | 2.595 | 0.959 | 2.706 | 0.054 | 0.312 | |
| | | | A2C | HCC | 52.530 | 49.259 | 3.270 | 0.925 | 3.535 | 0.004 | 0.393 | |
| | | | A2C | MCC | 52.530 | 50.232 | 2.298 | 1.042 | 2.205 | 0.179 | 0.276 | |
| | | | A2C | PPO | 52.530 | 52.511 | 0.018 | 0.959 | 0.019 | 0.900 | 0.002 | |
| | | | DQN | HCC | 49.935 | 49.259 | 0.676 | 0.925 | 0.730 | 0.900 | 0.081 | |
| | | | DQN | MCC | 49.935 | 50.232 | -0.297 | 1.042 | -0.285 | 0.900 | -0.036 | |
| | | | DQN | PPO | 49.935 | 52.511 | -2.576 | 0.959 | -2.687 | 0.057 | -0.310 | |
| | | | HCC | MCC | 49.259 | 50.232 | -0.973 | 1.011 | -0.962 | 0.860 | -0.117 | |
| | | | HCC | PPO | 49.259 | 52.511 | -3.252 | 0.925 | -3.515 | 0.004 | -0.391 | |
| | | | MCC | PPO | 50.232 | 52.511 | -2.280 | 1.042 | -2.187 | 0.186 | -0.274 | |
| i | % Long Rally | 0-5 Minutes | A2C | DQN | 12.722 | 10.195 | 2.527 | 0.631 | 4.004 | 0.001 | 0.461 | Tukey's |
| | | | A2C | HCC | 12.722 | 10.365 | 2.357 | 0.609 | 3.871 | 0.001 | 0.430 | |
| | | | A2C | MCC | 12.722 | 11.972 | 0.750 | 0.686 | 1.093 | 0.785 | 0.137 | |
| | | | A2C | PPO | 12.722 | 10.183 | 2.540 | 0.631 | 4.024 | 0.001 | 0.463 | |
| | | | DQN | HCC | 10.195 | 10.365 | -0.169 | 0.609 | -0.278 | 0.900 | -0.031 | |
| | | | DQN | MCC | 10.195 | 11.972 | -1.777 | 0.686 | -2.590 | 0.073 | -0.324 | |
| | | | DQN | PPO | 10.195 | 10.183 | 0.013 | 0.631 | 0.020 | 0.900 | 0.002 | |
| | | | HCC | MCC | 10.365 | 11.972 | -1.608 | 0.666 | -2.415 | 0.113 | -0.293 | |
| | | | HCC | PPO | 10.365 | 10.183 | 0.182 | 0.609 | 0.299 | 0.900 | 0.033 | |
| | | | MCC | PPO | 11.972 | 10.183 | 1.790 | 0.686 | 2.609 | 0.070 | 0.327 | |
| | | 6-20 Minutes | A2C | DQN | 11.266 | 9.629 | 1.637 | 0.589 | 2.777 | 0.044 | 0.226 | |
| | | | A2C | HCC | 11.266 | 7.444 | 3.823 | 0.569 | 6.721 | 0.001 | 0.529 | |
| | | | A2C | MCC | 11.266 | 9.645 | 1.621 | 0.641 | 2.530 | 0.085 | 0.224 | |
| | | | A2C | PPO | 11.266 | 9.793 | 1.474 | 0.589 | 2.500 | 0.091 | 0.204 | |
| | | | DQN | HCC | 9.629 | 7.444 | 2.186 | 0.569 | 3.843 | 0.001 | 0.302 | |
| | | | DQN | MCC | 9.629 | 9.645 | -0.016 | 0.641 | -0.025 | 0.900 | -0.002 | |
| | | | DQN | PPO | 9.629 | 9.793 | -0.163 | 0.589 | -0.277 | 0.900 | -0.023 | |
| | | | HCC | MCC | 7.444 | 9.645 | -2.202 | 0.622 | -3.540 | 0.004 | -0.305 | |
| | | | HCC | PPO | 7.444 | 9.793 | -2.349 | 0.569 | -4.130 | 0.001 | -0.325 | |

| | | | | | | | | | | | |
|---|---|---|---|---|---|---|---|---|---|---|---|
| | | | MCC | PPO | 9.645 | 9.793 | -0.147 | 0.641 | -0.230 | 0.900 | -0.020 | |
| p | Hit Counts | 0-5 Minutes | A2C | DQN | 0.722 | 0.719 | 0.003 | 0.027 | 0.124 | 0.900 | 0.007 | Tukey's |
| | | | A2C | HCC | 0.722 | 0.651 | 0.072 | 0.026 | 2.773 | 0.044 | 0.141 | |
| | | | A2C | MCC | 0.722 | 0.716 | 0.006 | 0.029 | 0.217 | 0.900 | 0.012 | |
| | | | A2C | PPO | 0.722 | 0.740 | -0.018 | 0.027 | -0.644 | 0.900 | -0.035 | |
| | | | DQN | HCC | 0.719 | 0.651 | 0.068 | 0.026 | 2.648 | 0.062 | 0.135 | |
| | | | DQN | MCC | 0.719 | 0.716 | 0.003 | 0.029 | 0.101 | 0.900 | 0.006 | |
| | | | DQN | PPO | 0.719 | 0.740 | -0.021 | 0.027 | -0.769 | 0.900 | -0.041 | |
| | | | HCC | MCC | 0.651 | 0.716 | -0.065 | 0.028 | -2.346 | 0.131 | -0.129 | |
| | | | HCC | PPO | 0.651 | 0.740 | -0.089 | 0.026 | -3.444 | 0.005 | -0.176 | |
| | | | MCC | PPO | 0.716 | 0.740 | -0.024 | 0.029 | -0.819 | 0.900 | -0.047 | |
| | | 6-20 Minutes | A2C | DQN | 0.724 | 0.741 | -0.017 | 0.018 | -0.947 | 0.868 | -0.029 | |
| | | | A2C | HCC | 0.724 | 0.854 | -0.131 | 0.017 | -7.488 | 0.001 | -0.220 | |
| | | | A2C | MCC | 0.724 | 0.852 | -0.129 | 0.020 | -6.378 | 0.001 | -0.217 | |
| | | | A2C | PPO | 0.724 | 0.727 | -0.004 | 0.018 | -0.218 | 0.900 | -0.007 | |
| | | | DQN | HCC | 0.741 | 0.854 | -0.114 | 0.017 | -6.567 | 0.001 | -0.192 | |
| | | | DQN | MCC | 0.741 | 0.852 | -0.112 | 0.020 | -5.570 | 0.001 | -0.189 | |
| | | | DQN | PPO | 0.741 | 0.727 | 0.013 | 0.018 | 0.727 | 0.900 | 0.022 | |
| | | | HCC | MCC | 0.854 | 0.852 | 0.002 | 0.020 | 0.100 | 0.900 | 0.003 | |
| | | | HCC | PPO | 0.854 | 0.727 | 0.127 | 0.017 | 7.259 | 0.001 | 0.214 | |
| | | | MCC | PPO | 0.852 | 0.727 | 0.125 | 0.020 | 6.181 | 0.001 | 0.211 | |
| q | % Aces | 0-5 Minutes | A2C | DQN | 51.318 | 53.675 | -2.356 | 1.437 | -1.640 | 0.473 | -0.189 | Tukey's |
| | | | A2C | HCC | 51.318 | 54.382 | -3.064 | 1.387 | -2.209 | 0.177 | -0.246 | |
| | | | A2C | MCC | 51.318 | 53.333 | -2.014 | 1.562 | -1.289 | 0.674 | -0.161 | |
| | | | A2C | PPO | 51.318 | 50.866 | 0.453 | 1.437 | 0.315 | 0.900 | 0.036 | |
| | | | DQN | HCC | 53.675 | 54.382 | -0.707 | 1.387 | -0.510 | 0.900 | -0.057 | |
| | | | DQN | MCC | 53.675 | 53.333 | 0.342 | 1.562 | 0.219 | 0.900 | 0.027 | |
| | | | DQN | PPO | 53.675 | 50.866 | 2.809 | 1.437 | 1.955 | 0.290 | 0.225 | |
| | | | HCC | MCC | 54.382 | 53.333 | 1.049 | 1.516 | 0.692 | 0.900 | 0.084 | |
| | | | HCC | PPO | 54.382 | 50.866 | 3.516 | 1.387 | 2.536 | 0.084 | 0.282 | |
| | | | MCC | PPO | 53.333 | 50.866 | 2.467 | 1.562 | 1.579 | 0.510 | 0.198 | |
| | | 6-20 Minutes | A2C | DQN | 52.596 | 51.199 | 1.397 | 0.907 | 1.540 | 0.532 | 0.177 | |
| | | | A2C | HCC | 52.596 | 49.259 | 3.337 | 0.875 | 3.813 | 0.001 | 0.424 | |
| | | | A2C | MCC | 52.596 | 50.232 | 2.364 | 0.986 | 2.398 | 0.117 | 0.300 | |
| | | | A2C | PPO | 52.596 | 51.658 | 0.938 | 0.907 | 1.034 | 0.818 | 0.119 | |
| | | | DQN | HCC | 51.199 | 49.259 | 1.940 | 0.875 | 2.217 | 0.175 | 0.246 | |
| | | | DQN | MCC | 51.199 | 50.232 | 0.968 | 0.986 | 0.981 | 0.848 | 0.123 | |
| | | | DQN | PPO | 51.199 | 51.658 | -0.459 | 0.907 | -0.506 | 0.900 | -0.058 | |
| | | | HCC | MCC | 49.259 | 50.232 | -0.973 | 0.957 | -1.017 | 0.828 | -0.124 | |
| | | | HCC | PPO | 49.259 | 51.658 | -2.399 | 0.875 | -2.742 | 0.049 | -0.305 | |
| | | | MCC | PPO | 50.232 | 51.658 | -1.427 | 0.986 | -1.447 | 0.584 | -0.181 | |
| r | % Long Rally | 0-5 Minutes | A2C | DQN | 9.519 | 9.710 | -0.191 | 0.965 | -0.198 | 0.900 | -0.023 | Tukey's |
| | | | A2C | HCC | 9.519 | 4.523 | 4.997 | 0.931 | 5.366 | 0.001 | 0.596 | |
| | | | A2C | MCC | 9.519 | 7.318 | 2.201 | 1.049 | 2.098 | 0.222 | 0.263 | |
| | | | A2C | PPO | 9.519 | 10.462 | -0.942 | 0.965 | -0.976 | 0.851 | -0.112 | |

| | | | | | | | | | | | | | |
|---|---|---|---|---|---|---|---|---|---|---|---|---|---|
| | | | | DQN | HCC | 9.710 | 4.523 | 5.188 | 0.931 | 5.571 | 0.001 | 0.619 | |
| | | | | DQN | MCC | 9.710 | 7.318 | 2.392 | 1.049 | 2.280 | 0.153 | 0.285 | |
| | | | | DQN | PPO | 9.710 | 10.462 | -0.752 | 0.965 | -0.779 | 0.900 | -0.090 | |
| | | | | HCC | MCC | 4.523 | 7.318 | -2.796 | 1.018 | -2.746 | 0.048 | -0.334 | |
| | | | | HCC | PPO | 4.523 | 10.462 | -5.939 | 0.931 | -6.378 | 0.001 | -0.709 | |
| | | | | MCC | PPO | 7.318 | 10.462 | -3.144 | 1.049 | -2.996 | 0.024 | -0.375 | |
| | | | 6-20 Minutes | A2C | DQN | 10.431 | 10.187 | 0.244 | 0.591 | 0.413 | 0.900 | 0.048 | |
| | | | | A2C | HCC | 10.431 | 10.365 | 0.066 | 0.570 | 0.116 | 0.900 | 0.013 | |
| | | | | A2C | MCC | 10.431 | 11.972 | -1.541 | 0.642 | -2.400 | 0.117 | -0.300 | |
| | | | | A2C | PPO | 10.431 | 10.049 | 0.382 | 0.591 | 0.646 | 0.900 | 0.074 | |
| | | | | DQN | HCC | 10.187 | 10.365 | -0.178 | 0.570 | -0.312 | 0.900 | -0.035 | |
| | | | | DQN | MCC | 10.187 | 11.972 | -1.785 | 0.642 | -2.780 | 0.044 | -0.348 | |
| | | | | DQN | PPO | 10.187 | 10.049 | 0.138 | 0.591 | 0.233 | 0.900 | 0.027 | |
| | | | | HCC | MCC | 10.365 | 11.972 | -1.608 | 0.623 | -2.579 | 0.075 | -0.313 | |
| | | | | HCC | PPO | 10.365 | 10.049 | 0.316 | 0.570 | 0.554 | 0.900 | 0.062 | |
| | | | | MCC | PPO | 11.972 | 10.049 | 1.923 | 0.642 | 2.994 | 0.024 | 0.375 | |
| S7 | a | Relative improvement (%) in the average hit counts – Ball Position Input | | A2C | DQN | 33.724 | 28.251 | 5.473 | 8.669 | 283.806 | 0.900 | 0.073 | Games Howell |
| | | | | A2C | HCC | 33.724 | 82.147 | -48.423 | 10.077 | 321.871 | 0.001 | -0.534 | |
| | | | | A2C | MCC | 33.724 | 50.755 | -17.031 | 10.274 | 238.311 | 0.464 | -0.207 | |
| | | | | A2C | PPO | 33.724 | 33.016 | 0.709 | 10.301 | 292.792 | 0.900 | 0.008 | |
| | | | | DQN | HCC | 28.251 | 82.147 | -53.896 | 9.206 | 304.784 | 0.001 | -0.651 | |
| | | | | DQN | MCC | 28.251 | 50.755 | -22.505 | 9.421 | 205.771 | 0.123 | -0.299 | |
| | | | | DQN | PPO | 28.251 | 33.016 | -4.765 | 9.450 | 266.029 | 0.900 | -0.058 | |
| | | | | HCC | MCC | 82.147 | 50.755 | 31.391 | 10.731 | 262.994 | 0.030 | 0.355 | |
| | | | | HCC | PPO | 82.147 | 33.016 | 49.131 | 10.756 | 317.852 | 0.001 | 0.508 | |
| | | | | MCC | PPO | 50.755 | 33.016 | 17.740 | 10.941 | 252.147 | 0.486 | 0.203 | |
| | b | Relative improvement (%) in the average hit counts – Paddle&Ball Position Input | | A2C | DQN | 21.717 | 24.949 | -3.232 | 8.194 | 291.151 | 0.900 | -0.045 | Games Howell |
| | | | | A2C | HCC | 21.717 | 82.147 | -60.429 | 9.165 | 303.151 | 0.001 | -0.733 | |
| | | | | A2C | MCC | 21.717 | 50.755 | -29.038 | 9.381 | 203.860 | 0.019 | -0.387 | |
| | | | | A2C | PPO | 21.717 | 14.690 | 7.027 | 7.082 | 292.773 | 0.842 | 0.114 | |
| | | | | DQN | HCC | 24.949 | 82.147 | -57.197 | 9.711 | 318.526 | 0.001 | -0.655 | |
| | | | | DQN | MCC | 24.949 | 50.755 | -25.806 | 9.915 | 226.675 | 0.073 | -0.326 | |
| | | | | DQN | PPO | 24.949 | 14.690 | 10.259 | 7.775 | 276.159 | 0.657 | 0.152 | |
| | | | | HCC | MCC | 82.147 | 50.755 | 31.391 | 10.731 | 262.994 | 0.030 | 0.355 | |

| | | | | | | | | | | | | | |
|---|---|---|---|---|---|---|---|---|---|---|---|---|---|
| | | | | HCC | PPO | 82.147 | 14.690 | 67.456 | 8.792 | 284.259 | 0.001 | 0.853 | |
| | | | | MCC | PPO | 50.755 | 14.690 | 36.065 | 9.017 | 184.981 | 0.001 | 0.501 | |
| S8 | e | Hit Counts | 0-5 Minutes | CL(3) | CL(7) | 0.696 | 0.682 | 0.014 | 0.050 | 0.281 | 0.900 | 0.027 | Tuckey's |
| | | | | CL(3) | HCC | 0.696 | 0.651 | 0.045 | 0.039 | 1.147 | 0.641 | 0.088 | |
| | | | | CL(3) | MCC | 0.696 | 0.716 | -0.020 | 0.042 | -0.484 | 0.900 | -0.039 | |
| | | | | CL(7) | HCC | 0.682 | 0.651 | 0.031 | 0.039 | 0.804 | 0.834 | 0.061 | |
| | | | | CL(7) | MCC | 0.682 | 0.716 | -0.034 | 0.041 | -0.827 | 0.821 | -0.066 | |
| | | | | HCC | MCC | 0.651 | 0.716 | -0.065 | 0.028 | -2.318 | 0.094 | -0.127 | |
| | | | 6-20 Minutes | CL(3) | CL(7) | 0.703 | 0.916 | -0.213 | 0.039 | -5.439 | 0.001 | -0.336 | |
| | | | | CL(3) | HCC | 0.703 | 0.854 | -0.151 | 0.030 | -4.972 | 0.001 | -0.239 | |
| | | | | CL(3) | MCC | 0.703 | 0.852 | -0.149 | 0.032 | -4.624 | 0.001 | -0.236 | |
| | | | | CL(7) | HCC | 0.916 | 0.854 | 0.061 | 0.030 | 2.017 | 0.182 | 0.097 | |
| | | | | CL(7) | MCC | 0.916 | 0.852 | 0.063 | 0.032 | 1.962 | 0.203 | 0.100 | |
| | | | | HCC | MCC | 0.854 | 0.852 | 0.002 | 0.021 | 0.094 | 0.900 | 0.003 | |
| | f | %Aces | 0-5 Minutes | CL(3) | CL(7) | 53.140 | 54.239 | -1.099 | 2.560 | -0.429 | 0.900 | -0.095 | |
| | | | | CL(3) | HCC | 53.140 | 54.382 | -1.242 | 2.008 | -0.618 | 0.900 | -0.108 | |
| | | | | CL(3) | MCC | 53.140 | 53.333 | -0.192 | 2.114 | -0.091 | 0.900 | -0.017 | |
| | | | | CL(7) | HCC | 54.239 | 54.382 | -0.143 | 2.008 | -0.071 | 0.900 | -0.012 | |
| | | | | CL(7) | MCC | 54.239 | 53.333 | 0.906 | 2.114 | 0.429 | 0.900 | 0.079 | |
| | | | | HCC | MCC | 54.382 | 53.333 | 1.049 | 1.395 | 0.752 | 0.863 | 0.091 | |
| | | | 6-20 Minutes | CL(3) | CL(7) | 55.605 | 47.256 | 8.349 | 1.712 | 4.876 | 0.001 | 1.080 | |
| | | | | CL(3) | HCC | 55.605 | 49.259 | 6.346 | 1.343 | 4.726 | 0.001 | 0.826 | |
| | | | | CL(3) | MCC | 55.605 | 50.232 | 5.373 | 1.414 | 3.800 | 0.001 | 0.698 | |
| | | | | CL(7) | HCC | 47.256 | 49.259 | -2.003 | 1.343 | -1.492 | 0.445 | -0.261 | |
| | | | | CL(7) | MCC | 47.256 | 50.232 | -2.976 | 1.414 | -2.105 | 0.154 | -0.387 | |
| | | | | HCC | MCC | 49.259 | 50.232 | -0.973 | 0.933 | -1.043 | 0.700 | -0.127 | |
| | g | %Long Rally | 0-5 Minutes | CL(3) | CL(7) | 7.692 | 6.923 | 0.769 | 1.432 | 0.537 | 0.900 | 0.119 | |
| | | | | CL(3) | HCC | 7.692 | 4.523 | 3.170 | 1.123 | 2.822 | 0.026 | 0.493 | |
| | | | | CL(3) | MCC | 7.692 | 7.318 | 0.374 | 1.183 | 0.316 | 0.900 | 0.058 | |
| | | | | CL(7) | HCC | 6.923 | 4.523 | 2.401 | 1.123 | 2.138 | 0.143 | 0.373 | |
| | | | | CL(7) | MCC | 6.923 | 7.318 | -0.395 | 1.183 | -0.334 | 0.900 | -0.061 | |
| | | | | HCC | MCC | 4.523 | 7.318 | -2.796 | 0.780 | -3.583 | 0.002 | -0.435 | |
| | | | 6-20 Minutes | CL(3) | CL(7) | 9.292 | 12.160 | -2.868 | 1.111 | -2.582 | 0.050 | -0.572 | |
| | | | | CL(3) | HCC | 9.292 | 10.365 | -1.073 | 0.871 | -1.232 | 0.594 | -0.215 | |
| | | | | CL(3) | MCC | 9.292 | 11.972 | -2.680 | 0.917 | -2.923 | 0.019 | -0.537 | |
| | | | | CL(7) | HCC | 12.160 | 10.365 | 1.795 | 0.871 | 2.061 | 0.168 | 0.360 | |
| | | | | CL(7) | MCC | 12.160 | 11.972 | 0.187 | 0.917 | 0.204 | 0.900 | 0.038 | |
| | | | | HCC | MCC | 10.365 | 11.972 | -1.608 | 0.605 | -2.657 | 0.041 | -0.323 | |
| | h | Relative improvement (%) in the average hit counts – Active Inference | | CL(3) | CL(7) | 20.341 | 54.109 | -33.768 | 15.953 | -2.117 | 0.157 | -0.469 | Games Howell |
| | | | | CL(3) | HCC | 20.341 | 82.147 | -61.806 | 14.023 | -4.407 | 0.001 | -0.770 | |
| | | | | CL(3) | MCC | 20.341 | 50.755 | -30.414 | 14.165 | -2.147 | 0.148 | -0.394 | |
| | | | | CL(7) | HCC | 54.109 | 82.147 | -28.038 | 13.000 | -2.157 | 0.144 | -0.377 | |

| | | | CL(7) | MCC | 54.109 | 50.755 | 3.353 | 13.154 | 0.255 | 0.900 | 0.047 | |
| | | | HCC | MCC | 82.147 | 50.755 | 31.391 | 10.731 | 2.925 | 0.019 | 0.355 | |

**Table S3.** Multivariate statistical tests and all results for tests done.

| Figure | Panel | Parameters | Source | DF1 | DF2 | MS | F | p-value | np2 | Method |
|--------|-------|------------|--------|-----|-----|----|----|---------|-----|--------|
| **2** | e | Average Rally Length | Group - all | 4 | 729 | 0.185 | 1.021 | 0.395 | 0.006 | Mixed ANOVA |
| | | | Time Interval - all | 1 | 729 | 2.134 | -21.944 | 1.000 | -0.031 | |
| | | | Interaction - all | 4 | 729 | 0.575 | -5.909 | 1.000 | -0.034 | |
| | f | % Aces | Group - all | 4 | 729 | 0.044 | 1.014 | 0.399 | 0.006 | Mixed ANOVA |
| | | | Time Interval - all | 1 | 729 | 0.124 | -5.589 | 1.000 | -0.008 | |
| | | | Interaction - all | 4 | 729 | 0.015 | -0.685 | 1.000 | -0.004 | |
| | g | % Long Rally | Group - all | 4 | 729 | 0.019 | 1.749 | 0.137 | 0.010 | Mixed ANOVA |
| | | | Time Interval - all | 1 | 729 | 0.063 | -11.125 | 1.000 | -0.015 | |
| | | | Interaction - all | 4 | 729 | 0.039 | -6.931 | 1.000 | -0.040 | |
| **3** | e | Average Rally Length | Group - all | 4 | 729 | 0.170 | 0.926 | 0.448 | 0.005 | Mixed ANOVA |
| | | | Time Intervals - all | 1 | 729 | 1.488 | -15.161 | 1.000 | -0.021 | |
| | | | Interaction - all | 4 | 729 | 0.704 | -7.170 | 1.000 | -0.041 | |
| | f | % Aces | Group - all | 4 | 729 | 0.061 | 1.332 | 0.256 | 0.007 | Mixed ANOVA |
| | | | Time Intervals - all | 1 | 729 | 0.022 | -0.957 | 1.000 | -0.001 | |
| | | | Interaction - all | 4 | 729 | 0.041 | -1.745 | 1.000 | -0.010 | |
| | g | % Long Rally | Group - all | 4 | 729 | 0.011 | 0.886 | 0.472 | 0.005 | Mixed ANOVA |
| | | | Time Intervals - all | 1 | 729 | 0.073 | -11.249 | 1.000 | -0.016 | |
| | | | Interaction - all | 4 | 729 | 0.033 | -5.038 | 1.000 | -0.028 | |
| **4** | e | Average Rally Length | Group - all | 4 | 729 | 0.499 | 2.589 | 0.036 | 0.014 | Mixed ANOVA |
| | | | Time Intervals- all | 1 | 729 | 1.934 | -18.645 | 1.000 | -0.026 | |
| | | | Interaction - all | 4 | 729 | 0.599 | -5.774 | 1.000 | -0.033 | |
| | f | % Aces | Group - all | 4 | 729 | 0.111 | 2.331 | 0.055 | 0.013 | Mixed ANOVA |
| | | | Time Intervals- all | 1 | 729 | 0.111 | -4.583 | 1.000 | -0.006 | |
| | | | Interaction - all | 4 | 729 | 0.021 | -0.871 | 1.000 | -0.005 | |
| | g | % Long Rally | Group - all | 4 | 729 | 0.018 | 1.523 | 0.194 | 0.008 | Mixed ANOVA |
| | | | Time Intervals - all | 1 | 729 | 0.081 | -12.847 | 1.000 | -0.018 | |
| | | | Interaction - all | 4 | 729 | 0.032 | -5.057 | 1.000 | -0.029 | |
| **S3** | d | Average Rally Length | Group - all | 5 | 478 | 1.645 | 8.293 | 0.0 | 0.080 | Mixed ANOVA |
| | | | Time Intervals - all | 1 | 478 | 2.153 | -18.414 | 1.0 | -0.040 | |
| | | | Interaction - all | 5 | 478 | 0.443 | -3.787 | 1.0 | -0.041 | |
| | e | % Aces | Group - all | 5 | 478 | 0.259 | 5.194 | 0.0 | 0.052 | Mixed ANOVA |
| | | | Time Intervals - all | 1 | 478 | 0.059 | -2.212 | 1.0 | -0.005 | |
| | | | Interaction - all | 5 | 478 | 0.027 | -1.008 | 1.0 | -0.011 | |

| | | | | | | | | | |
|---|---|---|---|---|---|---|---|---|---|
| | f | % Long Rally | Group - all | 5 | 478 | 0.049 | 4.611 | 0.0 | 0.046 | Mixed ANOVA |
| | | | Time Intervals - all | 1 | 478 | 0.113 | -18.808 | 1.0 | -0.041 | |
| | | | Interaction - all | 5 | 478 | 0.019 | -3.197 | 1.0 | -0.035 | |
| **S4** | d | Average Rally Length | Group - all | 5 | 478 | 0.765 | 4.206 | 0.001 | 0.042 | Mixed ANOVA |
| | | | Time Intervals - all | 1 | 478 | 1.873 | -17.980 | 1.000 | -0.039 | |
| | | | Interaction - all | 5 | 478 | 0.502 | -4.819 | 1.000 | -0.053 | |
| | e | % Aces | Group - all | 5 | 478 | 0.060 | 1.410 | 0.219 | 0.015 | Mixed ANOVA |
| | | | Time Intervals - all | 1 | 478 | 0.050 | -2.277 | 1.000 | -0.005 | |
| | | | Interaction - all | 5 | 478 | 0.029 | -1.306 | 1.000 | -0.014 | |
| | f | % Long Rally | Group - all | 5 | 478 | 0.032 | 2.926 | 0.013 | 0.030 | Mixed ANOVA |
| | | | Time Intervals - all | 1 | 478 | 0.081 | -13.550 | 1.000 | -0.029 | |
| | | | Interaction - all | 5 | 478 | 0.026 | -4.281 | 1.000 | -0.047 | |
| **S5** | d | Average Rally Length | Group - all | 5 | 478 | 2.177 | 10.721 | 0.0 | 0.101 | Mixed ANOVA |
| | | | Time Intervals - all | 1 | 478 | 1.503 | -12.236 | 1.0 | -0.026 | |
| | | | Interaction - all | 5 | 478 | 0.645 | -5.254 | 1.0 | -0.058 | |
| | e | % Aces | Group - all | 5 | 478 | 0.421 | 7.738 | 0.0 | 0.075 | Mixed ANOVA |
| | | | Time Intervals - all | 1 | 478 | 0.029 | -0.970 | 1.0 | -0.002 | |
| | | | Interaction - all | 5 | 478 | 0.046 | -1.526 | 1.0 | -0.016 | |
| | f | % Long Rally | Group - all | 5 | 478 | 0.046 | 4.651 | 0.0 | 0.046 | Mixed ANOVA |
| | | | Time Intervals - all | 1 | 478 | 0.095 | -16.734 | 1.0 | -0.036 | |
| | | | Interaction - all | 5 | 478 | 0.025 | -4.406 | 1.0 | -0.048 | |
| **S6** | d | Average Rally Length | Group - all | 4 | 729 | 0.260 | 1.372 | 0.242 | 0.007 | Mixed ANOVA |
| | | | Time Intervals - all | 1 | 729 | 2.355 | -23.160 | 1.000 | -0.033 | |
| | | | Interaction - all | 4 | 729 | 0.525 | -5.161 | 1.000 | -0.029 | |
| | e | % Aces | Group - all | 4 | 729 | 0.066 | 1.445 | 0.217 | 0.008 | Mixed ANOVA |
| | | | Time Intervals - all | 1 | 729 | 0.141 | -5.986 | 1.000 | -0.008 | |
| | | | Interaction - all | 4 | 729 | 0.021 | -0.897 | 1.000 | -0.005 | |
| | f | % Long Rally | Group - all | 4 | 729 | 0.017 | 1.370 | 0.243 | 0.007 | Mixed ANOVA |
| | | | Time Intervals - all | 1 | 729 | 0.113 | -17.552 | 1.000 | -0.025 | |
| | | | Interaction - all | 4 | 729 | 0.024 | -3.684 | 1.000 | -0.021 | |
| | p | Average Rally Length | Group - all | 4 | 729 | 0.136 | 0.756 | 0.554 | 0.004 | Mixed ANOVA |
| | | | Time Intervals - all | 1 | 729 | 1.690 | -17.577 | 1.000 | -0.025 | |
| | | | Interaction - all | 4 | 729 | 0.663 | -6.889 | 1.000 | -0.039 | |
| | q | % Aces | Group - all | 4 | 729 | 0.032 | 0.712 | 0.584 | 0.004 | Mixed ANOVA |
| | | | Time Intervals - all | 1 | 729 | 0.054 | -2.376 | 1.000 | -0.003 | |
| | | | Interaction - all | 4 | 729 | 0.042 | -1.838 | 1.000 | -0.010 | |
| | r | % Long Rally | Group - all | 4 | 729 | 0.009 | 0.763 | 0.55 | 0.004 | Mixed ANOVA |
| | | | Time Intervals - all | 1 | 729 | 0.073 | -11.682 | 1.00 | -0.016 | |

| Figure | Panel | Parameters | Source | DF | | MS | F | p-value | np2 | Method |
|---|---|---|---|---|---|---|---|---|---|---|
| | | | Interaction - all | 4 | 729 | 0.032 | -5.152 | 1.00 | -0.029 | |
| **S8** | a | Average Rally Length | Group - all | 3 | 360 | 0.160 | 0.792 | 0.499 | 0.007 | Mixed ANOVA |
| | | | Time Intervals - all | 1 | 360 | 4.486 | -38.506 | 1.000 | -0.120 | |
| | | | Interaction - all | 3 | 360 | 0.286 | -2.454 | 1.000 | -0.021 | |
| | b | % Aces | Group - all | 3 | 360 | 0.033 | 0.844 | 0.471 | 0.007 | Mixed ANOVA |
| | | | Time Intervals - all | 1 | 360 | 0.162 | -7.936 | 1.000 | -0.023 | |
| | | | Interaction - all | 3 | 360 | 0.031 | -1.503 | 1.000 | -0.013 | |
| | c | % Long Rally | Group - all | 3 | 360 | 0.012 | 1.004 | 0.391 | 0.008 | Mixed ANOVA |
| | | | Time Intervals - all | 1 | 360 | 0.234 | -36.162 | 1.000 | -0.112 | |
| | | | Interaction - all | 3 | 360 | 0.003 | -0.517 | 1.000 | -0.004 | |

**Table S4.** Multivariate statistical tests and all results for tests done.

| Figure | Panel | Parameters | Source | DF | MS | F | p-value | np2 | Method |
|---|---|---|---|---|---|---|---|---|---|
| **5** | a | Average Paddle Movement | Group - all | 4 | 1.064e+10 | 21.837 | 0.0 | 0.155 | ANOVA |
| | b | Relative improvement (%) in the average hit counts | Group - all | 4 | 104528.369 | 17.807 | 0.0 | 0.089 | ANOVA |
| | c | Average Paddle Movement | Group - all | 4 | 1.801e+10 | 49.523 | 0.0 | 0.293 | ANOVA |
| | d | Relative improvement (%) in the average hit counts | Group - all | 4 | 116698.296 | 16.243 | 0.0 | 0.082 | ANOVA |
| | e | Average Paddle Movement | Group - all | 4 | 1.009e+10 | 26.881 | 0.0 | 0.184 | ANOVA |
| | f | Relative improvement (%) in the average hit counts | Group - all | 4 | 79671.720 | 9.889 | 0.0 | 0.051 | ANOVA |
| **6** | a | Relative improvement (%) in the average hit counts - DQN | Group - all | 5 | 77257.903 | 10.241 | 0.0 | 0.097 | ANOVA |
| | b | Relative improvement (%) in the average hit counts – A2C | Group - all | 5 | 73239.513 | 11.211 | 0.0 | 0.105 | ANOVA |
| | c | Relative improvement (%) in the average hit counts - PPO | Group - all | 5 | 83698.926 | 9.517 | 0.0 | 0.091 | ANOVA |
| **S7** | a | Relative improvement (%) in the average hit counts - Ball Position Input | Group - all | 4 | 81200.989 | 10.941 | 0.0 | 0.057 | ANOVA |
| | b | Relative improvement (%) in the average hit counts - | Group - all | 4 | 125476.158 | 20.915 | 0.0 | 0.103 | ANOVA |

| | | | | | | | | | |
|---|---|---|---|---|---|---|---|---|---|
| | | Paddle&Ball Position Input | | | | | | | |
| **S8** | d | Relative improvement (%) in the average hit counts – Active Inference | Group - all | 3 | 52072.238 | 6.733 | 0.0 | 0.053 | ANOVA |

}
%   \label{extra_stats}
% \end{figure*}

\end{appendices}




% \end{thebibliography}

\end{document}